\documentclass[fleqn,usenatbib]{mnras}
\usepackage[T1]{fontenc}
\usepackage{ae,aecompl}

\usepackage{graphicx}
\usepackage{amsmath}
\usepackage{amssymb,bm}
\usepackage{tabularx}
\usepackage{longtable}
\usepackage[table]{xcolor}
\usepackage{booktabs}
\hypersetup{final}
\usepackage[normalem]{ulem}
\usepackage{subcaption}
\usepackage{mwe}

\newcommand{\Change}[1]{{\textcolor{black}{#1}}}
\newcommand{\smallspace}{\vspace{-.35cm}}

\newcommand{\eagle}{\textsc{eagle}}
\newcommand{\auriga}{\textsc{auriga}}
\newcommand{\agama}{\textsc{agama}}

\newcommand{\kpc}{\ensuremath{~\text{kpc}}}

\newcommand{\Msun}{\ensuremath{~\text{M}_\odot}}

\newcommand{\uJ}{\ensuremath{\,\mathrm{kpc}\,\mathrm{km}/\mathrm{s}}}

\newcommand{\uEe}{\ensuremath{10^{5}\,\mathrm{km}^2/\mathrm{s}^2}}

\newcommand{\J}{\ensuremath{\bf{J}}}

\newcommand{\JR}{\ensuremath{J_{\textnormal{R}}}}
\newcommand{\Jz}{\ensuremath{J_{\textnormal{z}}}}
\newcommand{\Lz}{\ensuremath{L_{\textnormal{z}}}}
\newcommand{\FJ}{\ensuremath{F\left(\mathbf{J}\right)}}
\newcommand{\Met}{\ensuremath{[\mathrm{Fe}/\mathrm{H}}]}
\newcommand{\pyield}{\ensuremath{p_{\mathrm{yield}}}}
\newcommand{\EJ}{\ensuremath{\left(E,\bf{J}\right)}}

\newcommand{\FBulge}{\ensuremath{F_{\mathrm{Bulge}}\left(\mathbf{J}\right)}}
\newcommand{\FDisc}{\ensuremath{F_{\mathrm{Disc}}\left(\mathbf{J}\right)}}

\newcommand{\FComX}{\ensuremath{F_{\mathrm{c}}\left(\mathbf{X}\right)}}

\newcommand{\FX}{\ensuremath{F\left(\mathbf{X}\right)}}

\newcommand{\X}{\ensuremath{\boldsymbol{X}}}

\newcommand{\OmC}{\ensuremath{\omega}-Centauri}
\newcommand{\MhNgc}{\ensuremath{M_{H}-N_{GC}}}
\newcommand{\Ngc}{\ensuremath{N_{GC}}}
\newcommand{\GES}{GES}

\newcommand{\Nfit}{\ensuremath{N_{\mathrm{Fit}}}}
\newcommand{\Ntrue}{\ensuremath{N_{\mathrm{True}}}}
\newcommand{\Nest}{\ensuremath{N_{\mathrm{Est}}}}

\defcitealias{CallinghamMass2019MNRAS.484.5453C}{Callingham19}
\defcitealias{CautunContract}{Cautun19}
\defcitealias{gaia_collaboration_gaia_2021}{Gaia Collaboration et al., 2021b}

\usepackage{ifthen}
\usepackage{helvet}
\makeatletter
\newcommand{\showfont}{encoding: \f@encoding{},
  family: \f@family{},
  series: \f@series{},
  shape: \f@shape{},
  size: \f@size{}
}
\newcommand{\iffont}[3]{\ifthenelse{\equal{\f@family}{#1}}{#2}{#3}}
\makeatother

\defcitealias{massari_origin_2019}{Massari19}
\defcitealias{forbes_reverse_2020}{Forbes20}
\defcitealias{horta_chemical_2020}{Horta20}
\newcommand{\massari}{\citetalias{massari_origin_2019}}
\newcommand{\forbes}{\citetalias{forbes_reverse_2020}}

\title{The chemo-dynamical groups of Galactic globular clusters}

\author[T. M. Callingham et al.]
{\parbox{\textwidth}{
Thomas M. Callingham,$^{1,2}$\thanks{E-mail: t.m.callingham@astro.rug.nl}
Marius Cautun,$^{3}$
Alis J. Deason,$^{1}$
Carlos S. Frenk,$^{1}$
Robert J.J Grand,$^{4,5}$
Federico Marinacci$^{6}$
\vspace{.20cm}} \\
$^{1}$Institute of Computational Cosmology, Department of Physics, University of Durham, South Road, Durham DH1 3LE, UK\\
$^{2}$Kapteyn Astronomical Institute, University of Groningen, Landeleven 12, 9747 AD Groningen, The Netherlands\\
$^{3}$Leiden Observatory, Leiden University, PO Box 9513, NL-2300 RA Leiden, the Netherlands\\
$^4$Instituto de Astrof\'isica de Canarias, Calle Vía L\'actea s/n, E-38205 La Laguna, Tenerife, Spain\\
$^5$Departamento de Astrof\'isica, Universidad de La Laguna, Av. del Astrof\'isico Francisco S\'anchez s/n, E-38206, La Laguna, Tenerife, Spain\\
$^6$Department of Physics \& Astronomy ``Augusto Righti'', University of Bologna, via Gobetti 93/2, 1-40129 Bologna, Italy
\vspace{.10cm}}

\pubyear{2022}

\begin{document}
\label{firstpage}
\pagerange{\pageref{firstpage}--\pageref{lastpage}}
\maketitle

\begin{abstract}
  We introduce a multi-component chemo-dynamical method for splitting the Galactic population of Globular Clusters (GCs) into three distinct constituents: bulge, disc, and stellar halo. The latter is further decomposed into the individual large accretion events that built up the Galactic stellar halo: the \textit{Gaia}-Enceladus-Sausage, Kraken and Sequoia structures, and the Sagittarius and Helmi streams. Our modelling is extensively tested using mock GC samples constructed from the \auriga{} suite of hydrodynamical simulations of Milky Way (MW)-like galaxies. We find that, on average, a proportion of the accreted GCs cannot be associated with their true infall group  and are left ungrouped, biasing our recovered population numbers to $\sim80\%$ of their true value. Furthermore, the identified groups have a completeness and a purity of only $\sim65\%$. This reflects the difficulty of the problem, a result of the large degree of overlap in energy-action space of the debris from past accretion events. We apply the method to the Galactic data to infer, in a statistically robust and easily quantifiable way, the GCs associated with each MW accretion event. The resulting groups' population numbers of GCs, corrected for biases, are then used to infer the halo and stellar masses of the now  defunct satellites that built up the halo of the MW.

\end{abstract}




\section{Introduction}
\label{Sec:Intro}

Our stellar halo is a cosmic graveyard populated by the stars and globular clusters (GCs) that were once part of
now destroyed dwarf galaxies.
Halo assembly stems from hierarchical growth - the hallmark of the $\Lambda$CDM cosmological model \citep{davis_evolution_1985}
- whereby massive galaxies like the Milky Way (MW) evolve by devouring many lower mass galaxies,
whose remains are mixed and spread into the stellar halo \citep[e.g.][]{bullock_tracing_2005,cooper_galactic_2010}.
Unravelling this galactic debris to reconstruct the assembly history of the MW is a difficult undertaking as
ancient mergers have long since phase-mixed, effectively erasing information in physical space.
However, simulation-based studies have shown that debris from the same progenitor remains localized,
preserving structure in the space of the integrals of motion \citep[e.g.][]{gomez_identification_2010}.
Combined with stellar age and chemistry information, which also persists over time,
this raises the prospect that we may be able to reconstruct our Galaxy's past.

Of the accreted material in the stellar halo,
GCs have long been recognised as sensitive probes of the accretion history of the MW \citep{searle_composition_1978}.
Several GCs are suspected of being the remnant nucleus of accreted dwarf galaxies (M54, M4, \OmC, NGC1851),
directly showing where the cores of fallen progenitors came to rest.
Furthermore, while major mergers dominate the stellar halo
\citep{cooper_galactic_2010, deason_eating_2016,fattahi_tale_2020}
it has been shown that GCs are generally associated with smaller accretion events in the MW's past
\citep[e.g.][]{harris_dark_2015, amorisco_globular_2019}.
When studying the origin of the MW's GC system,
it is necessary to identify which of them were born natively in our Galaxy (in-situ GCs) and
which formed in dwarf galaxies and were later accreted.

On average, in the MW there is a rough trend for metal-poor GCs to be located at a larger radius,
while metal-rich GCs are more centrally concentrated \citep{frenk_kinematics_1980}.
However, this is not enough to distinguish populations by chemistry alone \citep{trujillo-gomez_kinematics_2020}.
With precise age and metallicity data now available for many GCs,
it has been shown that the MW GC's age-metallicity relation (AMR) contains two branches:
a metal-poor one characterised by halo-like kinematics, and a metal-rich one whose GCs orbit the inner Galaxy,
suggesting an in-situ origin \citep{forbes_accreted_2010, marin-franch_acs_2009,leaman_bifurcated_2013}.
This behaviour can be understood using simple models, such as a leaky-box chemical enrichment model,
in which the stellar birth environment in smaller dwarf galaxies is enriched more slowly than in larger galaxies such as our own.

The recent explosion of Galactic data, such as those from
the \textit{Gaia} mission \citep{gaia-collaboration_gaia_2018}, APOGEE \citep{majewski_apache_2017},
the H3 survey \citep{conroy_mapping_2019}, and GALAH \citep{martell_galah_2017}
have revolutionised the field of Galactic astronomy.
In particular, they have revealed evidence of an ancient major merger,
\textit{Gaia}-Enceladus-Sausage (\GES) \citep{belokurov_co-formation_2018, helmi_formation_2018}.
Combined with previous discoveries such as the stellar stream of the Sagittarius dwarf galaxy \citep{ibata_dwarf_1994}
and the Helmi stream \citep{helmi_debris_1999},
there is a wealth of known structures present in the Galactic stellar halo \citep{naidu_evidence_2020}.
Characterising the properties of the progenitors of these structures is challenging since their debris consists of
extended, diffuse stellar distributions.
One solution is to identify the GCs associated with these structures, since GCs are compact,
bright objects whose properties and orbits can be measured accurately.

Arguably the easiest accreted group to identify is the set of GCs that belonged to the Sagittarius dwarf galaxy
\citep{ibata_dwarf_1994} as it is currently being disrupted and its stars and GCs can be found as an identifiable stream
\citep[e.g.][]{bellazzini_globular_2020, antoja_all-sky_2020, law_assessing_2010,penarrubia_identification_2021}.
Identifying members of other stellar halo structures remains a challenging problem.
The works of \cite{myeong_halo_2018, myeong_sausage_2018} have associated GCs to the \GES{} debris,
which is characterised by highly radial orbits.
Likely members of the Helmi stream were identified by \cite{koppelman_characterization_2019} from their proximity to
selection cuts in the phase space of the stellar halo.
The retrograde accretion event dubbed ``Sequoia'' was, in part,
born out of studies of notable retrograde GCs such as FSR1758 and \OmC{}
\citep{myeong_discovery_2018, myeong_evidence_2019,barba_sequoia_2019}, with other GCs similarly associated.

The recent work by \cite{massari_origin_2019}, hereafter
\massari,
was a significant development in this field.
These authors used a sample of 160 Galactic GCs to identify the major GC groups.
They did so by defining selection boxes in energy and angular momentum space that are based on `known' accretion groups
and expanding to include all likely GCs members.
GCs leftover from this process without a clear accretion origin were divided into a high energy group,
which is likely a collection of smaller accretion events,
and a lower energy group that was thought potentially to be a signature of an ancient accretion event.
This GC grouping has been refined by \cite{horta_chemical_2020}, hereafter \citetalias{horta_chemical_2020},
who have added APOGEE alpha element abundances for 46 inner GCs to make minor revisions.

The low energy group of \citetalias{massari_origin_2019} is consistent with the Kraken event predicted by
\cite{kruijssen_formation_2019,kruijssen_kraken_2020} to be the MW's most ancient merger.
This work identified the structure by comparing the observed distribution of MW GCs with the predictions of the
EMOSAICs hydrodynamic simulations of GC formation and evolution \citep{pfeffer_e-mosaics_2018,kruijssen_e-mosaics_2019}.
This merger is likely the same as or significantly overlapping with the one that gave rise to
the Koala structure of \cite{forbes_reverse_2020}, hereafter \citetalias{forbes_reverse_2020},
and the Inner Galaxy System (or later Heracles) of \citet{horta_evidence_2021}.
In this paper, we refer to this accretion event as Kraken.

Once the accretion groups of GCs have been identified, the number of GCs, and the age-metallicity and
the dynamical distributions of the GCs can all provide information about the progenitor galaxy.
The GC AMR relation provides clues to the formation time and the chemical enrichment of the progenitor dwarf
\citepalias{forbes_reverse_2020},
while groups of GCs with smaller apocentres indicate an ancient or massive merger \citep{pfeffer_predicting_2020}.
Using these techniques, the \citetalias{massari_origin_2019} GC group memberships have been used in studies such as
those by \citetalias{forbes_reverse_2020} and \cite{trujillo-gomez_kinematics_2020} to reverse engineer the assembly
history of the MW.
Combined with insights from the EMOSAICs project, \cite{kruijssen_kraken_2020} used these groups to suggest that
the MW has experienced 2-3 major mergers, and at least 15 smaller mergers contributing GCs in total.

The number of GCs in a progenitor galaxy is related to its mass.
For LMC-mass and more massive galaxies, observations have revealed a linear relationship between
the number of or total mass of GCs and the halo mass of the host galaxy \citep{forbes_globular_2018}.
Theoretical models reproduce this trend
\citep[e.g.][]{boylan-kolchin_globular_2017, burkert_high-precision_2020,bastian_globular_2020}.
However, it is unclear if this relation holds for dwarf galaxies with stellar masses below $10^9~\rm{M}_\odot$.
Observationally it is difficult to measure the halo mass of such systems,
and theoretical predictions in this range often do not agree with one another.
At lower masses, analytical models based on hierarchical clustering predict a continuation of the linear relation
between GC mass and total halo mass \citep[e.g.][]{boylan-kolchin_globular_2017},
while the EMOSAICs project predicts a linear relation with stellar mass instead of halo mass
\citep{bastian_globular_2020}.

One limitation of the current GC groupings is that they are defined in a rather subjective way, mostly by eye.
This methodology raises questions about whether the current groupings are statistically robust and physically relevant.
Furthermore, subjective methods are very difficult to test using mock catalogues,
but this represents an essential analysis step to trust the results \citep[e.g. see][]{wu_using_2021}.
Alternatively, recent work has seen the use of clustering algorithms to find structures in the halo
\citep[e.g.][]{ostdiek_cataloging_2020, necib_chasing_2020, koppelman_multiple_2019, helmi_box_2017,
myeong_discovery_2018}.
These should give more objective, quantifiable results, but as noted in \cite{naidu_evidence_2020},
it can be challenging to tune these clustering methods to the astrophysical problem
of identifying groups of accreted material.
A few studies have applied these sorts of techniques to GCs specifically.
Examples include the  use of a friends-of-friends clustering algorithm to associate GCs to the Sequoia merger
\citep{myeong_sausage_2018} and the
decomposition of GCs in the centre of our galaxy into bulge, disc, and halo components
\citep{perez-villegas_globular_2019}.
However, we know of no studies that have yet been applied to the total Galactic population of GCs.

In this paper, we develop an objective methodology combining chemo-dynamical information to identify the
likely progenitors of the full population of Galactic GCs.
By fitting models to both the dynamical distribution in action space and the age-metallicity relation of the accreted galaxy,
we calculate membership probabilities for each GC and statistically link them to
particular accretion events.
We do so by modelling the GCs as a combination of bulge, disc, and halo components,
the latter representing the focus of our study.
The stellar halo is further decomposed into the massive merger events that built it, such as \GES, Kraken,
and Sagittarius,
and an ungrouped component coming from lower mass mergers that did not contribute enough GCs to be robustly identified.
This methodology is extensively tested and characterised using mock GC catalogues built from the \auriga{} suite of
hydrodynamical simulations \citep{grand_auriga_2017}.
We apply the method to the Galactic GCs and fully account for observational errors to identify the most likely GCs
associated with each merger event.
Using these membership probabilities, properties of the progenitor galaxies, such as halo and stellar masses,
are derived.

The structure of the paper is as follows.
In Section~\ref{Sec:Model} we describe our chemo-dynamical mixture model.
Section~\ref{Sec:Mocks} describes the construction of our mock globular GCs catalogues from \auriga{} haloes.
In Section \ref{Sec:MockFits} we apply our method to the mocks.
In Section~\ref{Sec:MWFit} we apply our method to the MW and discuss the individual cluster fits.
We discuss the resulting implications for the MW's accretion history in Section~\ref{Sec:MWDiscuss}.
Finally, Section~\ref{Sec:Conclusion} summarises and concludes the paper.

\section{Multi-component model for the Galactic GC population}
\label{Sec:Model}
We model the MW population of GCs as a combination of a bulge, disc, and stellar halo components.
The latter is the main focus of our work and is further split into subgroups that correspond to all known major
accretion events, such as \GES{} and Kraken.
The decomposition is performed using an expectation-maximization algorithm applied to chemo-dynamical data,
that is combining age-metallicity information with orbital integrals of motions (i.e. action space).
This section presents a detailed description of the decomposition method and its motivation.

For a general space, \X, which represents a combination of metallicity and action quantities,
each GC component is modelled as a distribution, $\FComX{}\equiv F(\X|\boldsymbol{\theta}_{c})$,
specified in terms of a set of model parameters, $\boldsymbol{\theta}_{c}$,
whose details will be given when discussing each model component.
\FComX{} is normalised to integrate to $1$ over the space $\bf{X}$.
Then, the multi-component model describing the overall population of GCs is written as the sum over
each individual component:
\begin{equation}
    \FX = \sum^{\mathrm{Com}}_{c}W_{c} F_c(\X) \;,
\end{equation}
where $W_{c}$ denotes the weight of component $c$ and specifies the fraction of the GC population contributed by
each component.
The total distribution, \FX{}, is normalised to unity over the space, which implies that:
\begin{equation}
     \sum^{\mathrm{Com}}_{c}W_{c} = 1 \;.
\end{equation}

The probability that the $i$-th GC belongs to component $c$,
which is often referred to as the ``responsibility'' in multi-component models, such as Gaussian mixture models,
is given by:
\begin{equation}
    r_{ic} = \frac{W_c F_c(\X_i)}{\sum_{c'}W_{c'} F_{c'}(\X_i)} \equiv \frac{p_{ic}}{\sum_{c'}p_{ic'}}
    \label{Eq:Resp} \;,
\end{equation}
where $\X_i$ denotes the coordinates of the $i$-th GC in the chemo-dynamical space used to
identify the different populations.
For brevity, we also introduced the notation, $p_{ic}\equiv W_c F_c(\X_i)$,
which gives the value of the $F_c$ distribution at $\X_i$ multiplied by the weight of that component.
The total log-likelihood, $\ln\mathcal{L}$, of the mixture model is given as:
\begin{equation}
\ln\mathcal{L} = \sum_{i}^{\mathrm{GCs}}\
\ln{F(\X_i)} \equiv \sum_{i}^{\mathrm{GCs}} \ln{  \left(\sum_{c'}^{\mathrm{Com}}p_{ic'}\right)} \;,
\end{equation}
wherein the rightmost term the first sum is over all the GCs in the system
and the second sum is over all components of the model.
To find the maximum likelihood estimate, we need to find the maximum of $\mathcal{L}$ for the set of parameters
$\{ \boldsymbol{\theta}_{c}\}\equiv\{\boldsymbol{\theta}_{c=1},
\boldsymbol{\theta}_{c=2}, ..., \boldsymbol{\theta}_{c=K}\}$,
where $K$ is the number of components and each $\boldsymbol{\theta}_{c}$ is, in turn, a set of multiple parameters.
For example, if we model a component as a Gaussian distribution, then $\boldsymbol{\theta}_{c}$ is the combination of
peak position along each coordinate axis in \X-space and the corresponding covariance matrix.
The maximization procedure is further complicated by the fact that the $W_{c}$ weights that appear in the $p_{ic'}$
expression depends on the values of all the $\{ \boldsymbol{\theta}_{c}\}$ parameters which makes for
a very non-linear and multi-dimensional maximization procedure.

To solve this challenge, we use the expectation-maximization approach.
This algorithm is often used to fit Gaussian mixture models efficiently.
As explained below, our methodology is similar to this
but adapted to include relevant astrophysics such as the AMR of the component.
The algorithm corresponds to an iterative approach for finding the maximum likelihood and has the following steps:

\begin{enumerate}
\item \uline{Initialisation:}\\
  An initial guess is made for the responsibilities, $r_{ic}$.
  The outcome can be dependent on this initial choice.
  This dependence is tested and discussed in Sec. \ref{Sec:MockFits}.
\item \uline{Maximisation Step:}\\
  In this step, we assume that the responsibilities, $r_{ic}$, are known, and we find the
  $\{ \boldsymbol{\theta}_{c}\}$ parameters that maximize the log-likelihood, $\ln\mathcal{L}$, for fixed $r_{ic}$ values.
  The advantage is that once the $r_{ic}$ are known,
  maximising $\ln\mathcal{L}$ reduces to a much simpler problem in which the parameters of one component
  are independent of the parameters of the remaining components.
  For component $c$, $\ln\mathcal{L}$ is maximal for the $\boldsymbol{\theta}_{c}$ values that maximize the expression:
    \begin{equation}
        \sum_{i}^{\rm{GCs}} r_{ic}\ln{F_{c}(\X_i)}
        \label{Eq:MaxFit} \;.
    \end{equation}
    In the above equation, each data point contributes with a weight, $r_{ic}$,
    which is why $r_{ic}$ is called the responsibility.
    \item \uline{Expectation Step:}\\
    The values of the responsibilities are updated using the $\{ \boldsymbol{\theta}_{c}\}$ parameters
    found in the previous step.
    \item \uline{ Iteration:}\\
    Repeat the maximisation and expectation steps until $\ln\mathcal{L}$ is converged.
    In practice, we assume convergence when $\ln\mathcal{L}$ changes between consecutive steps by
    less than $0.001$ times the number of GCs.
\end{enumerate}

The space $\bf{X}$ we use to identify the components of the GC population is a combination of orbital dynamical quantities,
which we denote with $\bf{Y}$, and age-metallicity information, which we denote with $\bf{Z}$.
We assume that the orbital quantities are uncorrelated with the chemistry of GCs,
which implies that the distribution function of each component can be split into two independent distributions:
\begin{equation}
    F_{c}\left(\bf{X}\right)  =  F_{c}^{\mathrm{dyn}}\left(\bf{Y}\right) F_{c}^{\mathrm{AMR}}\left(\bf{Z}\right)
\end{equation}
In the following, we describe how we model the distribution of dynamical quantities,
$F_{c}^{\mathrm{dyn}}\left(\bf{Y}\right)$,
and of the age-metallicity relation, $F_{c}^{\mathrm{AMR}}\left(\bf{Z}\right)$, where we drop the superscripts for brevity.
These functions are independent, and so can be fit by maximising their respective likelihoods
(with Eq. \ref{Eq:MaxFit}) independently.

\subsection{Dynamical Modelling}
\label{Subsec:DFs}
In this work, we primarily consider a four-dimensional dynamical space consisting of the orbital energy and
the three orbital actions: the component of the angular momentum perpendicular on the disc plane, $L_{z}$,
the radial action, $J_{R}$, and vertical action, $J_{z}$.
The integrals of motion, $\bm{J}$, completely describe the orbit, which determines the orbital energy
(for more information see \citealt{binney_galactic_2008}).
This means that in the $(E, \bm{J})$ four-dimensional space all orbits, including those of our GCs,
lie on a three-dimensional surface.
This suggests that the energy only contains redundant information about the orbits;
however, tests on mock catalogues show that the combined $(E, \bm{J})$ space leads to a more accurate identification of
GC populations than $(\J)$ space, justifying our choice (more on this in Section~\ref{Sec:MockFits}).

\subsubsection{Accreted GCs}
\label{DF:Accreted}
The accreted components are modelled as multivariate Gaussian distributions in the $\bm{Y}=(E,L_z,J_R, J_z)$ space through,
\begin{equation}
\begin{aligned}
    F_{c}\left(\bf{Y}\right) &= N\left(\bf{Y}|\bm{\mu},\bm{\Sigma}\right)\\
    &=  \frac{1}{\sqrt{\left(2\pi\right)^{n_{\mathrm{dim}}}\left|\bm{\Sigma}\right|}}
    \exp{\left(-\frac{1}{2}\left(\bf{Y}-\bm{\mu}\right)^{T}\bm{\Sigma}^{-1}\left(\bf{Y}-\bm{\mu}\right)\right)}
\end{aligned}
\end{equation}
where $n_{\mathrm{dim}}=4$ is the number of dimensions of the space, $\bf{Y}$, $\bm{\mu}$ is the mean,
and $\bf{\Sigma}$ is the covariance matrix.
The values of these parameters that maximize the total model likelihood can be found analytically from
Eq. \eqref{Eq:MaxFit} by calculating moments of the distribution.

In reality,
not all the accreted material from a single merger event will necessarily be well represented by a Gaussian distribution.
Typically, the bulk of the material is often centred around the orbit of the accreting galaxy,
and can be well described by a single Gaussian component.
However, some of the material can be in more complex substructures formed in accretion,
such as leading or trailing stream arms of a stream, which can have a different dynamical distribution.
It should be noted that it is likely that material very near, or on, the boundaries of our chosen dynamical space \EJ{}
(such as the maximally circular orbits) can be poorly described.
However, we find that due to the relatively small number of GCs, alternative `assumption free' distributions,
such as density kernels, do not work effectively, and it is necessary to assume a form for the distribution.

If the number of points to which an unconstrained multivariate Gaussian distribution is fit, $N_{\mathrm{points}}$,
is equal to or less than $n_{\mathrm{dim}}$
then the covariance matrix becomes degenerate with some eigenvalues equalling zero (or infinitesimal).
The corresponding principle axes then have infinitesimal width, which can give unrealistically large probability values.
For example, in 2-dimensional space, two points will be fit as a line,
with the fit and grouping unable to develop further.
To prevent this, we fix the value of the smallest principal axis using the procedure described in
Appendix~\ref{Appendix:DegenFit}.

\subsubsection{Ungrouped GCs}
\label{Ungrouped}
Some GCs cannot be attributed to any known accretion event,
such as the High Energy group in the \citetalias{massari_origin_2019} analysis.
This could be because they fell in as small groups that do not contain enough information to be robustly identified.
Alternatively, the GC's orbit could have evolved such that it no longer resembles those of the rest of the group.
Our model accounts for such GCs which are classified as the `ungrouped' component.

The ungrouped component is modelled as a uniform background distribution,
normalised to integrate to one over the  \Change{convex hull volume}, $V$, of the dynamical space filled by \Change{all of} the GCs.
That is,
\begin{equation}
    F_{\mathrm{Ung}} = \frac{1}{V} \;,
\end{equation}
where $V$ is calculated using SciPy's convex hull module \citep{virtanen_scipy_2020}.

\subsubsection{In-situ Components}
\label{DF:InSitu}
In the MW, we cannot be certain if the GCs are accreted or have an in-situ origin.
Therefore, we need to include models of the bulge and disc components.
The dynamics of these components are not well described by Gaussians,
and instead, we model them as distribution functions in action space using the implementations in \agama{}
\citep{vasiliev_agama_2019}.

When modelling the bulge and disc components in \EJ{} space,
we assume that the energy distribution can be separated from the action distribution, that is:
\begin{equation}
    F\left(E,\bf{J}\right) = F\left(E\right) \FJ.
\end{equation}
The energy distribution is calculated numerically from the prescribed action distribution of the components.

We originally modelled the action distribution of the bulge as a double power-law with a cutoff as introduced in
\cite{posti_action-based_2015}.
In practice, we found that the fitting converges on values consistent with the simpler exponential fit:
\begin{equation}
    \FBulge = \frac{4}{J_{\mathrm{Cut}}^{3}\sqrt{3\pi^{3}}}\exp{[-\left(J_{\mathrm{Tot}}/J_{\mathrm{cut}}\right)^2]},
\end{equation}
where $J_{\mathrm{Tot}} = J_{R}+\left|L_{z}\right| + J_{z} $ and
$J_{\mathrm{cut}}$ is a free parameter that controls the steepness of the cutoff.

The disc is modelled using the quasi-isothermal disc, first described in \cite{binney_distribution_2010}.
This is also used to model GCs in \cite{posti_mass_2019}, whose assumptions we follow.
The action distribution is given as:
\begin{equation}
\begin{aligned}
    \FDisc &= \frac{\Sigma\nu\Omega}{2\pi^{2}\kappa\sigma_{R}^{2}\sigma_{z}^{2}}
    f_{\pm,d}\exp{\left(-\frac{\kappa\JR}{\sigma^{2}_{R}}-\frac{\nu\Jz}{\sigma^{2}_{z}}\right)}\\
    \Sigma &= \exp{[-R_{c}\left(\Lz\right)/R_{d}]}\\
    f_{\pm,d} &= \left\{ \begin{array}{rl}
                    1 &\mbox{$\Lz\geq0$} \\
                    \exp{\left(2\Omega\Lz/\sigma^{2}_{R}\right)} &\mbox{$\Lz<0$}
                    \end{array} \right.
\end{aligned} \;,
\end{equation}
where $\Sigma$ describes the disc surface density and $f_{\pm,d}$ controls the rotation of the disc.
The circular, radial, and vertical epicycle frequencies are denoted by $\Omega$, $\kappa$ and $\nu$ respectively,
and are evaluated at the radius of the circular orbit, $R_{c} = R_{c}\left(J_{\mathrm{Tot}}\right)$,
with angular momentum $J_{\mathrm{Tot}} = J_{R}+\left|L_{z}\right| + J_{z} $.
The radial velocity dispersion is given as $\sigma_{R} = \sigma_{R0}\exp{\left(-R_{c}/R_{\sigma}\right)}$,
and the vertical velocity dispersion is fixed at a constant scale-height, $\sigma_{z} = \sqrt{2}h_{d}\nu$.
The disc is chosen to match the thick disc of \cite{piffl_rave_2014} with
$R_{\sigma} = 13\kpc$ and $h_{d} = 0.2R_{d}$.
This leaves two free parameters: the disc scale-length, $R_{d}$, and the central radial dispersion, $\sigma_{R,0}$.

\subsection{Age-Metallicity Relation}
\label{Subsec:AMR}
We use the leaky-box chemical evolution model to describe the age-metallicity relation for GCs as given in
\cite{forbes_reverse_2020}:
\begin{equation}
    \Met = -p_{\mathrm{yield}} \log\left(\frac{t}{t_{f}}\right)
    \label{eq:AMR},
\end{equation}
where \pyield{} is a measure of how quickly the system enriches and $t_{f}$ is the formation time of the system.
Larger galaxies enrich in metallicity faster, giving a higher \pyield{} and steeper evolutionary track.
Note that this is equivalent to Eq.~4 of \cite{kruijssen_e-mosaics_2019}, with rearranged and renamed constants,
and is similar to the relation of \massari{} (Eq. 1).

We proceed by fitting Eq. \eqref{eq:AMR} to the GCs associated with each component taking into account the weights,
i.e. the responsibilities, associated with each object.
The fitted relation can be inverted to obtain the expected age as a function of metallicity,
which we denote as $t_{\mathrm{fit}}\left(\Met\right)$.
The probability of the GCs observed age being part of the modelled relation is then given by a normal distribution,
centred on the expected age with dispersion equal to the error in age, $\sigma_{t}$, i.e.
\begin{equation}
    F_{c}^{\mathrm{AMR}}\left(\left[t,\Met\right]\right)
    =
    N\left(t|\mu = t_{\mathrm{fit}}\left(\Met\right),\sigma=\sigma_{t}\right)
    \; .
\end{equation}
For the GCs that do not have age-metallicity data,
we assume that they have a constant probability to be assigned to the component in the age-metallicity space.
This is taken to be the inverse of the range of ages of the GCs (i.e. maximum age - minimum age),
similar to the uniform probability of the ungrouped component in dynamical space.
For the ungrouped component, we do not expect all group members to be from a single accretion event
or follow the same age-metallicity relation.
The probability is then taken as a constant value as if there were no age-metallicity data.

\subsection{Observational Errors}
\label{SubSec:ObsErrors}
To model the MW effectively it is necessary to include the statistical uncertainty from observational errors.
For this, we use the Monte Carlo method described in Sec.~\ref{SubSec:Obs} that samples the uncertainties in
the measured velocity and position of GCs.
The Monte Carlo samples of a single cluster are treated as independent points,
with their own responsibilities and are fit independently.
When the model has converged, the final probabilities of cluster $i$ is given as:
\begin{equation}
    p_{ic} = W_c \sum_{j}^{\mathrm{MC}} F_c\left(\bm{X}_{i}^{j}\right),
\end{equation}
where $\bf{X}_{i}^{j}$ is $j$-th Monte Carlo realisation of the $i$-th GC and
the sum is over all the Monte Carlo samples of the GC.
These probabilities are then used to calculate the responsibilities of the final results, according to Eq.~\eqref{Eq:Resp}.

\section{Mock Catalogues of Globular Clusters}
\label{Sec:Mocks}

We now describe our construction of mock GC catalogues from the \auriga{} hydrodynamical simulations.
The \auriga{} project consists of a suite of high-resolution cosmological zoom-in simulations of
individual MW-like haloes \citep{grand_auriga_2017} with halo masses between $1-2\times 10^{12} M_{\odot}$.
The haloes were selected from the $100^{3} \, \textnormal{Mpc}^{3}$ periodic cube of the \eagle{} project,
a $\Lambda$CDM cosmological hydrodynamical simulation \citep{crain_eagle_2015,schaye_eagle_2015}
adopting Planck1 \citep{planck_collaboration_planck_2014} cosmological parameters.
Using the N-body and moving mesh magnetohydrodynamic \textsc{arepo} code \citep{springel_moving-mesh_2011},
these haloes were resimulated to produce a zoom-in simulation of each halo.
We selected these simulations because they have been shown to reproduce many properties of the MW
and other MW-mass galaxies,
such as the satellite luminosity function \citep{shao_multiplicity_2018,simpson_quenching_2018},
stellar bulge and disc structures \citep{grand_auriga_2017,gomez_warps_2017},
and stellar halo \citep{monachesi_auriga_2019,fattahi_origin_2019, deason_mass_2021,grand_effects_2019}.
We use the level 4 resolution sample, with a DM particle mass of $\sim 3\times 10^{5} M_{\odot}$ and
an initial gas resolution element of mass $\sim 5\times 10^{4} M_{\odot}$.
This sample contains 30 haloes which we label Au1 to Au30.

Of the 30 level 4 \auriga{} haloes, 13 are unrelaxed at the present day according to the criteria of
\cite{neto_statistics_2007}.
These unrelaxed haloes are poorly modelled by static axisymmetric potentials.
This is typically because they are currently, or recently, undergoing a disruptive transient event such as a merger.
We therefore restrict our analysis to the 17 relaxed haloes.
There is some debate whether the presence of the LMC would cause the MW to be classified as unrelaxed
according to the same criteria \citep{cautun_aftermath_2019, erkal_equilibrium_2020, erkal_detection_2021}
and is, in fact, poorly modelled by a static axisymmetric potential.
We leave the effects of a time-dependent potential to future work.

The \auriga{} simulations do not `natively' contain GCs.
To represent groups of accreted GCs we select old accreted stars in the stellar halo
(c. f. \citealt{halbesma_globular_2020}).
For each accretion event we identify the accreted stars and randomly assigned GCs to a
subsample of them based on the properties of the progenitor galaxy.
To assign GCs, we select only accreted halo stars  older than $10$~Gyrs
and require them to be within $R_{200}$ of the host galaxy at the present day.
This is motivated by age estimates of the MW GCs which are, with a few exceptions, older than $10$ Gyrs
(see Fig.~\ref{fig:MW_AMR}).
To determine the origin of the stars, we use the accretion catalogue of stars as \cite{fattahi_origin_2019}.

The birthplace of the star is defined as the subhalo in which it resides at the first simulation snapshot
(as defined by the {\sc{subfind}} algorithm of \citealt{springel_gadget_2001}) after its formation.
If the star is born in the main halo, it is defined as an in-situ star.
If the star is born outside of the main halo,
its origin is defined to be the last subhalo to which it belonged before it fell into the main halo.
This prescription identifies the accreted stars that are associated with the accretion event
that brought them into the main halo.
The few stars that formed from the gas of infalling satellites in the main halo are classified as in-situ.

To create the in-situ GCs we generate test particles,
with positions and velocities randomly drawn from an action distribution of the in-situ components using AGAMA
(see Sec. \ref{Sec:Model}).
We use the bulge and disc action distributions described in Sec. \ref{Sec:Model},
fit to the GCs identified in the \massari{} groupings.
These action distributions are scaled appropriately by the mass of the \auriga{} galaxy ($M_{200}^{\mathrm{Au}}$),
such that $F_\mathrm{Au}\left({\bf{J}}\right) = F_\mathrm{MW}\left(\lambda{\bf{J}}\right)$,
where $\lambda = \left( M_{200}^{\mathrm{MW}}/ M_{200}^{\mathrm{Au}} \right)^{2/3}$.
We take $M_{200}^{\mathrm{MW}}=1.17\times10^{12}\Msun$ from \cite{callingham_mass_2019},
which also contains further discussion of this mass scaling technique.
We create 1000 mock catalogues of accreted and in-situ clusters for every relaxed \auriga{} halo.

\subsection{The GC populations}

To generate the mocks, we must choose the size of the membership of each GC group.
For the in-situ component, we assume fixed populations
of 40 GCs for the bulge and 20 for the disc, motivated by previous
groupings in the literature.
For the accreted groups, we adopt the
\cite{burkert_high-precision_2020} model in which the number of GCs is
proportional to the total mass of the host.
The mean expected number
of GCs, $N_{\mathrm{GC}}$, for an accretion event of mass,
$M_{\mathrm{Host}}$, is given by:
\begin{equation}
    N_{\mathrm{GC}} = \frac {M_{\mathrm{Host}}}{5\times 10^{9} M_{\odot}}
    \label{Eq:nGC} \;.
\end{equation}
From this mass - number of GC relation (\MhNgc) we generate 1000 GC mocks for each accreted satellite.
To keep the analysis as clear as possible,
each random realisation has an equal number of GCs given by the mean expectation, rounded to the nearest integer.
In principle, we could include scatter on this relation (as given in \citealt{burkert_high-precision_2020}).
However, in these tests, we are principally interested in the changes caused by the sampling of dynamics
of the accretion events,
not those caused by random variance in the population numbers.

Whilst the expected number of GCs from a single small accretion event
(objects of mass less than ${5\times 10^{9} M_{\odot}}$) is less than one,
we estimate that on average the expected \textit{total} number of GCs from small accretion events is typically ${\sim}5$.
This population of small accretion events bring in individual, ungrouped GCs.
To include them, we assign individual GCs starting from the largest `small' accretion event until
the expected population is accounted for.

\begin{figure}
	\includegraphics[width=\columnwidth]{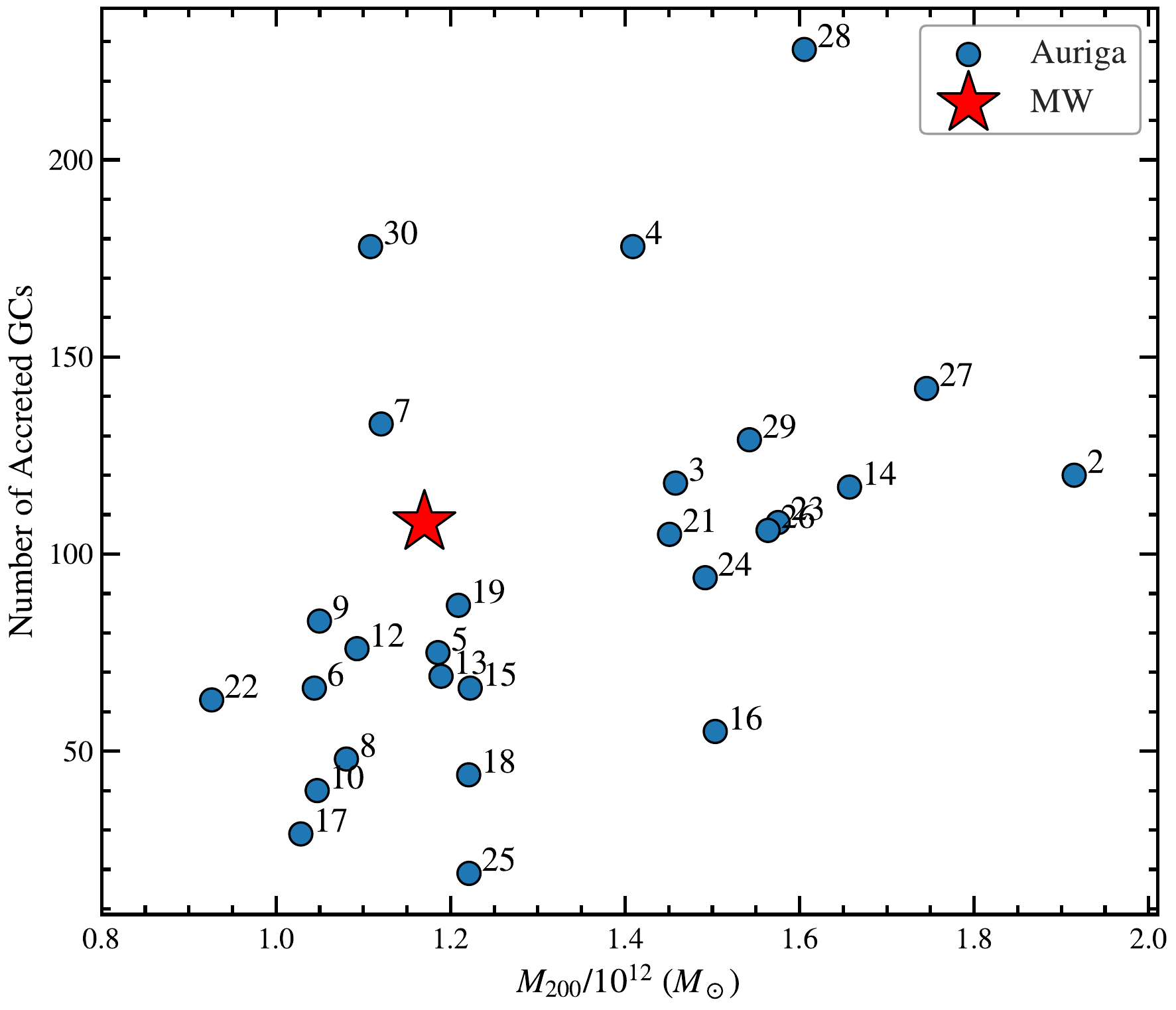}
    \smallspace{}
    \caption[The relation between total mass, and the number of
    accreted GCs for our \auriga{} mock catalogues and for the MW.]
    {The relation between total mass, $M_{200}$, and the number of
      accreted GCs for our \auriga{} mock catalogues (blue symbols)
      and for the MW (red star).}
    \label{fig:Ngc_M200}
\end{figure}

\begin{figure}
	\includegraphics[width=\columnwidth]{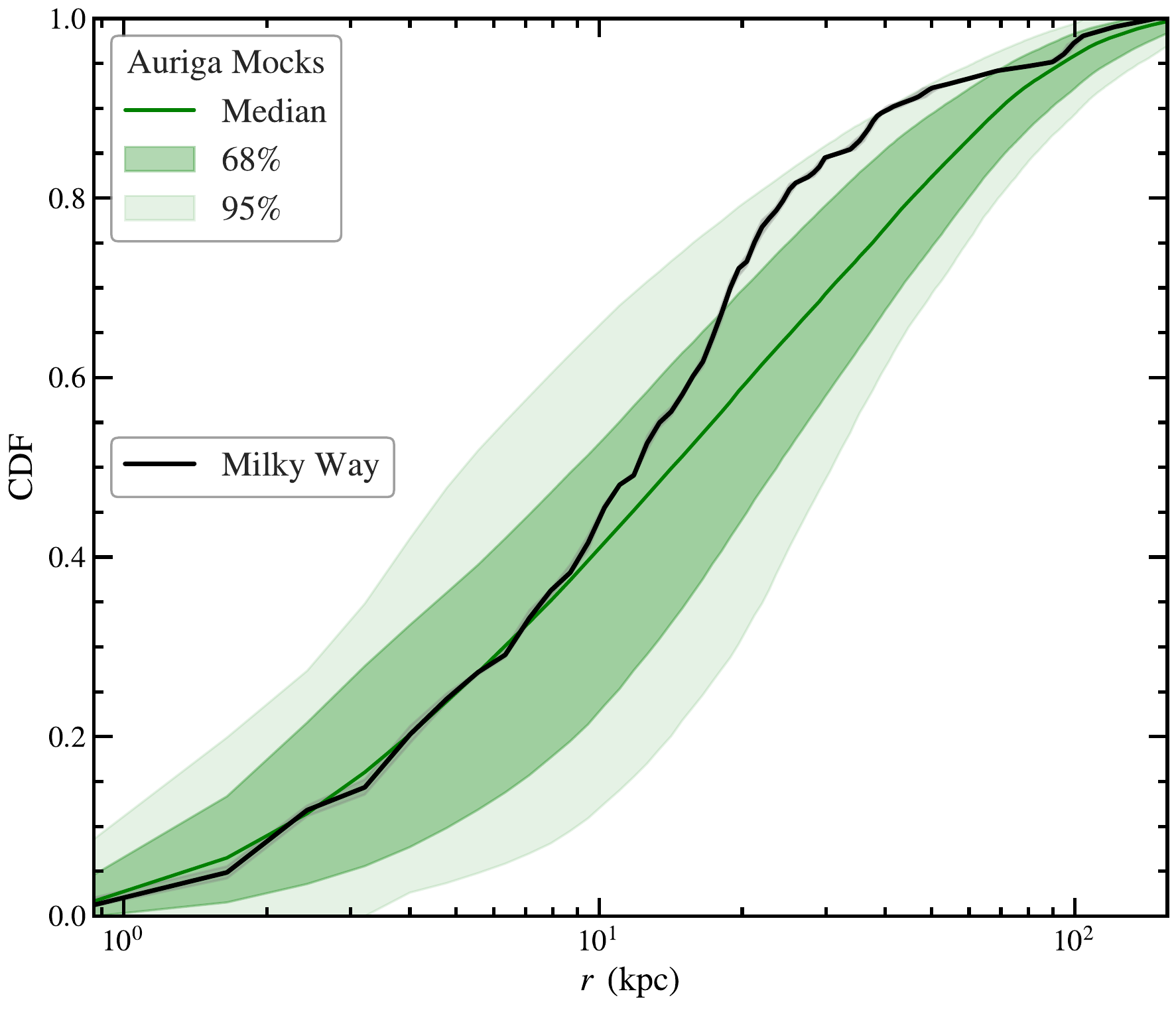}
    \smallspace{}
    \caption[The cumulative radial distribution of accreted GCs in our \auriga{} mock catalogues and in the MW.]
    {The cumulative radial distribution of accreted GCs in our
      \auriga{} mock catalogues and in the MW (black line). The green
      solid line shows the median in the mocks, and the shaded regions
      give the 68 and 95 percentile regions.  }
    \label{Fig:compMW-r}
\end{figure}

The resulting population of accreted GCs in our mocks is compared to the observed Galactic GCs in
Figs.~\ref{fig:Ngc_M200} and~\ref{Fig:compMW-r}.
For the MW data, we take the total mass estimate from \cite{callingham_mass_2019}
and the number of accreted GCs that we find in Sec.~\ref{Sec:MWFit}.
The number of GCs in the mocks increases with the mass of the host galaxy,
as expected from observations and theoretical models (see discussion in Sec.~\ref{Sec:Intro}).
Fig.~\ref{fig:Ngc_M200} shows that the number of accreted GCs in our mocks is consistent with the MW estimates.
The \auriga{} mocks with a total mass of ${\sim}1.2\times10^{12}~\rm{M}_\odot$ have slightly fewer GCs than the MW,
but the scatter is rather large and there are at least two systems with more GCs.

Fig. \ref{Fig:compMW-r} compares the radial distribution of GCs,
where the distance of the GCs in the \auriga{} mocks was scaled by $R^{\mathrm{MW}}_{200}/R^{\auriga}_{200}$
to account for the different sizes of the \auriga{} systems.
For this, we assumed $R^{\mathrm{MW}}_{200}=222\,\mathrm{kpc}$ from \cite{callingham_mass_2019}.
The radial distribution of GCs in our mocks is similar to the observed one,
although the MW is slightly more centrally concentrated in the $20-30\,\mathrm{kpc}$ region
than most of the \auriga{} sample.
This could potentially reflect that the Galactic stellar halo was mostly built from a few massive early accretion events
\citep[e.g.][]{kruijssen_formation_2019} whose remains are primarily found in the inner region of the MW.
Alternatively,
it has been suggested that the limited resolution of a simulation
can cause accreting satellites to disrupt before reaching the galaxies centre,
reducing the concentration of accreted stellar material
\citep[e.g.][]{springel_cosmological_2005, grand_determining_2021}.

The orbital dynamics of the GCs (including the energy, pericentres, apocentres, actions, angles, and frequencies)
for all stars in the main \auriga{} halo at the present day are calculated using the \agama{} package
\citep{vasiliev_agama_2019}.
The potential is modelled from the $z=0$ simulation snapshot;
representing the contribution of the hot gas and DM as a spherical harmonic expansion
and the contribution of the stars and cold gas as an azimuthal harmonic expansion (using \agama).

\subsection{The Age-Metallicity Relation}
Hydrodynamical simulations generally have difficulties reproducing the metallicity of dwarf galaxies
and their GCs \citep[e.g.][]{halbesma_globular_2020}, which is potentially due to uncertainties in stellar yields.
To mimic the observed age-metallicity relation of GCs we assign metallicity values to our mock GCs
using the relation given in Eq.~\eqref{eq:AMR}.
For each accretion event, we first choose an age-metallicity relation,
setting a formation time that is equal to the oldest star in that galaxy and a yield
determined by the yield-stellar mass relation described in \cite{forbes_reverse_2020}.
The \auriga{} galaxies have a somewhat high stellar mass for their halo mass,
and to mitigate this we recalculate the progenitor stellar masses using the halo mass at infall
and the stellar-mass-halo-mass relation of \cite{behroozi_universemachine_2019}.
This gives yields more comparable with those predicted for the MW than if we had used the original stellar mass of \auriga{}.
To mimic observational uncertainties, we add normally distributed errors with a mean of 1~Gyr to the age estimates,
which corresponds to the average errors for the MW GCs.
\Change{Note that these uncertainties are applied after determining the appropriate metallicity values
so that the final age-metallicity data does not lie exactly on the AMR relation.}

For the in-situ clusters, we randomly assign ages between $12$ and $14\, \mathrm{Gyrs}$,
motivated by the age distribution of MW in-situ clusters.
We then follow the same procedure used for generating the accreted metallicity values,
using the AMR fit to the MW in-situ clusters.
This process generates an age-metallicity distribution that is comparable with the MW's own,
with a distinctive steeper in-situ branch and a shallower, wider accreted branch.

\section{Mock tests of the mixture model}
\label{Sec:MockFits}
We proceed by testing our multi-component model for the GC population using our mock catalogues.
These tests will help select the optimal dynamical quantities to identify GC groups and characterise the
extent to which our modelling approach recovers the true GC groups
predicted by the cosmological simulations.

\begin{figure*}
	\includegraphics[width=2\columnwidth]{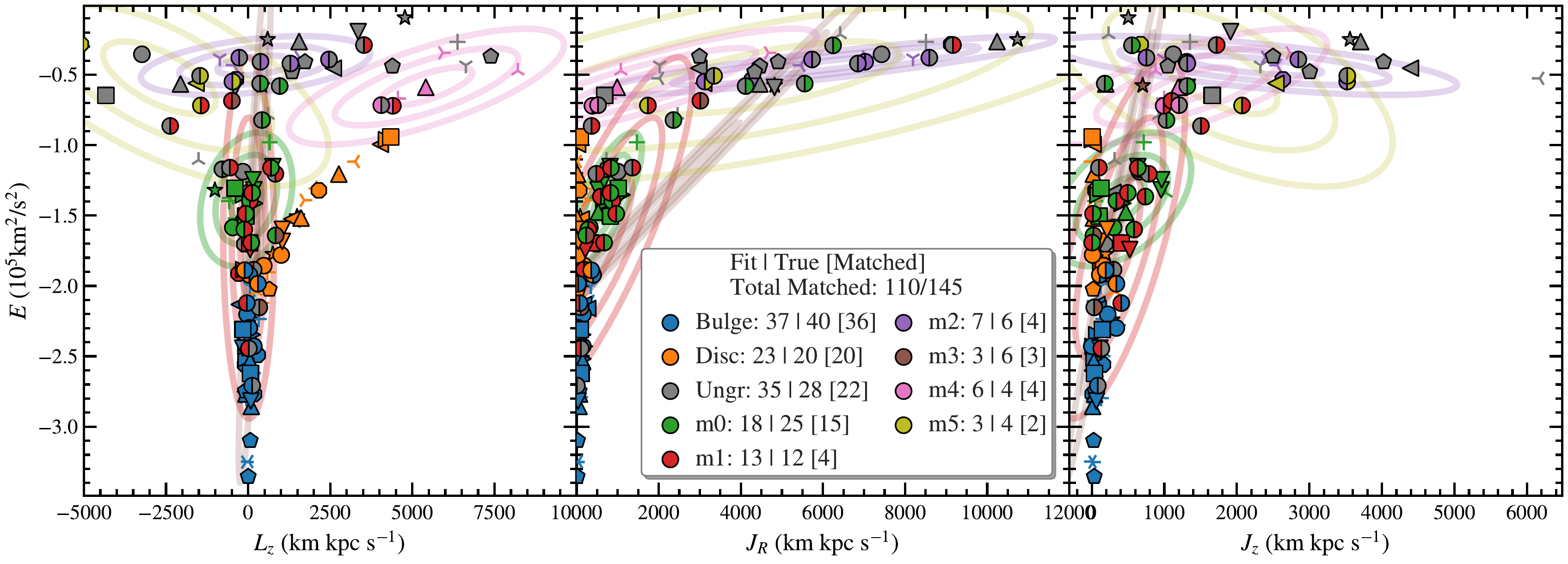}
        \caption{
	The chemo-dynamical model of Sec.~\ref{Sec:Model} fit to a mock GC sample from the \auriga{} 5 halo.
	The axes show the dynamical component of the fit in energy ($E$) and action space (angular momentum $L_{z}$,
	radial action $J_{R}$, vertical action $J_{z}$).
	The accompanying age-metallicity fit is shown in Fig.~\ref{fig:AU5_AMR}.
	GCs are represented by symbols, which are consistent for each object between the three panels,
	to help identify individual GCs.
	Different colours represent the different accretion groups, labelled m0 to m5 by decreasing accretion mass.
	The groups are modelled as Gaussian distributions, with the contours giving the 1, 2 and 3-$\sigma$ regions.
	The `ungrouped' group contains GCs that do not fit well in other groups
	or contain less than 4 members in the true groups.
	A GC is attributed to the group for which it has the highest probability of belonging.
	Solid symbol colours indicate where the model agrees with the true grouping.
	Symbols that are split in colour show the original true grouping on the right-hand side and
	the assigned group on the left-hand side.
	}
    \label{fig:AU5_EJ}
\end{figure*}

\begin{figure*}
	\includegraphics[width=2\columnwidth]{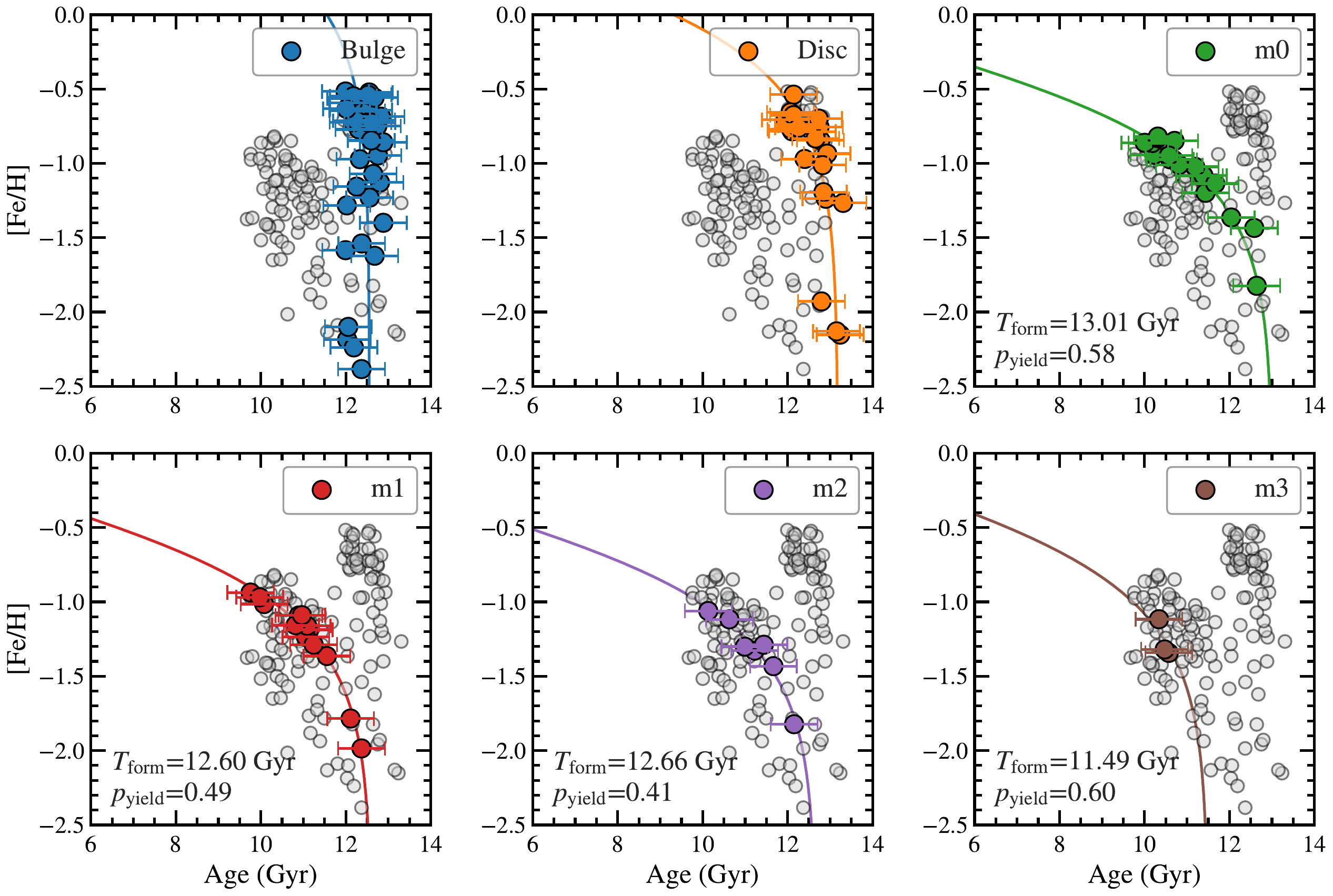}
	\smallspace{}
	\caption{
	The age-metallicity relation for a mock GC sample from the \auriga{} 5 halo.
	The different panels show the position of the different GC groups as identified by our method
	and the age-metallicity relation fit to them (see main text for details).
	The in-situ components (bulge and disc) can be seen to be steeper branches than the in-situ components.
	}
    \label{fig:AU5_AMR}
\end{figure*}

First, we illustrate an example of a chemo-dynamical fit for a mock catalogue.
The fit is obtained by following the steps described in detail in Sec.~\ref{Sec:Model}.
Fig.~\ref{fig:AU5_EJ} shows the $(E,\bm{J})$ distribution, and Fig.~\ref{fig:AU5_AMR} the AMR fits,
of the six most massive accretion events for the \auriga{} 5 halo.
These events are numbered from 0 to 5 in descending order of their total mass at accretion and
are the systems containing four or more GCs.
The remaining GCs, i.e. those from accretion events that brought three or fewer objects, are labelled as `ungrouped'.
In this fit, our model agrees with the groupings of $110$ out of the $145$ GCs,
where the GCs are associated with the group for which they have the highest probability of being a member.

We can see that in the dynamical space, the accreted distributions overlap in all three panels,
with some distributions very widely spread.
Some individual GCs of a group can be far from the rest of the group's GCs and the fitted distribution,
and have little chance of being correctly identified.
The accretion groups of these mock catalogues are undoubtedly more complex than our current picture of the MW,
highlighting the need for realistic testing to understand the feasibility of identifying these groupings.

The in-situ clusters are reasonably well-identified in both dynamical and age-metallicity space.
The accreted components seem to be more distinct and easy to identify at higher energy,
where the smaller groups can remain as compact distributions.
For the larger groups at lower energy, we see significant overlap with other groups,
making them difficult to identify confidently.

\subsection{Initial Groups}
\label{Subsec:MockInitial}

To apply our algorithm to the GCs, we must first make a choice of starting groups.
Through testing, we have found that with different initialisations it is possible to generate different final groupings
as the algorithm converges to different local maxima.
To overcome this, we apply our algorithm to many starting configurations.
The log-likelihood of the fits can then be compared, with the largest chosen as the best-fit grouping.

However, it is not feasible to try all possible starting groups.
We have experimented extensively with different methodologies to generate the initial groups,
including using other clustering algorithms and seeding the groups with random GCs and overdensities.
However, none of these alternatives returned satisfactory results, reflecting the difficulty of the problem.
Instead, we choose the `sensible' starting configurations described next and apply both a bootstrapping-based approach
and hand-selected variations to test.

In the mock catalogues, we know the true accretion groups of the GCs, and so we use them as the sensible starting point.
\Change{We tested the robustness of this initialisation step by re-assigning a fraction of the GCs to plausible alternative groups.
We find that the outcome is generally robust to such changes as long as the reassigned fraction is $\lesssim35\%$,
with the smaller groups being the most affected.
This is likely due to the average position and spread of the distribution describing the groups remaining similar until
a large fraction of the group members are lost.
From these distributions, the group can recover its members.
Further details of our tests on initial groupings can be found in Appendix \ref{Appendix:InitialGroups}.}

For the MW case discussed in Sec.~\ref{Sec:MWFit} we use selections from the literature as our starting points.
There is already a rough decomposition of GCs into accretion events with the main limitation being that
the boundary between these groups is rather subjectively defined \citepalias[e.g.][]{massari_origin_2019}.
When analysing the MW sample, we find all the known groups with more than 9 members,
which suggests that having a modest fraction of mislabelled GCs, does not strongly impact the outcome.
This is discussed  in Sec. \ref{Sec:MWFit}.

\subsection{\Change{Testing on Mock Samples}}
\label{sec:Testing_on_mocks}
We now apply our method to all the mock catalogues,
testing each of the $1000$ sets of mock GCs for each relaxed \auriga{} galaxy.

Arguably the most important quantity for inferring the properties of the progenitor galaxies
is our ability to estimate the population of each of the accreted groups.
\Change{In a first step,
we study how the population sizes of our recovered groups compares to the truth.
This test does not fully characterise our method, since it}
does not indicate whether the individual GCs have their correct groups identified.
To further quantify how well the method recovers each GC group, we define the purity, $P$, and completeness, $C$, as:
\begin{align}\label{Eq:Scoring}
     P &= N_{\cap} / N_{\mathrm{fit}}\\
     C &= N_{\cap} / N_{\mathrm{true}}.
\end{align}
Where \Ntrue{} is the true GC population of the group, \Nfit{} is the number of members identified by our fitting procedure,
and $N_{\cap}$ is the intersection of the fit and true groups.

\begin{figure}
	\includegraphics[width=\columnwidth]{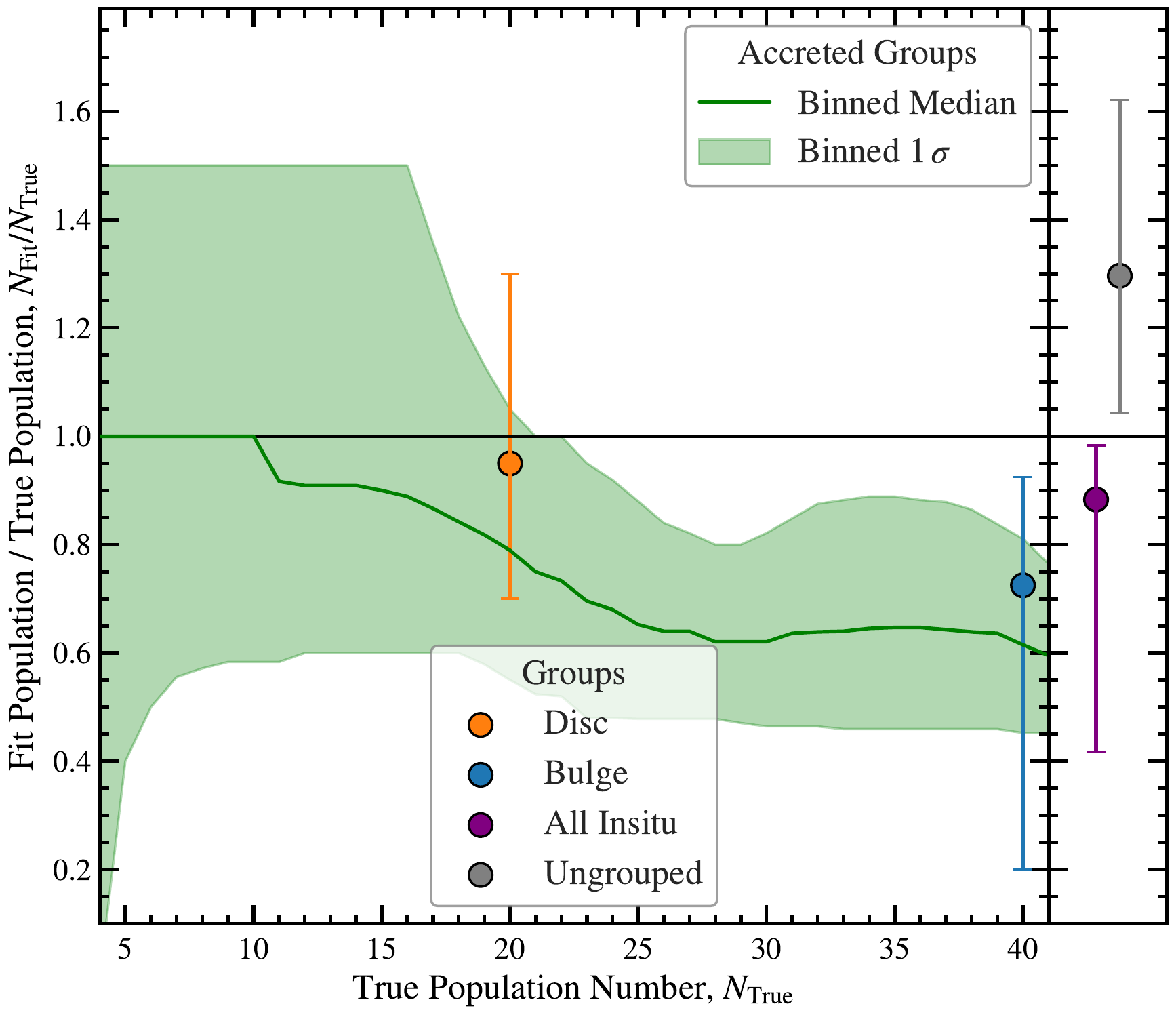}
	\smallspace{}
	\caption{
	The \Change{ratio, $N_{\mathrm{fit}}/N_{\mathrm{true}}$, between the group population size recovered by
	our method and the true value as a function of $N_{\mathrm{true}}$}
	for our mock GC catalogues.
	We bin the accreted groups
	\Change{(not including the ungrouped component)}
	in $N_{\mathrm{true}}$, and, using a Gaussian smoothing kernel of $\sigma=2$,
	show the median (solid green line) and $16\%-84\%$ range (shaded green region) for the distribution of $N_{\mathrm{fit}}/N_{\mathrm{true}}$.
	The symbols with error bars show the individual in-situ components, the total in-situ sample, and the ungrouped components (see legend).
	Note that the disc and bulge components of the mocks have populations of 20 and 40 by construction,
	whereas the population of the ungrouped component depends upon the accretion history of the halo.
	}
	\label{fig:Mock_NtrueNfit}
\end{figure}

First, {we study how the recovered number of GCs in each group compare against the true number of members. This is shown in Fig.~\ref{fig:Mock_NtrueNfit}.}
Across the total sample, we recover the accreted population numbers with an under-bias of $\sim10\%$.
\Change{This is the average value per component and, since there are more small groups than large ones, is biased towards small groups.}
There are clear trends if we look at the results as a function of the group richness.
The smallest groups ($\lessapprox 10$) are recovered without an under-bias,
but as the true size of the group increases so does a trend to systematically underestimate the population.
For smaller groups  there is a significant fractional scatter ($\sim50\%$),
as the changing membership of a single cluster corresponds to a larger proportion of the group.
This scatter decreases as the true population grows.
Furthermore, the smallest groups can be seen to have a small chance of going extinct (their population dropping to zero),
which can happen when the group is too spread out to be reliably identified.

\begin{figure}
	\includegraphics[width=1\columnwidth]{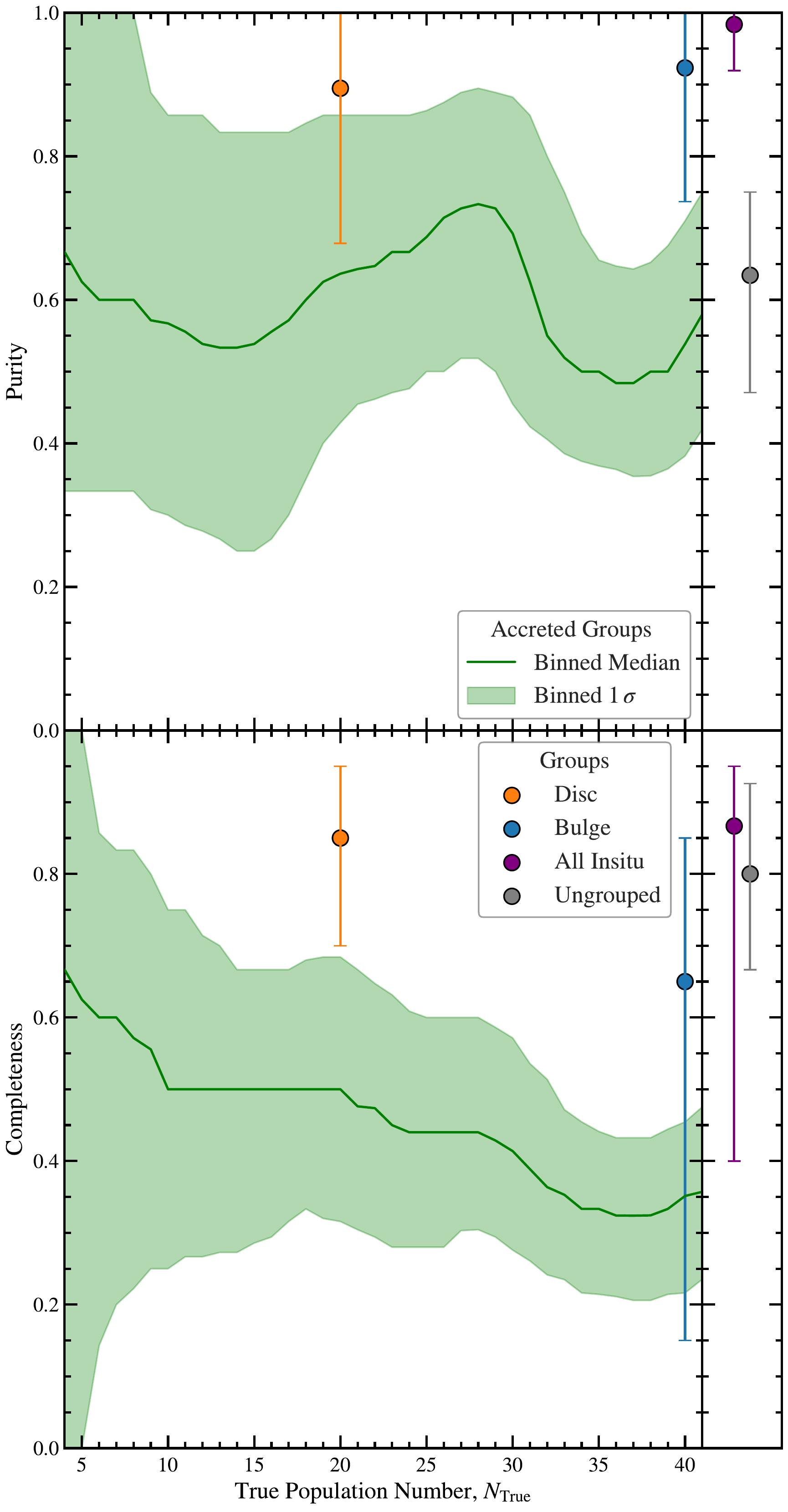}
	\smallspace{}
	\caption{
	The true group population, $N_{\mathrm{true}}$, against the purity, $P$ (top panel),
	and completeness, $C$ (bottom panel), of our fitted mock GC sample (see main text for definitions).
	We bin the results for the accreted groups in $N_{\mathrm{true}}$,
	using a Gaussian smoothing kernel of $\sigma=2$,
	and show the median and $16\%-84\%$ range
	with the shaded regions for the distribution of $N_{\mathrm{fit}}$.
	The symbols with error bars show the individual bulge and disc components, total in-situ sample, and ungrouped component.
	}
	\label{fig:Mock_NtruePurityComp}
\end{figure}

Similar to Fig.~\ref{fig:Mock_NtrueNfit},
we now consider the purity and completeness of the groups as a function of the true group richness (see Fig.~\ref{fig:Mock_NtruePurityComp}).
\Change{We find that, on average, for all groups we achieve a purity and a completeness of $\sim67\%$.
For the accreted groups, we find a purity of $\sim64\%$ and a completeness of $\sim55\%$.}
There are no clear trends in the purity against the true population number.
However, there is a clear decrease in the completeness of the groups as the richness increases,
matching the systematic under-bias in the fit population numbers seen in Fig.~\ref{fig:Mock_NtrueNfit}.

\Change{The dependence of the completeness on \Ntrue{} seen in Fig.~\ref{fig:Mock_NtruePurityComp} is}
likely a reflection of the \Change{characteristics of large versus small 
groups.}
Large groups have a greater spread in phase space due to their higher internal velocity dispersion before accretion
and tend to exist at lower energies because they experience greater dynamical friction.
These factors directly impact our ability to recover these groups.
Due to the wider spread in dynamical space, and the crowded nature of the lower energy regions,
more of the GCs are misattributed to other groups.
The smaller groups tend to be more compact in phase space, and typically exist at higher energies
(unless accreted at early times).
Providing the group itself has enough members to be reliably identified, they are recovered with greater confidence.

The total in-situ population is, in general, well recovered with very high purity.
Rarely does the methodology misidentify an accreted GC as an in-situ one in our mock tests,
with a median purity of $98\%$.
Compared to most of the accreted components, the in-situ components occupy distinct positions in the chemo-dynamical space.
Groups that do overlap with the in-situ populations tend to be older and
more massive mergers that can bring their material to the heart of the galaxy.
\Change{The purity and completeness of the bulge and disc components is marginally worse than for the in-situ population as a whole,
and the decrease is due to our method shuffling GCs between thee disc and bulge groups.}

We find that the richness of the ungrouped component is systematically overestimated by $\sim30\%$.
This is driven by the inclusion of GCs that could not be identified with their true groups
(effectively, the missing clusters that  cause the average $\sim10\%$ under-bias in the fit groups).
These are typically separated from the rest of their accreted group in phase space,
where they are difficult to identify, and so they fall into the ungrouped component.
This is also reflected in the low purity of this component, whilst the completeness is marginally better,
suggesting that ungrouped GCs are not normally being misattributed to other structures.

\subsection{Unbiased Population Estimates}
\label{Subsec:Unbiased}
\begin{figure}
\includegraphics[width=\columnwidth]{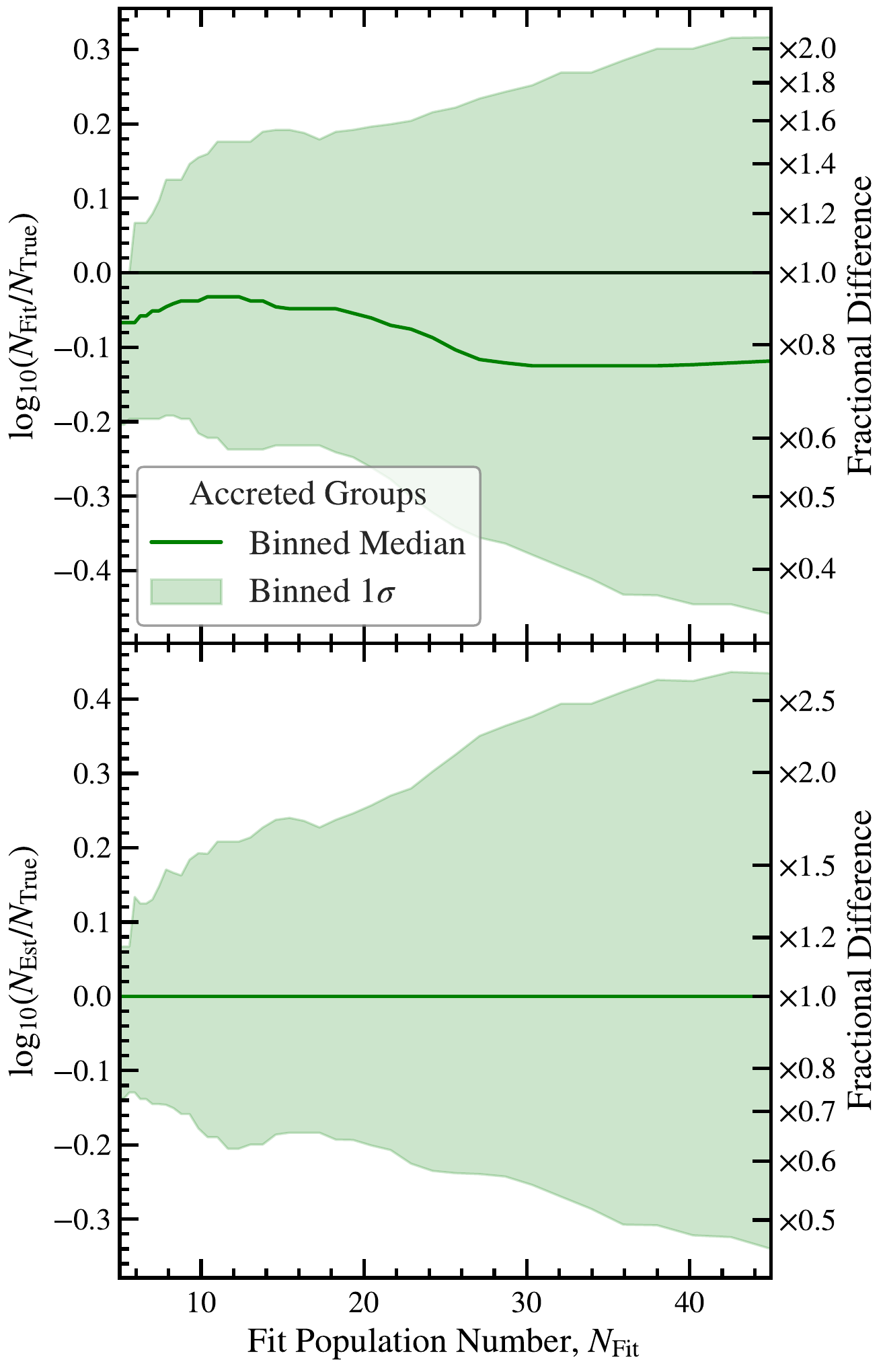}
\smallspace{}
\caption{
    Top panel: the ratio of the fit to true GC number, $\Nfit / \Ntrue$, for our mock catalogues.
	We bin the accreted groups in \Nfit, and,
	using a Gaussian smoothing kernel in log space of $\sigma=\log_{10}1.1$,
	show the median (solid green line) and $16\%-84\%$ range with shaded regions for the distribution of $N_{\mathrm{true}}$.
	These results are used to obtain an unbiased estimate, \Nest, of the likely number of GCs in each group.
	Bottom panel: the ratio of the unbiased estimate to the true GC number, $\Nest/\Ntrue$,
	following the format of the top panel.
	}
	\label{fig:Bias}
\end{figure}

From the results of our mock tests, we find that our fit population numbers, \Nfit,
is a biased estimator of the true population numbers, \Ntrue.
This can be seen explicitly in the top panel of Fig. \ref{fig:Bias},
where we consider the $\Nfit/\Ntrue$ ratio as a function of \Nfit.
Both the median bias (denoted $\beta\left(\Nfit\right)$) and the $16$ to $84$ percentiles change as a function of \Nfit{},
with an under-bias that increases for richer groups (as suggested by Fig. \ref{fig:Mock_NtrueNfit}).
Using these results we can correct our estimate
to obtain an unbiased estimate, \Nest, defined as
\begin{equation}
	\log_{10}\left(\Nest\right)= \log_{10}\left(\Nfit\right) - \beta\left(\Nfit\right).
	\label{eq:bias}
\end{equation}
as shown in the bottom panel of Fig. \ref{fig:Bias}.
The group-to-group scatter, whose $16-84$ percentiles are shown as a shaded region,
can then be used to give quantifiable uncertainty in \Nest.
We use this bias-corrected estimate and uncertainties, alongside the uncorrected estimates,
in our analysis of the MW's infall groups in Sec. \ref{Sec:MWDiscuss}.

\subsection{Choices of Chemo-Dynamical Spaces}

We also have used the mocks to investigate which dynamical spaces best recover the true GC groups,
which we define as the space that returns the highest purity and completeness.
When fitting in various dynamical spaces, we consider only the accreted components,
as our in-situ fitting scheme applies only in action-based space.
We found that the $(E,\bm{J})$ space is best at recovering the true groupings performing better than
$\bm{J}$ space alone (completeness of $\sim 50\%$),
or combinations between $E$ and angular momentum $\bm{L}$, components.
These include $(E,L_z, L_p)$, where $L_p$ is the $\bm{L}$ component in the disc plane (completeness of $\sim 50\%$),
which has been used by \massari,
and the 2-dimensional space $(E,L)$ (completeness of $\sim 43 \%$).

When fitting without the age-metallicity relation,
the purity and completeness of the accreted groups decrease by $\sim10\%$ in \EJ{} space.
This trend is similar across the other spaces tested.
Without the AMR relation, our ability to identify the in-situ components is significantly reduced.
The average total purity of the group decreases to $\sim 70\%$.
In areas where the dynamical distributions of the groups overlap,
it is this additional information that allows the memberships to be identified.
It should be noted that for our real sample of Galactic GCs,
only 96 of the 170 GCs have age-metallicity data, likely hindering our ability to confidently identify the groupings.

\section{Fitting the Galactic GCs}
\label{Sec:MWFit}
We now proceed to apply our multi-component model to the Galactic GC data.

\subsection{Observational Data}
\label{SubSec:Obs}
We make use of the largest Galactic GC sample to date, which consists of the 170 GCs studied by
\citet{vasiliev_gaia_2021} that have 6D phase space (i.e. position and velocity) data.
The GCs proper motions are based on the \textit{Gaia} Early Data Release 3 (EDR3) and
represent an improvement in precision by roughly a factor of 2
compared to the previous \textit{Gaia} Data Release 2 (DR2) measurements (\citetalias{gaia_collaboration_gaia_2021}).
Where available, we updated the \citeauthor{vasiliev_gaia_2021} GC distances with those from \citet{baumgardt_accurate_2021},
which are based on  \Change{the mean values of}
a combination of \textit{Gaia} EDR3, Hubble Space Telescope, and literature data.

To transform the observations to a Galactocentric reference frame we assume:
a Local Standard of Rest of $\mathrm{LSR} = 232.8 \, \mathrm{km}/\mathrm{s}$ \citep{mcmillan_mass_2017},
a solar radius of $R_{\odot}=8.2\,\textnormal{kpc}$,
a solar height of $z=0~\textnormal{pc}$ (assumed negligible),
and a local solar motion of $(U,V,W)=(11.1,12.24,7.25) \,\text{km/s}$ \citep{schonrich_local_2010}.

To calculate the dynamics of the GCs, we use the \agama{} package \citep{vasiliev_agama_2019} and
assume the \cite{mcmillan_mass_2017} potential of the MW, as implemented in \agama{}.
We have tried other potentials, such as that of \citet{cautun_milky_2020},
and we find that while the energy of the GCs shifts by an approximately constant value,
the individual groupings experience only minor changes.
We calculate a range of dynamical quantities, including the energy, actions, and angular momentum of the GCs' orbits.

To account for measurement errors in the positions and velocities of GCs, we create a Monte Carlo sample of 1000 points
in observed space (i.e. radial distance and velocity, and celestial proper motions)
using the quoted measurement errors which we model as Gaussians for each measured quantity.
These are then transformed into positions and velocities with respect to the Galactic Centre,
and fed into \agama{} to generate a Monte Carlo sample of dynamical quantities.
The precision of these phase-space coordinates is typically limited by distance uncertainties.

The age and chemistry data are taken from a compilation of literature data by \cite{kruijssen_formation_2019},
which provides ages and values of \Met{} for 96 GCs.
These are averaged from values derived by \cite{forbes_accreted_2010}, \cite{vandenberg_ages_2013},
\cite{dotter_acs_2010} and \cite{dotter_globular_2011}.
We neglect measurement uncertainties in metallicity since these are considerably smaller than the errors in the age.

\subsection{Fitting the Milky Way}

We now apply our model to the Milky Way.
The first step is to initialise our expectation-maximisation algorithm by postulating a set of starting groups.
We experimented with different initial groupings taken from the literature, primarily from
\massari, \citetalias{horta_chemical_2020} and \forbes.
We also tried a bootstrap-inspired approach,
relabelling one GC at a time as `ungrouped' and refitting the model to check for a higher likelihood.
In general, we find little dependence of the final groups on these small changes.

The results we present below are for the maximum likelihood model over all these variations
in the initialisation of the expectation-maximization algorithm.
The final fit is shown in Figs.~\ref{fig:MW_EJ} and~\ref{fig:MW_AMR}, and the fit parameters are listed in the Appendix.
The derived properties of the groups can be found in Table~\ref{Tab:GroupCompare} (Sec. \ref{Sec:MWDiscuss}),
group memberships are discussed in the next subsection.
In general, we find good agreement with previous work,
with all groups being distinct in either chemical or dynamical space.

At the centre of our Galaxy, we find significant overlap in dynamical space between
the two in-situ components (bulge and disc) and part of the Kraken group.
This is where we see the most change from previous literature groupings,
with a substantial increase in GCs identified as Kraken.
To separate these groups with confidence, we rely on the age-metallicity space for the GCs, where these data are available.
However, the in-situ and accreted tracks overlap for old, low metallicity GCs and cannot be distinguished.
In this region, we find that there is not enough information to separate all the GCs confidently into distinct groups.

Within the Kraken and in-situ groups, some GCs clusters can be identified with high membership probability.
We can be confident that there is an accreted group at low energy,
from both the distributions in the age-metallicity space and the dynamics.
It is the exact extent of the group that we find difficult to confidently ascertain,
and we caution against taking our proposed Kraken membership without considering these factors.

We find that Sequoia and \GES{} cannot be convincingly fit by a single group.
While there is no clear difference in the age-metallicity space, the dynamics of the two groups seem to be distinct.
The possibility that Kraken is the core of \GES{} was briefly discussed by \cite{horta_evidence_2021}.
We agree with their conclusions that Kraken and \GES{} are unlikely to have the same origin.
The dynamics of the two groups seem to be distinct,
and Kraken has a steeper metallicity-age relation (higher \pyield{}) than \GES.

We find no convincing evidence for additional subgroups,
\Change{such as the LMS-1/Wukong structure suggested to potentially contain GCs including ESO280, NGC5024, NGC5053, and Pal 5
\cite{naidu_evidence_2020, yuan_low-mass_2020, malhan_evidence_2021}.}
When we model these GCs as separate groups we find that the group becomes extinct as the GCs are absorbed into the \GES{} group.
However, we note that this group is in the regime where the number of points is less than the dimensions of the space,
and thus the groups are poorly modelled (see Sec~\ref{DF:Accreted} for details).
\Change{We can therefore not rule out the possibility of this substructure, or other small groups.}

\begin{figure*}
	\includegraphics[width=2.1\columnwidth]{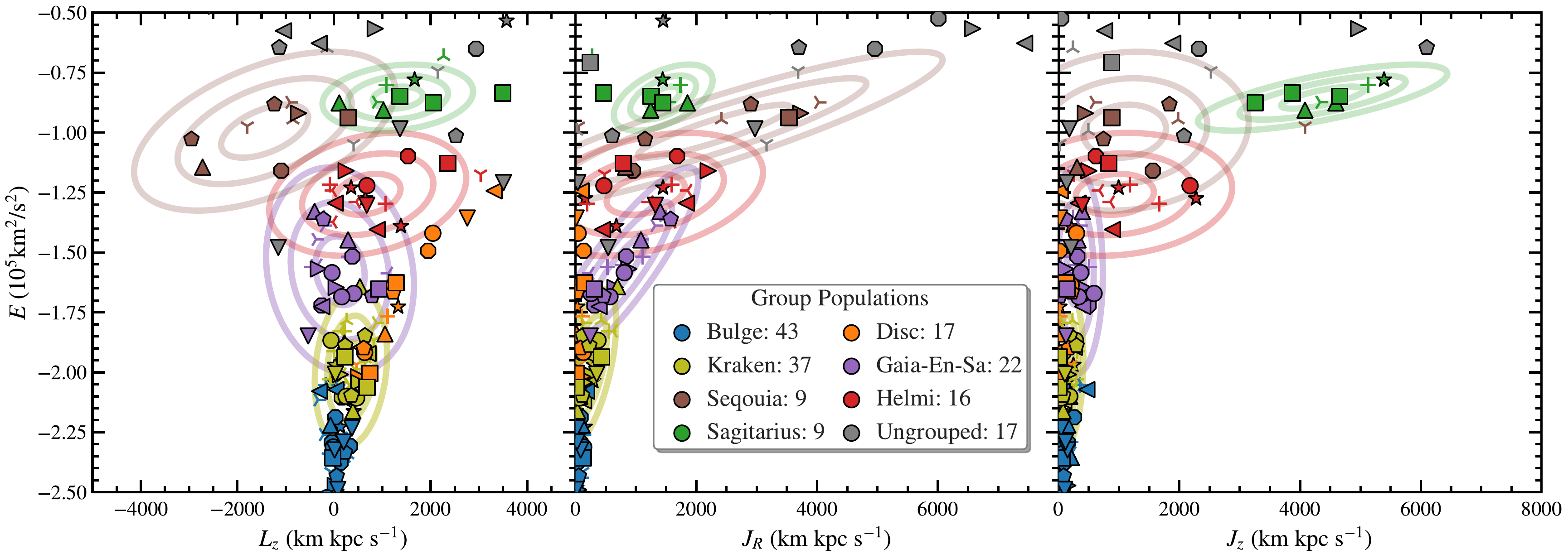}
	\caption{
	Dynamical groups in energy-action space as inferred by our chemo-dynamical model of the Galactic GC population.
	The companion age-metallicity modelling can be found in Fig.~\ref{fig:MW_AMR}.
	Symbols are the observed GCs and are consistent across panels to help identify individual GCs.
	Different colours indicate different groups, with each GC coloured by its most likely group.
	Accreted components are modelled as Gaussian distributions,
	with the 1-, 2- and 3-$\sigma$ intervals given by the contours.
	In-situ components, bulge and disc, are in the inner regions of the galaxy and typically contain the most bound GCs.
	The ungrouped component is modelled as a uniform distribution and
	contains objects accreted in small groups that cannot be reliably identified.
	Note that a few high energy GCs are beyond the axes limits and are not shown.
	}
    \label{fig:MW_EJ}
\end{figure*}

\begin{figure*}
	\includegraphics[width=2.1\columnwidth]{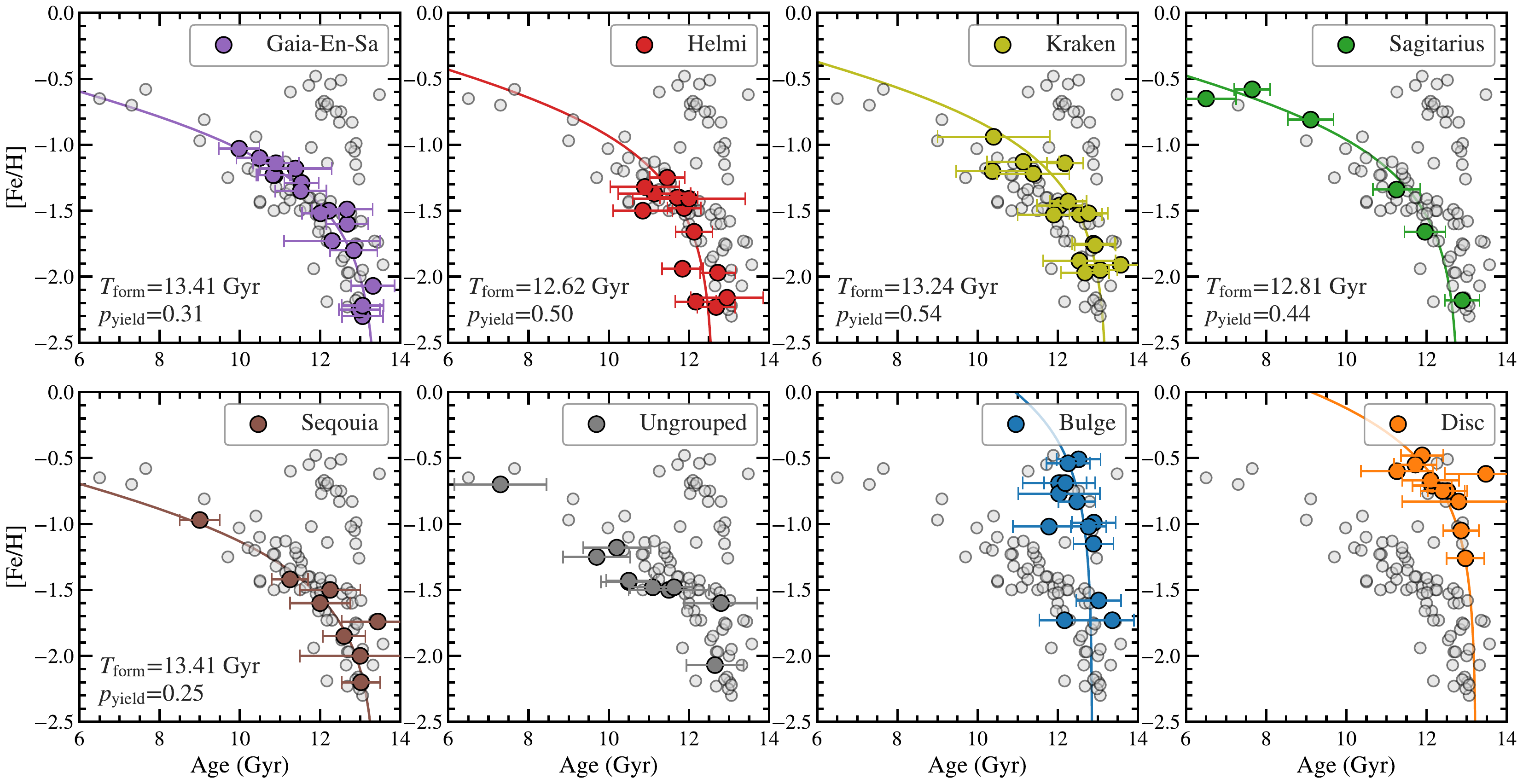}
	\smallspace{}
	\caption{
	The age-metallicity relation for the Galactic GCs split according to the component with which they are associated.
	The solid lines show the age-metallicity relation fit to the GCs associated with each component
	(see main text for details).
	}
	\label{fig:MW_AMR}
\end{figure*}

\subsection{Component Fits and Membership}
\label{Subsec:Groups}
We now discuss the groups individually.
The group membership can be found in Table~1, 
whilst the individual membership probabilities are compiled in Table A1 in Appendix \ref{Appendix:MemberProbs}.
To find the number of GCs associated with each accretion event, including uncertainties,
we use the GC membership probabilities.
We generate a Monte Carlo sample by drawing from the membership probability of each GC.
From this sample, we find the expected membership and the $68\%$ confidence interval.
It should be stressed that the mock tests of our methodology demonstrate that it is very difficult to
correctly identify the membership of each individual cluster
(although, on average, the population of accreted groups can be approximately recovered).
Therefore, we caution against placing undue emphasis on single GC memberships.

Note that when discussing the expected populations of the components in this section we refer to those returned by our fitted model,
not the bias-corrected estimates, as we are referring to the membership of the individual GCs.
The bias-corrected estimates are used in the following section (Sec. \ref{Sec:MWDiscuss}),
where we discuss the implications of the GC memberships for the properties of the accretion events of the MW.

\begin{table*}
    \label{Tab:GC_Group_Membership}
    \caption{The GC members of the accretion groups of the MW.
    Note that these are the most probable memberships.
    To see the membership probability of individual GCs, see Tab. 2.}
    \setlength\extrarowheight{2pt}
    \centering
    \begin{tabular}{lp{15cm}}
    Component & Membership\tabularnewline
    \hline
Bulge  &
Djorg2 (ESO456), Terzan6 (HP5), Terzan2 (HP3), NGC6380 (Ton1), NGC6440, Liller1, NGC6642, NGC6388, NGC6535, NGC6401, Terzan5 (11), NGC6638, NGC6528, 1636-283 (ESO452), Terzan9, NGC6624, NGC6558, Terzan4 (HP4), HP1 (BH229), NGC6325, NGC6453, NGC6626 (M28), NGC6304, NGC6522, Terzan1 (HP2), Pal6, NGC6652, NGC6266 (M62), NGC6342, NGC6637 (M69), NGC6355, NGC6540 (Djorg), NGC6717 (Pal9), NGC6293, NGC6256, NGC6517, NGC6144, VVVCL001, NGC6723, VVVCL002, NGC6171 (M107), NGC6093 (M80), Gran1,  \tabularnewline \hline

Disc  &
NGC6838 (M71), NGC5927, NGC104 (47Tuc), NGC6496, ESO93, NGC6362, NGC6366, NGC6352, BH176, Pal10, E3, NGC6218 (M12), NGC6441, Pal11, IC1276 (Pal7), Lynga7 (BH184), Pfleiderer,  \tabularnewline \hline

Gaia-En-Sa  &
NGC6205 (M13), NGC362, NGC6779 (M56), NGC7089 (M2), NGC2298, NGC1851, NGC2808, NGC7099 (M30), NGC6341 (M92), NGC5286, NGC1261, ESO-SC06 (ESO280), NGC288, NGC5139 (oCen), NGC6864 (M75), NGC5897, NGC6235, Ryu879 (RLGC2), BH140, NGC6656 (M22), NGC7078 (M15), IC1257,  \tabularnewline \hline

Helmi  &
NGC5904 (M5), NGC4147, NGC5634, NGC5272 (M3), NGC5053, Pal5, NGC7492, NGC5024 (M53), NGC6229, NGC4590 (M68), NGC6981 (M72), Rup106, NGC6584, Bliss1, NGC6426, NGC1904 (M79),  \tabularnewline \hline

Kraken  &
NGC6254 (M10), NGC6712, NGC6544, NGC5946, NGC6121 (M4), NGC6809 (M55), NGC4833, NGC6681 (M70), NGC6287, NGC5986, NGC6541, Terzan10, NGC6752, NGC6749, NGC6760, UKS1, NGC6284, Mercer5, NGC6397, Terzan3, FSR1716, FSR1735, NGC6539, Ton2 (Pismis26), Terzan12, NGC6402 (M14), Pal8, NGC6139, Djorg1, NGC6553, NGC6316, NGC4372, NGC6273 (M19), BH261 (AL3), NGC6569, NGC6333 (M9), NGC6356,  \tabularnewline \hline

Sagitarius  &
Arp2, NGC6715 (M54), Terzan8, Terzan7, Pal12, Whiting1, Munoz1, Kim3, Ko1,  \tabularnewline \hline

Seqouia  &
NGC5466, NGC6101, NGC7006, NGC3201, IC4499, Pal13, NGC5694, Pal15, AM4,  \tabularnewline \hline

Ungrouped  &
Ryu059 (RLGC1), Ko2, Pal3, NGC6934, Crater, Pyxis, Segue3, Pal14, AM1, Eridanus, Pal4, Pal1, NGC5824, NGC2419, Laevens3, Pal2, FSR1758,  \tabularnewline 
    \end{tabular}
\end{table*}

\subsubsection{In-Situ}
\label{Component:InSitu}
We find that $\sim60$ of our GCs are likely to have an in-situ origin,
with the bulge group containing an expected $42^{+2}_{-1}$ GCs, and the disc group an expected $17^{+2}_{-2}$ GCs.
This is comparable with the numbers of \massari{} who find $62$ in-situ GCs
and \cite{kruijssen_formation_2019} who predict $67$ out of their $157$ to have an in-situ origin.
The slightly lower total number is again likely the result of our larger Kraken component.
Individually, our bulge group is larger and our disc smaller than \massari's 36 bulge and 26 disc GCs.
We find that there is little information to distinguish the disc GCs at low radius and energy from the bulge component.

The bulge GCs typically have energies below $-2\times\uEe$ and apocentres below $5\,\mathrm{kpc}$.
This component does not have any significant rotation and has an AMR track that is slightly steeper than the disc.
The disc extends to higher energy, but all the GCs have
$z_{\mathrm{max}}<6~\rm{kpc}$, eccentricity, $e<0.6$, and $\mathrm{circularity}>0.5$.
In the very centre of the Galaxy, the disc overlaps with the bulge,
leaving a hole in the middle of the radial distribution, with no  disc GCs having an apocentre $<3\,\mathrm{kpc}$.

We find that VVVCL001, VVVCL002 and Gran1,
previously uncategorised by \massari,
are likely to be bulge members, but could also plausibly fit into the Kraken group.
This is in agreement with the work studying the individual GCs.
\cite{gran_hidden_2021}, find Gran1 either as in-situ or an ancient merger such as Kraken.
\cite{minniti_survival_2021} find VVVCL002 as the GC closest to the centre of our Galaxy, strongly suggesting that likely it is of in-situ origin.
However,
\cite{fernandez-trincado_vvv_2021} find VVVCL001 to be very metal poor GC on an eccentric orbit, and tentatively suggest an accretion origin,
likely Sequoia or \GES.

Other noteworthy observations of in-situ GCs include:
\begin{itemize}
\item We find that 10 GCs previously associated with the disc are, instead, probable Kraken members.
	All these GCs lack age/metallicity data or have $\Met{}<-1.5$.
	This highlights the overlapping nature of the in-situ and lower energy components.
\item Liller1 and NGC6388 are very likely part of the bulge, in agreement with \citetalias{horta_chemical_2020}.
\item ESO93, previously uncategorised, is almost certainly a member of the disc.
\item E3 (ESO37-1) has previously been associated with the Helmi streams \citep{koppelman_characterization_2019}.
  We find that it has over $94\%$ probability of being a disc member,
  reflecting its position on the in-situ AMR track (in agreement with \citealt{kruijssen_kraken_2020}).
  It does however reach the highest height above the plane of any disc cluster, $z_{\mathrm{max}}\approx5.5\,\mathrm{kpc}$.
\end{itemize}

\subsubsection{Kraken}
\label{Acc:Kraken}

We expect $37_{-1}^{+2}$ GCs in the Kraken group, a substantial increase from the $25$ in \massari,
with the difference consisting of a net contribution of 9 from the disc and 4 from the \GES{} component.

We find that two GCs previously unclassified by \massari, UKS1 and Mercer5, are highly likely to be Kraken members.
However, previous works studying the individual clusters believed that they were likely members of the bulge.
Notably  UKS1, as an old but metal poor GC, was suggested to  belong to the bulge in \cite{fernandez-trincado_enigmatic_2020},
but the result was highly dependent on its then very uncertain distance.
With our more recent distance estimates of \cite{baumgardt_accurate_2021}, we find that it is a likely Kraken member.

On its discovery in \cite{longmore_mercer_2011},
Mercer5 was believed to be a typical bulge GC due to its position in the inner galaxy ($\sim5.5\,\mathrm{kpc}$)
and in subsequent extensive chemical follow up, \cite{penaloza_chemical_2015}.
With the chemo-dynamical information used in our methodology,
the GC is classified as likely part of the Kraken group.

Our model predicts that the energy distribution of the Kraken group is approximately normally distributed with
a mean of $-2\times\uEe$ and dispersion values of $0.1\times\uEe$.
This is higher Energy distribution than other selections in the literature, which typically give values of $E<-2\times\uEe$
(such as \massari, \citealt{horta_evidence_2021}).
Notably, our Kraken group seems to have bridged the gap seen in stars by \citet{horta_evidence_2021} at energies $-2<E/\uEe<-1.85$.
Furthermore, unlike previous results, our Kraken group has net prograde motion,
with angular momentum  $L_z$ distributed with a mean of $\sim 350\uJ$ and a dispersion of $\sim250\uJ$.
This prograde bias suggests that perhaps some disc GCs have been included in the group.

For GCs on lower energy orbits, and without age-metallicity information,
or for those that have low metallicity where the in-situ and accreted branches overlap,
we have found that distinguishing between membership of Kraken or the in-situ groups is difficult.
In future,
further chemistry information may allow us to distinguish better between the accreted and in-situ components at low energy.

\subsubsection{Sagittarius}
\label{Acc:Sagittarius}

Our Sagittarius group contains an expected population of $9^{+1}_{-0}$ GCs.
Due to recent accretion and tidal stripping, much of its material is in an easily identifiable stream.
This allows 7 GCs to be identified with a high degree of certainty as being associated with the Sagittarius dwarf:
Terzan7, Arp2, Terzan8, Pal12, Whiting1 and M54 (NGC6715), which is believed to possibly be the nucleus of Sagittarius.
\citep{bellazzini_globular_2020, antoja_all-sky_2020, law_assessing_2010,penarrubia_identification_2021}.
We find that these GCs have a near-certain membership.
We also find that two uncategorised GCs, Munoz1 and Kim3, also have over $90\%$ probability of membership,
and Koposov1 has $\sim70\%$ probability.
Koposov1 has been previously noted to lie close to a distant branch of the Sagittarius stream \citep{koposov_discovery_2007, paust_reinvestigating_2014}.
These high probabilities are driven by the Sagittarius group's high group density in dynamical space and
a distinct age-metallicity branch.

Several other GCs have been tentatively linked to Sagittarius in the literature, but we find no other likely members.
Compared to the literature, we find:
\begin{itemize}
\item Pal2 has been proposed to lie on the trailing arm of the stream \citep{bellazzini_globular_2020, law_assessing_2010}.
However, we find that it has a $81\%$ probability of being ungrouped
and a $14\%$ probability of being associated with Sequoia
\item  NGC2419 and NGC5824 are commonly linked to Sagittarius
\citep{antoja_all-sky_2020,bellazzini_globular_2020,penarrubia_identification_2021},
but we find them almost certainly to be ungrouped.
The orbit of NGC2419 is more radial than the average, more vertical Sagittarius orbit,
and NGC5824 is at lower energy than the other GCs.
\item  NGC5634 and NGC5053 have been proposed as lying on ancient wraps of the stream \citep{bellazzini_globular_2020}.
We find that they are not likely members (in agreement with \citealt{law_assessing_2010});
they are near certain members of the Helmi group.
\item  AM4 was attributed to Sagittarius by \forbes{} based on chemistry since, at the time,
	AM4 did not have {\it Gaia} kinematics
	(for this reason \massari{} did not assign the cluster to a group).
We find that, as a prograde cluster, its orbit is  incompatible with the Sagittarius orbit.
Instead, we find that it is a likely member of Sequoia, but has a $17\%$ chance of being ungrouped.
\item Before Koposov 1 and 2 (Ko1, Ko2) had measured radial velocities,
	\cite{paust_reinvestigating_2014} suggested that they could plausibly lie on the Sagittarius stream.
Improved observations by \cite{vasiliev_gaia_2021} have placed Ko1 as a likely member, but Ko2 is almost certainly ungrouped.
\end{itemize}

\subsubsection{Gaia-Enceladus-Sausage}
\label{Acc:GES}
Our analysis gives $23^{+2}_{-1}$ GCs in the \GES{} structure,
in good agreement with \massari{} ($25$) and \forbes{} ($28$).
The \GES{} group is consistent with having no net rotation and has an energy distribution
with a mean of $-1.58\times\uEe$ and a dispersion of $0.15\times\uEe$
This approximately agrees with previous literature selections:
$-1.75<E/\uEe<-1.3$ in \cite{horta_evidence_2021},
and $-1.86<E/\uEe<-0.9$ in \massari.
The latter also notes that the apocentres are mostly less than 25 kpc,
in good agreement with \cite{deason_apocenter_2018}.
We also find that our \GES{} GC apocentres lie between $10\,\mathrm{kpc}$ and $20\,\mathrm{kpc}$.

\begin{itemize}
\item The previously unclassified clusters, Ryu879 (RLGC2) and BH140, are likely members of \GES.
    \item  In contrast to \cite{myeong_evidence_2019},
	    we infer that NGC4147  and NGC6981 (M72) are part of the Helmi Streams and
NGC7006, Pal15 and  NGC5694 are associated with Sequoia.
\item We find that four GCs that have been previously associated with \GES{} are now associated with the Kraken structure
	(NGC4833, NGC6284, Djorg1 and Terzan10).
\item Pal2 is likely to be an ungrouped GC, despite being linked to \GES{} by \massari{} and \forbes.
  We find it is at higher energy ($-1.1\times\uEe$) than the rest of the \GES{} group.
\item Our method classifies \OmC{} as an almost certainly \GES{} member,
	in agreement with a tentative classification by \massari.
  This cluster has been claimed to be the nucleus of Sequoia by \cite{myeong_evidence_2019}
  and is discussed more in the following section.

\end{itemize}

\subsubsection{Sequoia}
\label{Acc:Sequoia}
We predict $9^{+1}_{-0}$ members in the Sequoia group,
comparable to the 7 attributed by \cite{myeong_evidence_2019} and \massari,
and the 9 by \forbes.

The Sequoia group has a narrow energy distribution, with a mean of $-1\times\uEe$ and a dispersion of $0.1\times\uEe$.
However, the angular momentum has a wide distribution, with a mean of $-1400\uJ$ and a dispersion of $900\uJ$.
This is noticeably smaller than other selections in the literature,
such as that by \cite{myeong_evidence_2019} (and \massari)
of $-1.5<E/\uEe<-0.7$ and $-3700<L_{z}\uJ<-850$.
The distribution in $J_{z}$ and $J_{R}$ is very broad, stretching across the space.

\begin{itemize}
\item FSR1758 was characterised by \cite{barba_sequoia_2019},
	where its unusually large size (for a GC) lead to it being dubbed as ``a Sequoia in the garden".
  Later \cite{myeong_evidence_2019}, suggested that it is the nucleus of a retrograde accretion event,
  with the entire accretion event being named Sequoia after the first paper.
  However, other works, such as \cite{romero-colmenares_capos_2021},
  have found the links of this GC with the main structure tenuous,
  making it more likely that it belongs to a different group such as \GES.
  We find that FSR1758 has a $28\%$ chance of being associated with Sequoia and a $62\%$ chance of being ungrouped.
  This is primarily driven by its position at lower energy than the rest of Sequoia.
  It should be noted that, as the heart of an accretion event,
  it is plausible that FSR1758 has suffered more dynamical friction and fallen to lower energy
  than other accreted material.

\item \OmC{} is another cluster that has been previously classified as
  a key Sequoia member.
  Due to its peculiar chemistry,
  \OmC{} has long been suspected by some to be the nucleus of a dwarf galaxy \citep{bekki_formation_2003},
  thought to have a mass of $\sim10^{10}M_{\odot}$ \citep{valcarce_formation_2011}.
  \forbes{} and \cite{myeong_evidence_2019} believed this to be Sequoia, based on its retrograde orbit.
  As noted in \cite{myeong_discovery_2018}, \OmC{} could have sunk to lower energy by dynamical friction.
  We find it has a near-certain \GES{} membership.

\item As briefly discussed in the Sagittarius Section, the AM4 cluster has been tentatively linked to Sagittarius before,
	but we find it is likely a Sequoia member.
  If true, it is the youngest Sequoia member, with an age of ${\sim}9\,\mathrm{Gyr}$,
  approximately $2\,\mathrm{Gyr}$ younger than the rest of the group.
  Its position in energy-action space is also unusual for Sequoia;
  while at the centre of the angular momentum distribution, it has negligible radial action and
  is primarily on a vertical orbit, in agreement with Sagittarius.
  We flag this cluster as a potential outlier.
\end{itemize}

\subsubsection{Helmi Streams}
\label{Acc:Helmi}
\cite{koppelman_characterization_2019} identified 7 GC members,
(NGC4590, NGC5272, NGC5904, NGC5024, NGC5053, NGC5634, NGC6981).
This group was tentatively expanded by \massari{} to include 3 additional members,
suggesting a total of 10 members.
We find that these are indeed very likely members, inferring $15\pm1$ GCs in the Helmi streams group.
Our Helmi stream structure has a narrow energy distribution,
with a mean of $\sim -1.25\times\uEe$ and a dispersion of $0.08\times\uEe$.
The distribution in angular momentum is broader,
with a mean of $\sim700\uJ$ and a dispersion of $700\uJ$.
This is in approximate agreement with values in the literature
\citep{koppelman_characterization_2019,massari_origin_2019,naidu_evidence_2020}.

\begin{itemize}
\item Our Helmi group includes 4 probable members that were previously associated with \GES,
	NGC4147, NGC7492, NGC6229 and NGC1904 \citep{myeong_evidence_2019,forbes_accreted_2010}.
\item We find that the previously unclassified cluster, Bliss1, is likely a Helmi member.
\item NGC6441 is almost certainly a member of the disc and not of a Helmi Stream or Kraken,
	as suggested by \massari.
\end{itemize}

\subsubsection{Ungrouped}
\label{Acc:Ung}
We find $17\pm1$ GCs that are ungrouped or do not fall into any of the other accretion groups.
This is likely to be a collection of GCs from different low mass dwarfs
that have otherwise left no significant stellar material to be identified.
In their equivalent high energy group,
\massari{} (and \forbes) identified 11 members.

Pal1 has previously been linked with the disc (\massari) and \GES{} (\forbes),
but instead we find that it has a very high probability of being ungrouped (in agreement with \citealt{kruijssen_kraken_2020}).
It is on a circular orbit compatible with the outskirts of the disc, but it is young and
has high \Met{} similar to the young Sagittarius GCs Whiting1 and Terzan7.
Other hints from its chemistry support this view \citep{sakari_detailed_2011}.
\cite{naidu_evidence_2020} associated Pal1 with a newly identified Aleph structure due to chemo-dynamical similarities.

Clusters NGC5824 and NGC2419 have been previously associated with Sagittarius,
but we find it is highly likely that they have a different accretion origin; we associate them with the ungrouped component.
\Change{NGC5824 has also been associated with the Cetus stream \cite{yuan_revealing_2019, chang_is_2020}.
As the only associated GC with the structure, it would be correct to categorise it as Ungrouped.}

Pal2 and NGC6934 lie close together in \EJ{} space, at an energy just below that of the Sagittarius group.
These clusters have previously been associated with \GES{} \citepalias{massari_origin_2019},
but we find that they are at higher energy than other \GES{} clusters.
The fit of the \GES{} group is improved by their removal.

The rest of the group members are all at high energy ($E>-0.75\times\uEe$).
We find 4 GCs uncategorised by \massari{} (Ryu059, Ko2, Segue3 and Laevens3) that are highly likely to be ungrouped.
We find no obvious subgroups in these GCs.

\subsection{Completeness of In-Situ Sample}
We expect the GC members of the in-situ components to be phase mixed and consistent with having axisymmetric distributions.
This can be tested by checking that $\phi$, the angle in the plane of the disc, is uniformly distributed with a Kuiper test.
Similar to the more commonly used KS test,
the Kuiper test can be used to quantify if the cumulative of two distributions are statistically compatible,
but it is particularly suited to test distributions of modular variables as the statistic is
invariant under cyclic transformations of the random variable \citep{kuiper_tests_1960}.

We find that the in-situ components are not consistent with being axisymmetric,
with  p-values of $0.039$ for the bulge, $0.154$ for the disc, and $0.03$ for the combined sample.
This is consistent with an overabundance of GCs on the near side of the Galactic Centre.
Binning the GCs into quarter slices with the Sun at $\phi=0$,
we find that our 60 in-situ clusters are distributed in angle as:
24 in $-\pi/4<\phi\leq\pi/4$, 11 in $\pi/4<\phi\leq3\pi/4$, 14 in $-3\pi/4<\phi\leq-\pi/4$,
11 in $\phi\leq-3\pi/4 ,3\pi/4<\phi$.
In the grouping of \massari, out of 62 in-situ clusters, 30 fall in $-\pi/4<\phi<\pi/4$,
with the rest evenly distributed.
These results suggest either incomplete observations, with on the order $30$ missing in-situ clusters,
or an otherwise undiscovered structure in the GCs identified as in-situ.

The Kuiper test can also be applied to the angles of the orbital actions of the GCs,
which should also be uniformly distributed between $0$ and $2\pi$ if the group is phase mixed.
We find that the accreted components are all consistent with being phase mixed,
apart from Sagittarius (with a combined p-value of less than $10^{-4}$).

\section{Inferring the Properties of Accreted Galaxies}
\label{Sec:MWDiscuss}

\begin{table*}
	\caption[Properties of the Galactic GC accretion groups.]
	{Properties of the Galactic GC accretion groups, as derived in this work and other works in the literature.
	  The first section gives our results; the second column gives the expected number of GCs (\Ngc),
	  including 68\% confidence interval, as inferred from our chemo-dynamical model.
	  From this, using the halo mass-number of GCs relation of \protect\cite{burkert_high-precision_2020},
	  we find the halo mass of the accretion event.
	  The halo mass is used to further infer the stellar mass (fourth column)
	  from the stellar mass-halo mass relation of \protect\cite{behroozi_universemachine_2019}.
	  The halo mass and stellar masses of the ungrouped (and in-situ components)
	  cannot be estimated in the same way as for the accreted components.
	  The tests on mock catalogues have showed a bias in our method.
	  The fifth and sixth column give the halo and stellar masses, $\hat{\rm{M}}_{\rm halo}$ and $\hat{\rm{M}}_\star$,
	  corrected for this bias and also including the considerable group-to-group dispersion's in recovering the true number of GCs
	  (details in the main text).
	  The second section gives relevant values from the literature, with references given below.
	  Our summed total stellar mass is the amount accreted by the named groups,
	  which are assumed to make up the bulk of the contribution to the stellar halo.
	}
	\label{Tab:GroupCompare}
    \setlength\extrarowheight{5pt}
\makebox[\textwidth][c]{
\begin{tabularx}{2\columnwidth}{l  l c c c c  c c c}
\toprule
& \multicolumn{5}{c}{This Work} & \multicolumn{3}{c}{Literature}\\
\cmidrule(lr){2-6}\cmidrule(lr){7-9}
Acc. Event & \Ngc &  $\log_{10}\rm{M}_{\rm halo}$ & $\log_{10}\rm{M}_\star$
&  $\log_{10}\hat{\rm{M}}_{\rm halo}$ & $\log_{10}\hat{\rm{M}}_\star$
& \Ngc & $\log_{10}\rm{M}_{\rm halo}$ & $\log_{10}\rm{M}_\star$ \tabularnewline[.1cm]
\midrule
Gaia-En-Sa & $23_{-1}^{+2}$ & $11.00_{-0.13}^{+0.13}$ & $8.51_{-0.27}^{+0.27}$
& $11.07_{-0.22}^{+0.20}$ & $8.64_{-0.40}^{+0.44}$
& ${20}$, ${28}$ & ${10.98\pm0.08}$, ${\left(11{\text -}11.7\right)}$
&  ${8.43\pm{0.15}}$, ${\left(8.7{\text-}9.7\right)}$  \tabularnewline
Helmi & $15_{-1}^{+1}$ & $10.80_{-0.15}^{+0.16}$ & $8.13_{-0.32}^{+0.32}$
& $10.87_{-0.19}^{+0.21}$ & $8.26_{-0.42}^{+0.38}$
& ${5}$, ${10}$  & ${10.74\pm0.1}$ &  ${7.96\pm{0.18}}$, $8$ \tabularnewline
Kraken & $37_{-1}^{+2}$ & $11.22_{-0.10}^{+0.10}$ & $8.95_{-0.22}^{+0.22}$
& $11.38_{-0.22}^{+0.27}$ & $9.25_{-0.50}^{+0.45}$
& ${13}$, ${25}$ & ${10.92\pm0.1}$ &  ${8.28\pm{0.18}}$, ${8.7}$ \tabularnewline
Sagittarius & $9_{-1}^{+0}$ & $10.49_{-0.20}^{+0.20}$ & $7.51_{-0.41}^{+0.41}$
& $10.58_{-0.23}^{+0.21}$ & $7.69_{-0.47}^{+0.43}$
& ${7}$, ${8}$ & ${10.94\pm0.1}$, ${>10.8}$ &  ${8.44\pm{0.22}}$ \tabularnewline
Sequoia & $9_{-0}^{+1}$ & $10.52_{-0.20}^{+0.20}$ & $7.58_{-0.41}^{+0.41}$
& $10.61_{-0.22}^{+0.21}$ & $7.74_{-0.46}^{+0.42}$
& ${3}$, ${7}$ & ${10.70\pm0.06}$,  ${\left(10{\text -}10.7\right)}$ &
${7.90\pm{0.11}}$, ${\left(6.7{\text-}6.9\right)}$ \tabularnewline
Ungrouped & $17_{-1}^{+1}$ &  - &  - & - & - &${11}$ & - & - \tabularnewline
Bulge & $42_{-1}^{+2}$ &  - &  - & - & - & $36$ & - & - \tabularnewline
Disc & $17_{-2}^{+2}$ &  - &  - & - & - & $26$ & - & -  \tabularnewline
\midrule
TOTAL & $170$ &  - &  $9.19_{-0.17}^{+0.17}$ & - & $9.42_{-0.39}^{+0.44}$ & $151$ & - & ${9.15_{-0.15}^{+0.11}}$, ${8.95_{-0.05}^{+0.09}}$  \tabularnewline
\bottomrule
\end{tabularx}
}
\footnotesize{
References:
\cite{kruijssen_kraken_2020},
\cite{massari_origin_2019},
\cite{myeong_evidence_2019},
\cite{horta_evidence_2021},
\cite{laporte_footprints_2019,laporte_influence_2018},\\
\cite{koppelman_characterization_2019},
\cite{deason_total_2019} (Total),
\cite{mackereth_weighing_2020} (Accreted)
}
\end{table*}

From the fit to the MW data,
we have found the likely population numbers of each of the GC groups that have been accreted onto the MW.
Using the results from the mock tests,
we correct these fit populations to give
an unbiased estimate of the true population numbers with estimates of the uncertainty (see Sec. \ref{Subsec:Unbiased}).
We now use the \MhNgc{} relation (Eq.~\ref{Eq:nGC}) to estimate the mass of the progenitor dwarf galaxies.
We also include the theoretical uncertainties in this relation (normally distributed as $\sigma_{\mathrm{Ngc}}$),
given in Figure 2 of \cite{burkert_high-precision_2020} as
$\sigma_{\mathrm{Ngc}}/N_{\mathrm{GC}} \approx \left( N_{\mathrm{GC}}/2\right)^{-1/2}$.
This relationship then gives a probability density function (PDF):
\begin{equation}
    p\left(\Ngc|M\right) = N\left(\mu=\frac{M}{5\times{10}^{9}\Msun}, \sigma= \sqrt{\frac{M}{2.5\times{10}^{9}\Msun}}\right),
\end{equation}
where N is the normal distribution.

Using Bayes theorem, this relationship can be inverted:
\begin{equation}
    p\left(M|\Ngc\right) =
    \frac{p\left(M\right)}{p\left(\Ngc\right)}p\left(\Ngc|M\right).
\end{equation}
For the mass prior, $p\left(M\right)$,
we adopt the mass function of
\citet[][Eqs. 7,8; Fig. 5]{boylan-kolchin_theres_2010}.
This relation gives the distribution in terms of the ratio of the mass of the accreted satellite to
the virial mass of the host galaxy at present day.
For this we take the mass of the MW as $1.17 \times 10^{12} M_{\odot}$ \citep{callingham_mass_2019}.
The $\Ngc$ prior, $p\left(\Ngc\right)$, is effectively the normalisation factor.

\begin{figure}
	\includegraphics[width=\columnwidth]{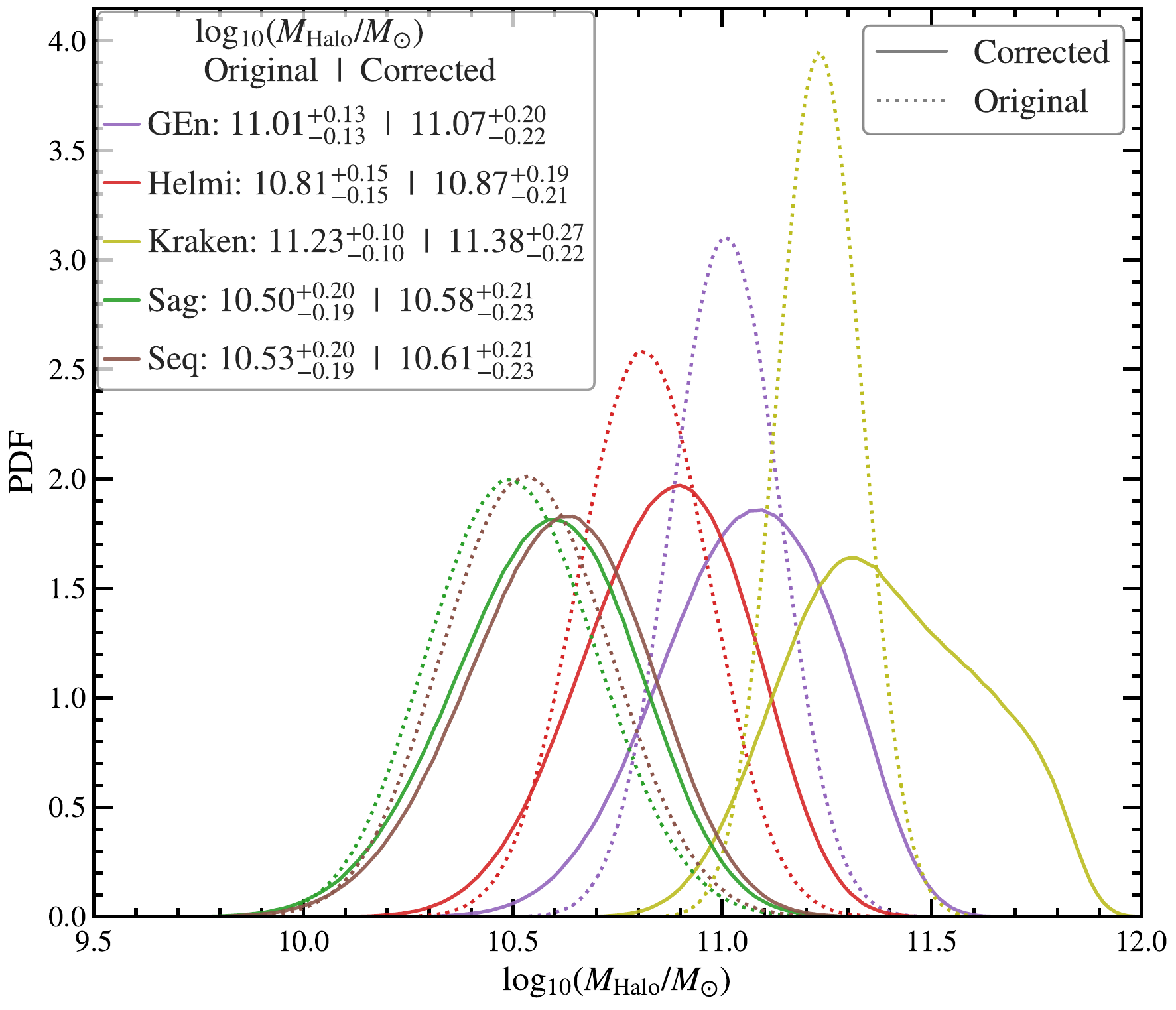}
	\smallspace{}
	\caption{
	The halo mass likelihood for accreted satellites.
	This is derived from their probable populations of GCs using the relation between halo mass and number of GCs of
	\protect\cite{burkert_high-precision_2020}.
	This includes uncertainties from grouping the clusters and theoretical uncertainties from the relation.
	The dotted lines show the estimates using the original group population numbers as found from the fitting method.
	The solid lines show the estimates when correcting for bias and including the group-to-group scatter in recovering the true number of GCs
	(see main text for details).
	}
	\label{fig:MW-Mass}
\end{figure}

To account for the probabilistic nature of our GC memberships,
we randomly draw population samples of the accretion groups from the membership probabilities.
Each drawn population, \Ngc{}, is then used to derive a PDF, $p\left(M|\Ngc\right)$.
The total mass PDF is then the sum over this sample.
These results are given as probability density functions (PDFs) in Fig. \ref{fig:MW-Mass}.
We note that this methodology has significant limitations, as discussed by \cite{kruijssen_formation_2019}.
We do not include any redshift dependence in the \MhNgc{} relation, assuming that this is sufficiently flat.
The errors assumed in this relation are theoretical, and they, as well as the underlying relation,
are the subject of extensive debate in the literature.

\begin{figure}
\includegraphics[width=\columnwidth]{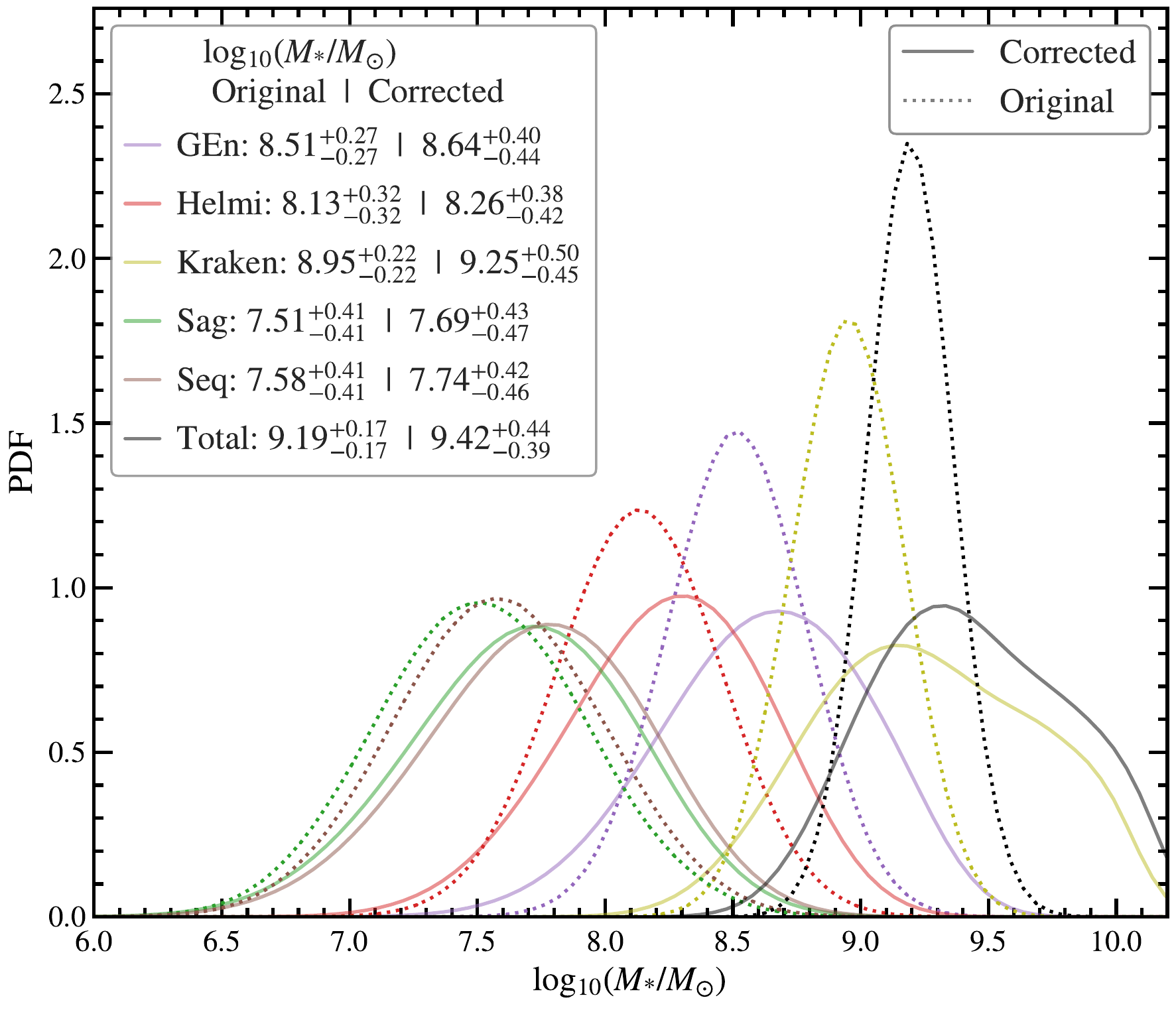}
\smallspace{}
\caption{
	The stellar mass likelihoods of accreted galaxies.
	These were calculated by assuming the stellar mass-to-halo mass relation of \protect\cite{behroozi_universemachine_2019}
	(assuming $z=0$) to transform the halo mass PDFs of Fig.~\ref{fig:MW-Mass}.
	The effect of accounting for the redshift of accretion is considered in Appendix \ref{Appendix:zSMHM}.
	The dotted lines show the estimates using the original group population numbers as found from the fitting method,
	the solid lines the corrected unbiased estimates (see main text for details).
	}
\label{fig:MW-StellarMass}
\end{figure}

The estimated halo masses (including uncertainties) can be combined with a stellar mass-halo mass (SMHM) relation to infer
the likely stellar masses of the accretion events.
We use the stellar mass-halo mass relation of \cite{behroozi_universemachine_2019}, including the given uncertainties.
This relation has a non-negligible dependence on redshift; here we assume the $z=0$ relation.
The resulting stellar-mass PDFs are presented in Fig. \ref{fig:MW-StellarMass};
the median and $68\%$ confidence limits of these results are summarised in Table~\ref{Tab:GroupCompare}.
Alternatively, one could assume that star formation in a galaxy stops approximately around the redshift of accretion.
This assumption has the effect of lowering the stellar masses, particularly of the older accretion events such as Kraken.
Further discussion of this effect can be found in Appendix \ref{Appendix:zSMHM}.

In general, we find good agreement with results in the literature, particularly for \GES, the Helmi streams and Sequoia.
We have more GCs than earlier work, and so we find slightly higher halo and stellar masses,
but nonetheless consistent within the uncertainty interval.
The greatest difference between our results and those in the literature is the higher mass for Kraken that we infer.
This reflects the considerable increase in \Ngc{} that we attribute to the Kraken event,
but we note the difficulty in distinguishing between Kraken and the in-situ component.
As a result, we find better agreement with the higher mass estimates,
our stellar mass of $\sim 10^{9} \Msun$ being closer to that of \cite{horta_evidence_2021},
who estimated a log stellar mass of $8.7$, approximately twice the stellar mass of \GES.

From dynamical arguments, the total mass of Sagittarius is thought to be greater than $6\times{10}^{10}\Msun$
\citep{laporte_footprints_2019,laporte_influence_2018},
around twice our estimate of its halo mass, $\sim 3.4\times{10}^{10}\Msun$.
However, recent work has claimed to find up to an additional 20 plausible GCs in the body of Sagittarius
\citep{minniti_eight_2021,minniti_discovery_2021}.
In our analysis, this would suggest a log halo mass of $\sim11.16$.
Whilst this revised estimate is high compared to the majority of the literature,
it agrees with recent work by \cite{bland-hawthorn_galactic_2021},
who suggested that the infall mass of Sagittarius has been underestimated because of rapid tidal stripping.
This could make its halo mass comparable to the LMC (with halo mass $\sim 10^{11}M_{\odot}$).

We estimate the total stellar mass accreted to be $\sim{2.6}_{-1.5}^{+4.6}\times{10}^{9}\Msun$
(or $\sim{1.5}_{-0.4}^{+0.7}\times{10}^{9}\Msun$ with the uncorrected mass estimates),
obtained by summing the stellar masses of the fitted named groups.
This does not include estimates of the stellar mass of the ungrouped component,
which we assume to be subdominant to the larger accretion events.
In comparison with results in the literature,
our estimate is higher than the results of \cite{deason_total_2019}
who estimate the total stellar mass in the halo as $1.4\pm0.4\times10^{9}\Msun$,
but within one sigma.
Similarly, \cite{mackereth_weighing_2020} estimate a total stellar mass of $1.3^{+0.3}_{-0.2}\times10^9\Msun$,
but then conclude that only $\sim70\%$ ($0.9^{+0.2}_{-0.1}\times10^9\Msun$) has been accreted.
We note that including a redshift dependence in the stellar mass-to-halo mass
relation reduces the individual and total stellar masses (see Appendix~\ref{Appendix:zSMHM}).
However, the systematic uncertainties on both the redshift dependence of the SMHM relation and
the uncertainties on the accretion time make the extent of this effect unclear.

\section{Conclusions}
\label{Sec:Conclusion}

We have introduced a multi-component model for the GCs in the MW that splits the population into
three individual constituents: bulge, disc, and stellar halo.
The latter is further decomposed into the individual large accretion events that built up the Galactic stellar halo.
The identification of the components has been performed in a chemo-dynamical space for GCs that combines information
on the age-metallicity relation with the orbital energy, $E$, and the action, $\bm{J}$.
Our study has aimed to obtain an objective and statistically robust identification of accreted GCs groups.
These have been modelled as multivariate Gaussian distributions in $(E,\bm{J})$ space
that follow the age-metallicity relation proposed by \citet{forbes_reverse_2020}.

We have extensively tested our methodology using mock GC catalogues built from
the \auriga{} suite of zoom-in simulations of MW-like galaxies.
The mocks roughly reproduce the number, radial distribution and, by construction,
the age-metallicity relation of GCs in our galaxy.
These show that the best space for our modelling is the combined energy-action space;
including the age-metallicity information improves our results by $\sim 10\%$.

Our approach recovers, on average, the population numbers of the GCs associated with each merger event in the simulations
\Change{with an under-bias of $\sim10-20\%$},
but with considerable group-to-group scatter.
We find that a proportion of `missing' clusters causing the under-bias cannot be associated with their true groups and are identified as ungrouped.
Using our mock test results, we create an unbiased estimate for the true population of the fitted groups with realistic uncertainties of our methodology.
However, the grouped GCs are not always associated with the actual objects brought in by a particular merger event;
the fit accretion groups have an average purity and completeness of only \Change{${\sim}60\%$}.
These relatively low values reflect the large overlap between various accretion events,
which makes it difficult unequivocally to associate many GCs to a single accretion group.

We then have applied this methodology to the Galactic GC data, accounting for measurement errors.
The result is a decomposition of the GC population into bulge, disc, and the following halo components:
GES, Kraken, Sagittarius, Sequoia and Helmi groups and an ungrouped component.
This ungrouped component contains $17$ `left over' GCs that have a uniform background distribution and are
likely associated with many small accretion events that do not contain enough members to be robustly identified.

We find that it is difficult to separate some of the low energy GCs into a contribution from Kraken and the in-situ components.
Where these groups overlap in dynamical space, age-metallicity information is needed to help identify the groups.
However, in the region where the age-metallicity relations of the in-situ and Kraken components overlap (high age and low metallicity),
or for GCs without age-metallicity information,
there is not enough information to identify them confidently.
This likely leads to an overestimate of the membership of the Kraken component.
This is supported by our Krakens slight net rotation, possibly indicating the inclusion of disc GCs.

Combining the resulting groupings of GCs with the relation between
halo mass and GC number of \citet{burkert_high-precision_2020},
we have inferred the halo mass of the progenitor of each accretion event.
Combining these halo masses with the stellar mass-halo mass relation of \citet{behroozi_universemachine_2019},
we then inferred the progenitor stellar masses.
We find the Kraken group to be the most massive accreted galaxy ($M_{\mathrm{halo}}\sim{2.4}_{-1.0}^{+2.0}\times10^{11}\Msun$),
likely slightly larger than \GES{} ($M_{\mathrm{halo}}\sim1.2_{-0.4}^{+0.7}\times10^{11}\Msun$).

We find evidence in the phase distribution that the sample of in-situ MW GCs are probably incomplete,
with on the order of $20-30$ GCs `missing' from the far side of the Galaxy.
These are likely to be obscured by the Galactic centre and disc.

There are two relatively straightforward possible improvements of our study:
\begin{itemize}
\item
Our tests with mock catalogues indicate that increasing the sample of GCs with age-metallicity data would
improve our ability to identify groups, particularly for the in-situ component.
Furthermore, the age-chemistry modelling used in our method could, in principle,
be refined by including more detailed models of chemical evolution.
The obvious choice for this would be to include the $\alpha$ abundances which are currently available
only for a small subset of GCs \citep{horta_chemical_2020}.
This could be very useful for disentangling the in-situ and Kraken groups.

\item
Our methodology could be extended to include stellar halo stars.
The automated and statistical nature of our method makes it straightforward to handle large samples,
as well as the much larger observational errors of stellar samples.
Increasing the number of dynamic tracers by several orders of magnitude would allow
a much more accurate inference of the MW's accretion history.

\end{itemize}

This work has developed a GC grouping methodology that combines dynamical and chemical data in a statistically robust manner.
A crucial part of our analysis has been the tests using mocks which have highlighted the difficulties
inherent in this kind of study.
In the face of considerable uncertainties due to the messy nature of accretion,
we believe that this philosophy represents an improvement on previous work.
In the future,
with further development and more data, our method should allow a stronger inference of the MW accretion history.

\section*{Acknowledgements}
\Change{We thank the anonymous referee for their comments, which have improved our paper.}
TC and CSF were supported by the Science and Technology Facilities
Council (STFC) [grant number ST/F001166/1, ST/I00162X/1,ST/P000541/1]
MC also acknowledges support by the EU Horizon 2020 research and
innovation programme under a Marie Sk{\l}odowska-Curie grant agreement 794474 (DancingGalaxies).
AD is supported by a Royal Society University Research Fellowship.
CSF acknowledges European Research Council (ERC) Advanced Investigator grant DMIDAS (GA 786910).
This work used the DiRAC Data Centric system at Durham University,
operated by the ICC on behalf of the STFC DiRAC HPC Facility (www.dirac.ac.uk).
This equipment was funded by BIS National E-infrastructure capital grant ST/K00042X/1, STFC capital grant
ST/H008519/1, and STFC DiRAC Operations grant ST/K003267/1 and Durham University.
DiRAC is part of the National E-Infrastructure.
RG acknowledges financial support from the Spanish Ministry of Science and Innovation (MICINN) through the Spanish State Research Agency, under the Severo Ochoa Program 2020-2023 (CEX2019-000920-S).
FM acknowledges support through the program ``Rita Levi Montalcini'' of the Italian MUR.

\section*{Data availability}
The used and data produced in this paper are available upon reasonable
request to the corresponding author. The membership probabilities of the GCs and the chemo-dynamical
  fits of the groups are available at
\href{https://github.com/TomCallingham/Callingham22_MW_GCs}{https://github.com/TomCallingham/Callingham22\_MW\_GCs}.




\bibliographystyle{mnras}
\bibliography{references.bib}

\begin{thebibliography}{}
\makeatletter
\relax
\def\mn@urlcharsother{\let\do\@makeother \do\$\do\&\do\#\do\^\do\_\do\%\do\~}
\def\mn@doi{\begingroup\mn@urlcharsother \@ifnextchar [ {\mn@doi@}
  {\mn@doi@[]}}
\def\mn@doi@[#1]#2{\def\@tempa{#1}\ifx\@tempa\@empty \href
  {http://dx.doi.org/#2} {doi:#2}\else \href {http://dx.doi.org/#2} {#1}\fi
  \endgroup}
\def\mn@eprint#1#2{\mn@eprint@#1:#2::\@nil}
\def\mn@eprint@arXiv#1{\href {http://arxiv.org/abs/#1} {{\tt arXiv:#1}}}
\def\mn@eprint@dblp#1{\href {http://dblp.uni-trier.de/rec/bibtex/#1.xml}
  {dblp:#1}}
\def\mn@eprint@#1:#2:#3:#4\@nil{\def\@tempa {#1}\def\@tempb {#2}\def\@tempc
  {#3}\ifx \@tempc \@empty \let \@tempc \@tempb \let \@tempb \@tempa \fi \ifx
  \@tempb \@empty \def\@tempb {arXiv}\fi \@ifundefined
  {mn@eprint@\@tempb}{\@tempb:\@tempc}{\expandafter \expandafter \csname
  mn@eprint@\@tempb\endcsname \expandafter{\@tempc}}}

\bibitem[\protect\citeauthoryear{Amorisco}{Amorisco}{2019}]{amorisco_globular_2019}
Amorisco N.~C.,  2019, \mn@doi [Monthly Notices of the Royal Astronomical
  Society] {10.1093/mnras/sty2927}, 482, 2978

\bibitem[\protect\citeauthoryear{Antoja, Ramos, Mateu, Helmi, Anders, Jordi  \&
  Carballo-Bello}{Antoja et~al.}{2020}]{antoja_all-sky_2020}
Antoja T.,  Ramos P.,  Mateu C.,  Helmi A.,  Anders F.,  Jordi C.,
  Carballo-Bello J.~A.,  2020, \mn@doi [Astronomy and Astrophysics]
  {10.1051/0004-6361/201937145}, 635, L3

\bibitem[\protect\citeauthoryear{Barba, Minniti, Geisler, Alonso-Garcia,
  Hempel, Monachesi, Arias  \& Gomez}{Barba et~al.}{2019}]{barba_sequoia_2019}
Barba R.~H.,  Minniti D.,  Geisler D.,  Alonso-Garcia J.,  Hempel M.,
  Monachesi A.,  Arias J.~I.,   Gomez F.~A.,  2019, \mn@doi [The Astrophysical
  Journal] {10.3847/2041-8213/aaf811}, 870, L24

\bibitem[\protect\citeauthoryear{Bastian, Pfeffer, Kruijssen, Crain,
  Trujillo-Gomez  \& Reina-Campos}{Bastian
  et~al.}{2020}]{bastian_globular_2020}
Bastian N.,  Pfeffer J.,  Kruijssen J. M.~D.,  Crain R.~A.,  Trujillo-Gomez S.,
    Reina-Campos M.,  2020, \mn@doi [Monthly Notices of the Royal Astronomical
  Society] {10.1093/mnras/staa2453}, 498, 1050

\bibitem[\protect\citeauthoryear{Baumgardt \& Vasiliev}{Baumgardt \&
  Vasiliev}{2021}]{baumgardt_accurate_2021}
Baumgardt H.,  Vasiliev E.,  2021, \mn@doi [Monthly Notices of the Royal
  Astronomical Society] {10.1093/mnras/stab1474}, 505, 5957

\bibitem[\protect\citeauthoryear{Behroozi, Wechsler, Hearin  \&
  Conroy}{Behroozi et~al.}{2019}]{behroozi_universemachine_2019}
Behroozi P.,  Wechsler R.~H.,  Hearin A.~P.,   Conroy C.,  2019, \mn@doi
  [Monthly Notices of the Royal Astronomical Society] {10.1093/mnras/stz1182},
  488, 3143

\bibitem[\protect\citeauthoryear{Bekki \& Freeman}{Bekki \&
  Freeman}{2003}]{bekki_formation_2003}
Bekki K.,  Freeman K.~C.,  2003, \mn@doi [Monthly Notices of the Royal
  Astronomical Society, Volume 346, Issue 2, pp. L11-L15.]
  {10.1046/j.1365-2966.2003.07275.x}, 346, L11

\bibitem[\protect\citeauthoryear{Bellazzini, Ibata, Malhan, Martin, Famaey  \&
  Thomas}{Bellazzini et~al.}{2020}]{bellazzini_globular_2020}
Bellazzini M.,  Ibata R.,  Malhan K.,  Martin N.,  Famaey B.,   Thomas G.,
  2020, \mn@doi [Astronomy \& Astrophysics] {10.1051/0004-6361/202037621}, 636,
  A107

\bibitem[\protect\citeauthoryear{Belokurov, Erkal, Evans, Koposov  \&
  Deason}{Belokurov et~al.}{2018}]{belokurov_co-formation_2018}
Belokurov V.,  Erkal D.,  Evans N.~W.,  Koposov S.~E.,   Deason A.~J.,  2018,
  \mn@doi [Monthly Notices of the Royal Astronomical Society]
  {10.1093/mnras/sty982}, 478, 611

\bibitem[\protect\citeauthoryear{Binney}{Binney}{2010}]{binney_distribution_2010}
Binney J.,  2010, \mn@doi [Monthly Notices of the Royal Astronomical Society]
  {10.1111/j.1365-2966.2009.15845.x}, 401, 2318

\bibitem[\protect\citeauthoryear{Binney \& Tremaine}{Binney \&
  Tremaine}{2008}]{binney_galactic_2008}
Binney J.,  Tremaine S.,  2008, Galactic {Dynamics}: {Second} {Edition}.
Princeton University Press, \url
  {https://ui.adsabs.harvard.edu/abs/2008gady.book.....B}

\bibitem[\protect\citeauthoryear{Bland-Hawthorn \&
  Tepper-García}{Bland-Hawthorn \&
  Tepper-García}{2021}]{bland-hawthorn_galactic_2021}
Bland-Hawthorn J.,  Tepper-García T.,  2021, \mn@doi [Monthly Notices of the
  Royal Astronomical Society] {10.1093/mnras/stab704}, 504, 3168

\bibitem[\protect\citeauthoryear{Boylan-Kolchin}{Boylan-Kolchin}{2017}]{boylan-kolchin_globular_2017}
Boylan-Kolchin M.,  2017, \mn@doi [Monthly Notices of the Royal Astronomical
  Society] {10.1093/mnras/stx2164}, 472, 3120

\bibitem[\protect\citeauthoryear{Boylan-Kolchin, Springel, White  \&
  Jenkins}{Boylan-Kolchin et~al.}{2010}]{boylan-kolchin_theres_2010}
Boylan-Kolchin M.,  Springel V.,  White S. D.~M.,   Jenkins A.,  2010, \mn@doi
  [Monthly Notices of the Royal Astronomical Society]
  {10.1111/j.1365-2966.2010.16774.x}, 406, 896

\bibitem[\protect\citeauthoryear{Bullock \& Johnston}{Bullock \&
  Johnston}{2005}]{bullock_tracing_2005}
Bullock J.~S.,  Johnston K.~V.,  2005, \mn@doi [The Astrophysical Journal]
  {10.1086/497422}, 635, 931

\bibitem[\protect\citeauthoryear{Burkert \& Forbes}{Burkert \&
  Forbes}{2020}]{burkert_high-precision_2020}
Burkert A.,  Forbes D.~A.,  2020, \mn@doi [The Astronomical Journal]
  {10.3847/1538-3881/ab5b0e}, 159, 56

\bibitem[\protect\citeauthoryear{Callingham et~al.,}{Callingham
  et~al.}{2019}]{callingham_mass_2019}
Callingham T.~M.,  et~al., 2019, \mn@doi [Monthly Notices of the Royal
  Astronomical Society] {10.1093/mnras/stz365}, 484, 5453

\bibitem[\protect\citeauthoryear{Cautun, Deason, Frenk  \& McAlpine}{Cautun
  et~al.}{2019}]{cautun_aftermath_2019}
Cautun M.,  Deason A.~J.,  Frenk C.~S.,   McAlpine S.,  2019, \mn@doi [Monthly
  Notices of the Royal Astronomical Society] {10.1093/mnras/sty3084}, 483, 2185

\bibitem[\protect\citeauthoryear{Cautun et~al.,}{Cautun
  et~al.}{2020}]{cautun_milky_2020}
Cautun M.,  et~al., 2020, \mn@doi [Monthly Notices of the Royal Astronomical
  Society] {10.1093/mnras/staa1017}, 494, 4291

\bibitem[\protect\citeauthoryear{Chang, Yuan, Xue, Simion, Kang, Li, Zhao  \&
  Zhao}{Chang et~al.}{2020}]{chang_is_2020}
Chang J.,  Yuan Z.,  Xue X.-X.,  Simion I.~T.,  Kang X.,  Li T.~S.,  Zhao
  J.-K.,   Zhao G.,  2020, \mn@doi [The Astrophysical Journal]
  {10.3847/1538-4357/abc338}, 905, 100

\bibitem[\protect\citeauthoryear{Collaboration et~al.,}{Collaboration
  et~al.}{2021}]{gaia_collaboration_gaia_2021}
Collaboration G.,  et~al., 2021, arXiv:2101.05811 [astro-ph]

\bibitem[\protect\citeauthoryear{Conroy et~al.,}{Conroy
  et~al.}{2019}]{conroy_mapping_2019}
Conroy C.,  et~al., 2019, \mn@doi [The Astrophysical Journal]
  {10.3847/1538-4357/ab38b8}, 883, 107

\bibitem[\protect\citeauthoryear{Cooper et~al.,}{Cooper
  et~al.}{2010}]{cooper_galactic_2010}
Cooper A.~P.,  et~al., 2010, \mn@doi [Monthly Notices of the Royal Astronomical
  Society] {10.1111/j.1365-2966.2010.16740.x}, 406, 744

\bibitem[\protect\citeauthoryear{Crain et~al.,}{Crain
  et~al.}{2015}]{crain_eagle_2015}
Crain R.~A.,  et~al., 2015, \mn@doi [Monthly Notices of the Royal Astronomical
  Society] {10.1093/mnras/stv725}, 450, 1937

\bibitem[\protect\citeauthoryear{Davis, Efstathiou, Frenk  \& White}{Davis
  et~al.}{1985}]{davis_evolution_1985}
Davis M.,  Efstathiou G.,  Frenk C.~S.,   White S. D.~M.,  1985, \mn@doi [The
  Astrophysical Journal] {10.1086/163168}, 292, 371

\bibitem[\protect\citeauthoryear{Deason, Mao  \& Wechsler}{Deason
  et~al.}{2016}]{deason_eating_2016}
Deason A.~J.,  Mao Y.-Y.,   Wechsler R.~H.,  2016, \mn@doi [The Astrophysical
  Journal] {10.3847/0004-637X/821/1/5}, 821, 5

\bibitem[\protect\citeauthoryear{Deason, Belokurov, Koposov  \&
  Lancaster}{Deason et~al.}{2018}]{deason_apocenter_2018}
Deason A.~J.,  Belokurov V.,  Koposov S.~E.,   Lancaster L.,  2018, \mn@doi
  [The Astrophysical Journal Letters, Volume 862, Issue 1, article id. L1,
  {\textless}NUMPAGES{\textgreater}5{\textless}/NUMPAGES{\textgreater} pp.
  (2018).] {10.3847/2041-8213/aad0ee}, 862, L1

\bibitem[\protect\citeauthoryear{Deason, Belokurov  \& Sanders}{Deason
  et~al.}{2019}]{deason_total_2019}
Deason A.~J.,  Belokurov V.,   Sanders J.~L.,  2019, \mn@doi [Monthly Notices
  of the Royal Astronomical Society] {10.1093/mnras/stz2793}, 490, 3426

\bibitem[\protect\citeauthoryear{Deason et~al.,}{Deason
  et~al.}{2021}]{deason_mass_2021}
Deason A.~J.,  et~al., 2021, \mn@doi [Monthly Notices of the Royal Astronomical
  Society] {10.1093/mnras/staa3984}, 501, 5964

\bibitem[\protect\citeauthoryear{Dotter et~al.,}{Dotter
  et~al.}{2010}]{dotter_acs_2010}
Dotter A.,  et~al., 2010, \mn@doi [The Astrophysical Journal]
  {10.1088/0004-637X/708/1/698}, 708, 698

\bibitem[\protect\citeauthoryear{Dotter, Sarajedini  \& Anderson}{Dotter
  et~al.}{2011}]{dotter_globular_2011}
Dotter A.,  Sarajedini A.,   Anderson J.,  2011, \mn@doi [The Astrophysical
  Journal] {10.1088/0004-637X/738/1/74}, 738, 74

\bibitem[\protect\citeauthoryear{Erkal, Belokurov  \& Parkin}{Erkal
  et~al.}{2020}]{erkal_equilibrium_2020}
Erkal D.,  Belokurov V.~A.,   Parkin D.~L.,  2020, \mn@doi [Monthly Notices of
  the Royal Astronomical Society] {10.1093/mnras/staa2840}, 498, 5574

\bibitem[\protect\citeauthoryear{Erkal et~al.,}{Erkal
  et~al.}{2021}]{erkal_detection_2021}
Erkal D.,  et~al., 2021, \mn@doi [Monthly Notices of the Royal Astronomical
  Society] {10.1093/mnras/stab1828}

\bibitem[\protect\citeauthoryear{Fattahi et~al.,}{Fattahi
  et~al.}{2019}]{fattahi_origin_2019}
Fattahi A.,  et~al., 2019, \mn@doi [Monthly Notices of the Royal Astronomical
  Society] {10.1093/mnras/stz159}, 484, 4471

\bibitem[\protect\citeauthoryear{Fattahi et~al.,}{Fattahi
  et~al.}{2020}]{fattahi_tale_2020}
Fattahi A.,  et~al., 2020, \mn@doi [Monthly Notices of the Royal Astronomical
  Society] {10.1093/mnras/staa2221}, 497, 4459

\bibitem[\protect\citeauthoryear{Fernández-Trincado
  et~al.,}{Fernández-Trincado
  et~al.}{2020}]{fernandez-trincado_enigmatic_2020}
Fernández-Trincado J.~G.,  et~al., 2020, \mn@doi [Astronomy and Astrophysics]
  {10.1051/0004-6361/202039328}, 643, A145

\bibitem[\protect\citeauthoryear{Fernández-Trincado
  et~al.,}{Fernández-Trincado et~al.}{2021}]{fernandez-trincado_vvv_2021}
Fernández-Trincado J.~G.,  et~al., 2021, \mn@doi [The Astrophysical Journal
  Letters] {10.3847/2041-8213/abdf47}, 908, L42

\bibitem[\protect\citeauthoryear{Forbes}{Forbes}{2020}]{forbes_reverse_2020}
Forbes D.~A.,  2020, \mn@doi [Monthly Notices of the Royal Astronomical
  Society] {10.1093/mnras/staa245}, 493, 847

\bibitem[\protect\citeauthoryear{Forbes \& Bridges}{Forbes \&
  Bridges}{2010}]{forbes_accreted_2010}
Forbes D.~A.,  Bridges T.,  2010, \mn@doi [Monthly Notices of the Royal
  Astronomical Society] {10.1111/j.1365-2966.2010.16373.x}, 404, 1203

\bibitem[\protect\citeauthoryear{Forbes et~al.,}{Forbes
  et~al.}{2018}]{forbes_globular_2018}
Forbes D.~A.,  et~al., 2018, \mn@doi [Proceedings of the Royal Society A:
  Mathematical, Physical and Engineering Sciences] {10.1098/rspa.2017.0616},
  474, 20170616

\bibitem[\protect\citeauthoryear{Frenk \& White}{Frenk \&
  White}{1980}]{frenk_kinematics_1980}
Frenk C.~S.,  White S. D.~M.,  1980, \mn@doi [Monthly Notices of the Royal
  Astronomical Society] {10.1093/mnras/193.2.295}, 193, 295

\bibitem[\protect\citeauthoryear{{Gaia-Collaboration}
  et~al.,}{{Gaia-Collaboration} et~al.}{2018}]{gaia-collaboration_gaia_2018}
{Gaia-Collaboration} et~al., 2018, \mn@doi [Astronomy and Astrophysics]
  {10.1051/0004-6361/201833051}, 616, A1

\bibitem[\protect\citeauthoryear{Gran et~al.,}{Gran
  et~al.}{2021}]{gran_hidden_2021}
Gran F.,  et~al., 2021, arXiv:2108.11922 [astro-ph]

\bibitem[\protect\citeauthoryear{Grand et~al.,}{Grand
  et~al.}{2017}]{grand_auriga_2017}
Grand R. J.~J.,  et~al., 2017, \mn@doi [Monthly Notices of the Royal
  Astronomical Society] {10.1093/mnras/stx071}, 467, 179

\bibitem[\protect\citeauthoryear{Grand, Deason, White, Simpson, Gómez,
  Marinacci  \& Pakmor}{Grand et~al.}{2019}]{grand_effects_2019}
Grand R. J.~J.,  Deason A.~J.,  White S. D.~M.,  Simpson C.~M.,  Gómez F.~A.,
  Marinacci F.,   Pakmor R.,  2019, \mn@doi [Monthly Notices of the Royal
  Astronomical Society] {10.1093/mnrasl/slz092}, 487, L72

\bibitem[\protect\citeauthoryear{Grand et~al.,}{Grand
  et~al.}{2021}]{grand_determining_2021}
Grand R. J.~J.,  et~al., 2021, arXiv:2105.04560 [astro-ph]

\bibitem[\protect\citeauthoryear{Gómez, Helmi, Brown  \& Li}{Gómez
  et~al.}{2010}]{gomez_identification_2010}
Gómez F.~A.,  Helmi A.,  Brown A. G.~A.,   Li Y.-S.,  2010, \mn@doi [Monthly
  Notices of the Royal Astronomical Society]
  {10.1111/j.1365-2966.2010.17225.x}, 408, 935

\bibitem[\protect\citeauthoryear{Gómez, White, Grand, Marinacci, Springel  \&
  Pakmor}{Gómez et~al.}{2017}]{gomez_warps_2017}
Gómez F.~A.,  White S. D.~M.,  Grand R. J.~J.,  Marinacci F.,  Springel V.,
  Pakmor R.,  2017, \mn@doi [Monthly Notices of the Royal Astronomical Society]
  {10.1093/mnras/stw2957}, 465, 3446

\bibitem[\protect\citeauthoryear{Halbesma, Grand, Gómez, Marinacci, Pakmor,
  Trick, Busch  \& White}{Halbesma et~al.}{2020}]{halbesma_globular_2020}
Halbesma T. L.~R.,  Grand R. J.~J.,  Gómez F.~A.,  Marinacci F.,  Pakmor R.,
  Trick W.~H.,  Busch P.,   White S. D.~M.,  2020, \mn@doi [Monthly Notices of
  the Royal Astronomical Society] {10.1093/mnras/staa1380}, 496, 638

\bibitem[\protect\citeauthoryear{Harris, Harris  \& Hudson}{Harris
  et~al.}{2015}]{harris_dark_2015}
Harris W.~E.,  Harris G.~L.,   Hudson M.~J.,  2015, \mn@doi [The Astrophysical
  Journal] {10.1088/0004-637X/806/1/36}, 806, 36

\bibitem[\protect\citeauthoryear{Helmi, White, de Zeeuw  \& Zhao}{Helmi
  et~al.}{1999}]{helmi_debris_1999}
Helmi A.,  White S. D.~M.,  de Zeeuw P.~T.,   Zhao H.,  1999, \mn@doi [Nature,
  Volume 402, Issue 6757, pp. 53-55 (1999).] {10.1038/46980}, 402, 53

\bibitem[\protect\citeauthoryear{Helmi, Veljanoski, Breddels, Tian  \&
  Sales}{Helmi et~al.}{2017}]{helmi_box_2017}
Helmi A.,  Veljanoski J.,  Breddels M.~A.,  Tian H.,   Sales L.~V.,  2017,
  \mn@doi [Astronomy \& Astrophysics] {10.1051/0004-6361/201629990}, 598, A58

\bibitem[\protect\citeauthoryear{Helmi, Babusiaux, Koppelman, Massari,
  Veljanoski  \& Brown}{Helmi et~al.}{2018}]{helmi_formation_2018}
Helmi A.,  Babusiaux C.,  Koppelman H.~H.,  Massari D.,  Veljanoski J.,   Brown
  A. G.~A.,  2018, ArXiv e-prints

\bibitem[\protect\citeauthoryear{Horta et~al.,}{Horta
  et~al.}{2020}]{horta_chemical_2020}
Horta D.,  et~al., 2020, \mn@doi [Monthly Notices of the Royal Astronomical
  Society] {10.1093/mnras/staa478}, 493, 3363

\bibitem[\protect\citeauthoryear{Horta et~al.,}{Horta
  et~al.}{2021}]{horta_evidence_2021}
Horta D.,  et~al., 2021, \mn@doi [Monthly Notices of the Royal Astronomical
  Society] {10.1093/mnras/staa2987}, 500, 1385

\bibitem[\protect\citeauthoryear{Ibata, Gilmore  \& Irwin}{Ibata
  et~al.}{1994}]{ibata_dwarf_1994}
Ibata R.~A.,  Gilmore G.,   Irwin M.~J.,  1994, \mn@doi [Nature]
  {10.1038/370194a0}, 370, 194

\bibitem[\protect\citeauthoryear{Koposov et~al.,}{Koposov
  et~al.}{2007}]{koposov_discovery_2007}
Koposov S.,  et~al., 2007, \mn@doi [The Astrophysical Journal]
  {10.1086/521422}, 669, 337

\bibitem[\protect\citeauthoryear{Koppelman, Helmi, Massari, Roelenga  \&
  Bastian}{Koppelman et~al.}{2019a}]{koppelman_characterization_2019}
Koppelman H.~H.,  Helmi A.,  Massari D.,  Roelenga S.,   Bastian U.,  2019a,
  \mn@doi [Astronomy and Astrophysics] {10.1051/0004-6361/201834769}, 625, A5

\bibitem[\protect\citeauthoryear{Koppelman, Helmi, Massari, Price-Whelan  \&
  Starkenburg}{Koppelman et~al.}{2019b}]{koppelman_multiple_2019}
Koppelman H.~H.,  Helmi A.,  Massari D.,  Price-Whelan A.~M.,   Starkenburg
  T.~K.,  2019b, \mn@doi [Astronomy \&amp; Astrophysics, Volume 631, id.L9,
  {\textless}NUMPAGES{\textgreater}6{\textless}/NUMPAGES{\textgreater} pp.]
  {10.1051/0004-6361/201936738}, 631, L9

\bibitem[\protect\citeauthoryear{Kruijssen, Pfeffer, Crain  \&
  Bastian}{Kruijssen et~al.}{2019a}]{kruijssen_e-mosaics_2019}
Kruijssen J. M.~D.,  Pfeffer J.~L.,  Crain R.~A.,   Bastian N.,  2019a, \mn@doi
  [Monthly Notices of the Royal Astronomical Society] {10.1093/mnras/stz968},
  486, 3134

\bibitem[\protect\citeauthoryear{Kruijssen, Pfeffer, Reina-Campos, Crain  \&
  Bastian}{Kruijssen et~al.}{2019b}]{kruijssen_formation_2019}
Kruijssen J. M.~D.,  Pfeffer J.~L.,  Reina-Campos M.,  Crain R.~A.,   Bastian
  N.,  2019b, \mn@doi [Monthly Notices of the Royal Astronomical Society]
  {10.1093/mnras/sty1609}, 486, 3180

\bibitem[\protect\citeauthoryear{Kruijssen et~al.,}{Kruijssen
  et~al.}{2020}]{kruijssen_kraken_2020}
Kruijssen J. M.~D.,  et~al., 2020, \mn@doi [Monthly Notices of the Royal
  Astronomical Society] {10.1093/mnras/staa2452}, 498, 2472

\bibitem[\protect\citeauthoryear{Kuiper}{Kuiper}{1960}]{kuiper_tests_1960}
Kuiper N.~H.,  1960, \mn@doi [Indagationes Mathematicae (Proceedings)]
  {10.1016/S1385-7258(60)50006-0}, 63, 38

\bibitem[\protect\citeauthoryear{Laporte, Johnston, Gómez, Garavito-Camargo
  \& Besla}{Laporte et~al.}{2018}]{laporte_influence_2018}
Laporte C. F.~P.,  Johnston K.~V.,  Gómez F.~A.,  Garavito-Camargo N.,   Besla
  G.,  2018, \mn@doi [Monthly Notices of the Royal Astronomical Society]
  {10.1093/mnras/sty1574}, 481, 286

\bibitem[\protect\citeauthoryear{Laporte, Minchev, Johnston  \& Gómez}{Laporte
  et~al.}{2019}]{laporte_footprints_2019}
Laporte C. F.~P.,  Minchev I.,  Johnston K.~V.,   Gómez F.~A.,  2019, \mn@doi
  [Monthly Notices of the Royal Astronomical Society] {10.1093/mnras/stz583},
  485, 3134

\bibitem[\protect\citeauthoryear{Law \& Majewski}{Law \&
  Majewski}{2010}]{law_assessing_2010}
Law D.~R.,  Majewski S.~R.,  2010, \mn@doi [The Astrophysical Journal]
  {10.1088/0004-637X/718/2/1128}, 718, 1128

\bibitem[\protect\citeauthoryear{Leaman, VandenBerg  \& Mendel}{Leaman
  et~al.}{2013}]{leaman_bifurcated_2013}
Leaman R.,  VandenBerg D.~A.,   Mendel J.~T.,  2013, \mn@doi [Monthly Notices
  of the Royal Astronomical Society] {10.1093/mnras/stt1540}, 436, 122

\bibitem[\protect\citeauthoryear{Longmore et~al.,}{Longmore
  et~al.}{2011}]{longmore_mercer_2011}
Longmore A.~J.,  et~al., 2011, \mn@doi [Monthly Notices of the Royal
  Astronomical Society] {10.1111/j.1365-2966.2011.19056.x}, 416, 465

\bibitem[\protect\citeauthoryear{Mackereth \& Bovy}{Mackereth \&
  Bovy}{2020}]{mackereth_weighing_2020}
Mackereth J.~T.,  Bovy J.,  2020, \mn@doi [Monthly Notices of the Royal
  Astronomical Society] {10.1093/mnras/staa047}, 492, 3631

\bibitem[\protect\citeauthoryear{Majewski et~al.,}{Majewski
  et~al.}{2017}]{majewski_apache_2017}
Majewski S.~R.,  et~al., 2017, \mn@doi [The Astronomical Journal]
  {10.3847/1538-3881/aa784d}, 154, 94

\bibitem[\protect\citeauthoryear{Malhan, Yuan, Ibata, Arentsen, Bellazzini  \&
  Martin}{Malhan et~al.}{2021}]{malhan_evidence_2021}
Malhan K.,  Yuan Z.,  Ibata R.,  Arentsen A.,  Bellazzini M.,   Martin N.~F.,
  2021, arXiv:2104.09523 [astro-ph]

\bibitem[\protect\citeauthoryear{Martell et~al.,}{Martell
  et~al.}{2017}]{martell_galah_2017}
Martell S.~L.,  et~al., 2017, \mn@doi [Monthly Notices of the Royal
  Astronomical Society] {10.1093/mnras/stw2835}, 465, 3203

\bibitem[\protect\citeauthoryear{Marín-Franch et~al.,}{Marín-Franch
  et~al.}{2009}]{marin-franch_acs_2009}
Marín-Franch A.,  et~al., 2009, \mn@doi [The Astrophysical Journal]
  {10.1088/0004-637X/694/2/1498}, 694, 1498

\bibitem[\protect\citeauthoryear{Massari, Koppelman  \& Helmi}{Massari
  et~al.}{2019}]{massari_origin_2019}
Massari D.,  Koppelman H.~H.,   Helmi A.,  2019, \mn@doi [Astronomy \&
  Astrophysics] {10.1051/0004-6361/201936135}, 630, L4

\bibitem[\protect\citeauthoryear{McMillan}{McMillan}{2017}]{mcmillan_mass_2017}
McMillan P.~J.,  2017, \mn@doi [Monthly Notices of the Royal Astronomical
  Society] {10.1093/mnras/stw2759}, 465, 76

\bibitem[\protect\citeauthoryear{Minniti, Fernández-Trincado, Gómez, Smith,
  Lucas  \& Ramos}{Minniti et~al.}{2021b}]{minniti_discovery_2021}
Minniti D.,  Fernández-Trincado J.~G.,  Gómez M.,  Smith L.~C.,  Lucas P.~W.,
    Ramos R.~C.,  2021b, arXiv:2106.01383 [astro-ph]

\bibitem[\protect\citeauthoryear{Minniti, Gómez, Alonso-García, Saito  \&
  Garro}{Minniti et~al.}{2021a}]{minniti_eight_2021}
Minniti D.,  Gómez M.,  Alonso-García J.,  Saito R.~K.,   Garro E.~R.,
  2021a, arXiv:2106.03605 [astro-ph]

\bibitem[\protect\citeauthoryear{Minniti, Fernández-Trincado, Smith, Lucas,
  Gómez  \& Pullen}{Minniti et~al.}{2021c}]{minniti_survival_2021}
Minniti D.,  Fernández-Trincado J.~G.,  Smith L.~C.,  Lucas P.~W.,  Gómez M.,
    Pullen J.~B.,  2021c, \mn@doi [Astronomy and Astrophysics]
  {10.1051/0004-6361/202039820}, 648, A86

\bibitem[\protect\citeauthoryear{Monachesi et~al.,}{Monachesi
  et~al.}{2019}]{monachesi_auriga_2019}
Monachesi A.,  et~al., 2019, \mn@doi [Monthly Notices of the Royal Astronomical
  Society] {10.1093/mnras/stz538}, 485, 2589

\bibitem[\protect\citeauthoryear{Myeong, Evans, Belokurov, Amorisco  \&
  Koposov}{Myeong et~al.}{2018a}]{myeong_halo_2018}
Myeong G.~C.,  Evans N.~W.,  Belokurov V.,  Amorisco N.~C.,   Koposov S.~E.,
  2018a, \mn@doi [Monthly Notices of the Royal Astronomical Society]
  {10.1093/mnras/stx3262}, 475, 1537

\bibitem[\protect\citeauthoryear{Myeong, Evans, Belokurov, Sanders  \&
  Koposov}{Myeong et~al.}{2018b}]{myeong_discovery_2018}
Myeong G.~C.,  Evans N.~W.,  Belokurov V.,  Sanders J.~L.,   Koposov S.~E.,
  2018b, \mn@doi [Monthly Notices of the Royal Astronomical Society]
  {10.1093/mnras/sty1403}, 478, 5449

\bibitem[\protect\citeauthoryear{Myeong, Evans, Belokurov, Sanders  \&
  Koposov}{Myeong et~al.}{2018c}]{myeong_sausage_2018}
Myeong G.~C.,  Evans N.~W.,  Belokurov V.,  Sanders J.~L.,   Koposov S.~E.,
  2018c, \mn@doi [The Astrophysical Journal Letters]
  {10.3847/2041-8213/aad7f7}, 863, L28

\bibitem[\protect\citeauthoryear{Myeong, Vasiliev, Iorio, Evans  \&
  Belokurov}{Myeong et~al.}{2019}]{myeong_evidence_2019}
Myeong G.~C.,  Vasiliev E.,  Iorio G.,  Evans N.~W.,   Belokurov V.,  2019,
  \mn@doi [Monthly Notices of the Royal Astronomical Society]
  {10.1093/mnras/stz1770}, 488, 1235

\bibitem[\protect\citeauthoryear{Naidu, Conroy, Bonaca, Johnson, Ting,
  Caldwell, Zaritsky  \& Cargile}{Naidu et~al.}{2020}]{naidu_evidence_2020}
Naidu R.~P.,  Conroy C.,  Bonaca A.,  Johnson B.~D.,  Ting Y.-S.,  Caldwell N.,
   Zaritsky D.,   Cargile P.~A.,  2020, \mn@doi [The Astrophysical Journal]
  {10.3847/1538-4357/abaef4}, 901, 48

\bibitem[\protect\citeauthoryear{Necib, Ostdiek, Lisanti, Cohen, Freytsis  \&
  Garrison-Kimmel}{Necib et~al.}{2020}]{necib_chasing_2020}
Necib L.,  Ostdiek B.,  Lisanti M.,  Cohen T.,  Freytsis M.,   Garrison-Kimmel
  S.,  2020, \mn@doi [The Astrophysical Journal] {10.3847/1538-4357/abb814},
  903, 25

\bibitem[\protect\citeauthoryear{Neto et~al.,}{Neto
  et~al.}{2007}]{neto_statistics_2007}
Neto A.~F.,  et~al., 2007, \mn@doi [{\textbackslash}mnras]
  {10.1111/j.1365-2966.2007.12381.x}, 381, 1450

\bibitem[\protect\citeauthoryear{Ostdiek et~al.,}{Ostdiek
  et~al.}{2020}]{ostdiek_cataloging_2020}
Ostdiek B.,  et~al., 2020, \mn@doi [Astronomy and Astrophysics]
  {10.1051/0004-6361/201936866}, 636, A75

\bibitem[\protect\citeauthoryear{Paust, Wilson  \& van Belle}{Paust
  et~al.}{2014}]{paust_reinvestigating_2014}
Paust N.,  Wilson D.,   van Belle G.,  2014, \mn@doi [The Astronomical Journal]
  {10.1088/0004-6256/148/1/19}, 148, 19

\bibitem[\protect\citeauthoryear{Peñaloza, Pessev, Vaśquez, Borissova, Kurtev
   \& Zoccali}{Peñaloza et~al.}{2015}]{penaloza_chemical_2015}
Peñaloza F.,  Pessev P.,  Vaśquez S.,  Borissova J.,  Kurtev R.,   Zoccali
  M.,  2015, \mn@doi [Publications of the Astronomical Society of the Pacific]
  {10.1086/680597}, 127, 329

\bibitem[\protect\citeauthoryear{Peñarrubia \& Petersen}{Peñarrubia \&
  Petersen}{2021}]{penarrubia_identification_2021}
Peñarrubia J.,  Petersen M.~S.,  2021, arXiv:2106.11984 [astro-ph]

\bibitem[\protect\citeauthoryear{Pfeffer, Kruijssen, Crain  \& Bastian}{Pfeffer
  et~al.}{2018}]{pfeffer_e-mosaics_2018}
Pfeffer J.,  Kruijssen J. M.~D.,  Crain R.~A.,   Bastian N.,  2018, \mn@doi
  [Monthly Notices of the Royal Astronomical Society] {10.1093/mnras/stx3124},
  475, 4309

\bibitem[\protect\citeauthoryear{Pfeffer, Trujillo-Gomez, Kruijssen, Crain,
  Hughes, Reina-Campos  \& Bastian}{Pfeffer
  et~al.}{2020}]{pfeffer_predicting_2020}
Pfeffer J.~L.,  Trujillo-Gomez S.,  Kruijssen J. M.~D.,  Crain R.~A.,  Hughes
  M.~E.,  Reina-Campos M.,   Bastian N.,  2020, \mn@doi [Monthly Notices of the
  Royal Astronomical Society] {10.1093/mnras/staa3109}, 499, 4863

\bibitem[\protect\citeauthoryear{Piffl et~al.,}{Piffl
  et~al.}{2014}]{piffl_rave_2014}
Piffl T.,  et~al., 2014, \mn@doi [{\textbackslash}aap]
  {10.1051/0004-6361/201322531}, 562, A91

\bibitem[\protect\citeauthoryear{{Planck Collaboration} et~al.,}{{Planck
  Collaboration} et~al.}{2014}]{planck_collaboration_planck_2014}
{Planck Collaboration} et~al., 2014, \mn@doi [Astronomy \& Astrophysics]
  {10.1051/0004-6361/201321529}, 571, A1

\bibitem[\protect\citeauthoryear{Posti \& Helmi}{Posti \&
  Helmi}{2019}]{posti_mass_2019}
Posti L.,  Helmi A.,  2019, \mn@doi [Astronomy \& Astrophysics]
  {10.1051/0004-6361/201833355}, 621, A56

\bibitem[\protect\citeauthoryear{Posti, Binney, Nipoti  \& Ciotti}{Posti
  et~al.}{2015}]{posti_action-based_2015}
Posti L.,  Binney J.,  Nipoti C.,   Ciotti L.,  2015, \mn@doi [Monthly Notices
  of the Royal Astronomical Society] {10.1093/mnras/stu2608}, 447, 3060

\bibitem[\protect\citeauthoryear{Pérez-Villegas, Barbuy, Kerber, Ortolani,
  Souza  \& Bica}{Pérez-Villegas et~al.}{2019}]{perez-villegas_globular_2019}
Pérez-Villegas A.,  Barbuy B.,  Kerber L.,  Ortolani S.,  Souza S.~O.,   Bica
  E.,  2019, \mn@doi [Monthly Notices of the Royal Astronomical Society]
  {10.1093/mnras/stz3162}, p. stz3162

\bibitem[\protect\citeauthoryear{Romero-Colmenares et~al.,}{Romero-Colmenares
  et~al.}{2021}]{romero-colmenares_capos_2021}
Romero-Colmenares M.,  et~al., 2021, arXiv:2106.00027 [astro-ph]

\bibitem[\protect\citeauthoryear{Sakari, Venn, Irwin, Aoki, Arimoto  \&
  Dotter}{Sakari et~al.}{2011}]{sakari_detailed_2011}
Sakari C.~M.,  Venn K.~A.,  Irwin M.,  Aoki W.,  Arimoto N.,   Dotter A.,
  2011, \mn@doi [The Astrophysical Journal] {10.1088/0004-637X/740/2/106}, 740,
  106

\bibitem[\protect\citeauthoryear{Schaye et~al.,}{Schaye
  et~al.}{2015}]{schaye_eagle_2015}
Schaye J.,  et~al., 2015, \mn@doi [Monthly Notices of the Royal Astronomical
  Society] {10.1093/mnras/stu2058}, 446, 521

\bibitem[\protect\citeauthoryear{Schönrich, Binney  \& Dehnen}{Schönrich
  et~al.}{2010}]{schonrich_local_2010}
Schönrich R.,  Binney J.,   Dehnen W.,  2010, \mn@doi [Monthly Notices of the
  Royal Astronomical Society] {10.1111/j.1365-2966.2010.16253.x}, 403, 1829

\bibitem[\protect\citeauthoryear{Searle \& Zinn}{Searle \&
  Zinn}{1978}]{searle_composition_1978}
Searle L.,  Zinn R.,  1978, \mn@doi [The Astrophysical Journal]
  {10.1086/156499}, 225, 357

\bibitem[\protect\citeauthoryear{Shao, Cautun, Frenk, Grand, Gómez, Marinacci
  \& Simpson}{Shao et~al.}{2018}]{shao_multiplicity_2018}
Shao S.,  Cautun M.,  Frenk C.~S.,  Grand R. J.~J.,  Gómez F.~A.,  Marinacci
  F.,   Simpson C.~M.,  2018, \mn@doi [Monthly Notices of the Royal
  Astronomical Society] {10.1093/mnras/sty343}, 476, 1796

\bibitem[\protect\citeauthoryear{Simpson, Grand, Gómez, Marinacci, Pakmor,
  Springel, Campbell  \& Frenk}{Simpson et~al.}{2018}]{simpson_quenching_2018}
Simpson C.~M.,  Grand R. J.~J.,  Gómez F.~A.,  Marinacci F.,  Pakmor R.,
  Springel V.,  Campbell D. J.~R.,   Frenk C.~S.,  2018, \mn@doi [Monthly
  Notices of the Royal Astronomical Society] {10.1093/mnras/sty774}, 478, 548

\bibitem[\protect\citeauthoryear{Springel}{Springel}{2005}]{springel_cosmological_2005}
Springel V.,  2005, \mn@doi [Monthly Notices of the Royal Astronomical Society]
  {10.1111/j.1365-2966.2005.09655.x}, 364, 1105

\bibitem[\protect\citeauthoryear{Springel}{Springel}{2011}]{springel_moving-mesh_2011}
Springel V.,  2011, \mn@doi [Proceedings of the International Astronomical
  Union] {10.1017/S1743921311000378}, 270, 203

\bibitem[\protect\citeauthoryear{Springel, Yoshida  \& White}{Springel
  et~al.}{2001}]{springel_gadget_2001}
Springel V.,  Yoshida N.,   White S. D.~M.,  2001, \mn@doi [New Astronomy]
  {10.1016/S1384-1076(01)00042-2}, 6, 79

\bibitem[\protect\citeauthoryear{Trujillo-Gomez, Kruijssen, Reina-Campos,
  Pfeffer, Keller, Crain, Bastian  \& Hughes}{Trujillo-Gomez
  et~al.}{2020}]{trujillo-gomez_kinematics_2020}
Trujillo-Gomez S.,  Kruijssen J. M.~D.,  Reina-Campos M.,  Pfeffer J.~L.,
  Keller B.~W.,  Crain R.~A.,  Bastian N.,   Hughes M.~E.,  2020,
  arXiv:2005.02401 [astro-ph]

\bibitem[\protect\citeauthoryear{Valcarce \& Catelan}{Valcarce \&
  Catelan}{2011}]{valcarce_formation_2011}
Valcarce A. a.~R.,  Catelan M.,  2011, \mn@doi [Astronomy \&amp; Astrophysics,
  Volume 533, id.A120,
  {\textless}NUMPAGES{\textgreater}15{\textless}/NUMPAGES{\textgreater} pp.]
  {10.1051/0004-6361/201116955}, 533, A120

\bibitem[\protect\citeauthoryear{VandenBerg, Brogaard, Leaman  \&
  Casagrande}{VandenBerg et~al.}{2013}]{vandenberg_ages_2013}
VandenBerg D.~A.,  Brogaard K.,  Leaman R.,   Casagrande L.,  2013, \mn@doi
  [The Astrophysical Journal] {10.1088/0004-637X/775/2/134}, 775, 134

\bibitem[\protect\citeauthoryear{Vasiliev}{Vasiliev}{2019}]{vasiliev_agama_2019}
Vasiliev E.,  2019, \mn@doi [Monthly Notices of the Royal Astronomical Society]
  {10.1093/mnras/sty2672}, 482, 1525

\bibitem[\protect\citeauthoryear{Vasiliev \& Baumgardt}{Vasiliev \&
  Baumgardt}{2021}]{vasiliev_gaia_2021}
Vasiliev E.,  Baumgardt H.,  2021, \mn@doi [Monthly Notices of the Royal
  Astronomical Society] {10.1093/mnras/stab1475}, 505, 5978

\bibitem[\protect\citeauthoryear{Virtanen et~al.,}{Virtanen
  et~al.}{2020}]{virtanen_scipy_2020}
Virtanen P.,  et~al., 2020, \mn@doi [Nature Methods]
  {10.1038/s41592-019-0686-2}, 17, 261

\bibitem[\protect\citeauthoryear{Wu, Valluri, Panithanpaisal, Sanderson,
  Freese, Wetzel  \& Sharma}{Wu et~al.}{2021}]{wu_using_2021}
Wu Y.,  Valluri M.,  Panithanpaisal N.,  Sanderson R.~E.,  Freese K.,  Wetzel
  A.,   Sharma S.,  2021, \mn@doi [Monthly Notices of the Royal Astronomical
  Society] {10.1093/mnras/stab3306}, 509, 5882

\bibitem[\protect\citeauthoryear{Yuan, Smith, Xue, Li, Liu, Wang, Li  \&
  Chang}{Yuan et~al.}{2019}]{yuan_revealing_2019}
Yuan Z.,  Smith M.~C.,  Xue X.-X.,  Li J.,  Liu C.,  Wang Y.,  Li L.,   Chang
  J.,  2019, \mn@doi [The Astrophysical Journal] {10.3847/1538-4357/ab2e09},
  881, 164

\bibitem[\protect\citeauthoryear{Yuan, Chang, Beers  \& Huang}{Yuan
  et~al.}{2020}]{yuan_low-mass_2020}
Yuan Z.,  Chang J.,  Beers T.~C.,   Huang Y.,  2020, \mn@doi [The Astrophysical
  Journal] {10.3847/2041-8213/aba49f}, 898, L37

\makeatother
\end{thebibliography}



\appendix
\section{Membership Probabilities}
\label{Appendix:MemberProbs}

\onecolumn
\begin{table}
	\centering
	\label{Tab:GC_Membership}
    \setlength\extrarowheight{1pt}
    \caption{The membership probability of individual GCs, as found by our chemo-dynamical model.
    We give the most likely group and probability of each cluster,
    and the second most probable alternate group.
    We also give the groupings from the literature where possible:
    M19 corresponds to \protect\cite{massari_origin_2019},
    F20 corresponds to \protect\cite{forbes_reverse_2020}, and 
    H20 corresponds to \protect\cite{horta_chemical_2020}.
    }
    \rowcolors{2}{gray!25}{white}
    \begin{center}
    \begin{tabularx}{\columnwidth}{l l c c c c c r r} 
    \hline
    \rowcolor{white}
    Name & Alternative & Main Group & Prob & Alt Group & Alt Prob & M19 & F20 & H20 \tabularnewline[.1cm] 
\hline  
Djorg2 & ESO456 & Bulge & 1.00 & - & - & Bulge& -& - \tabularnewline
Terzan6 & HP5 & Bulge & 1.00 & - & - & Bulge& -& - \tabularnewline
Terzan2 & HP3 & Bulge & 1.00 & - & - & Bulge& -& Bulge \tabularnewline
NGC6380 & Ton1 & Bulge & 1.00 & - & - & Bulge& -& Bulge \tabularnewline
NGC6440 & - & Bulge & 1.00 & - & - & Bulge& -& - \tabularnewline
Liller1 & - & Bulge & 1.00 & - & - & -& -& Ungr \tabularnewline
NGC6642 & - & Bulge & 1.00 & - & - & Bulge& -& - \tabularnewline
NGC6388 & - & Bulge & 1.00 & - & - & Bulge& -& Seq/Bulge  \tabularnewline
NGC6535 & - & Bulge & 1.00 & - & - & Kraken/Seq & Seq& - \tabularnewline
NGC6401 & - & Bulge & 1.00 & - & - & Kraken& Kraken& - \tabularnewline
Terzan5 & 11 & Bulge & 1.00 & - & - & Bulge& -& - \tabularnewline
NGC6638 & - & Bulge & 1.00 & - & - & Bulge& -& - \tabularnewline
NGC6528 & - & Bulge & 1.00 & - & - & Bulge& -& - \tabularnewline
1636-283 & ESO452 & Bulge & 1.00 & - & - & Bulge& -& - \tabularnewline
Terzan9 & - & Bulge & 1.00 & - & - & Bulge& -& - \tabularnewline
NGC6624 & - & Bulge & 1.00 & - & - & Bulge& -& - \tabularnewline
NGC6558 & - & Bulge & 1.00 & - & - & Bulge& -& - \tabularnewline
Terzan4 & HP4 & Bulge & 1.00 & - & - & Bulge& -& - \tabularnewline
HP1 & BH229 & Bulge & 1.00 & - & - & Bulge& -& Bulge \tabularnewline
NGC6325 & - & Bulge & 1.00 & - & - & Bulge& -& - \tabularnewline
NGC6453 & - & Bulge & 0.99 & - & - & Kraken& Kraken& - \tabularnewline
NGC6626 & M28 & Bulge & 0.99 & Disc & 0.01 & Bulge& -& - \tabularnewline
NGC6304 & - & Bulge & 0.99 & Disc & 0.01 & Bulge& -& - \tabularnewline
NGC6522 & - & Bulge & 0.99 & Disc & 0.01 & Bulge& -& Bulge \tabularnewline
Terzan1 & HP2 & Bulge & 0.99 & Disc & 0.01 & Bulge& -& - \tabularnewline
Pal6 & - & Bulge & 0.99 & Kraken & 0.01 & Kraken& -& Kraken \tabularnewline
NGC6652 & - & Bulge & 0.99 & Disc & 0.01 & Bulge& -& - \tabularnewline
NGC6266 & M62 & Bulge & 0.99 & Disc & 0.01 & Bulge& -& - \tabularnewline
NGC6342 & - & Bulge & 0.99 & Disc & 0.01 & Bulge& -& - \tabularnewline
NGC6637 & M69 & Bulge & 0.98 & Disc & 0.02 & Bulge& -& - \tabularnewline
NGC6355 & - & Bulge & 0.98 & Kraken & 0.02 & Bulge& -& - \tabularnewline
NGC6540 & Djorg & Bulge & 0.97 & Disc & 0.03 & Bulge& -& Bulge \tabularnewline
NGC6717 & Pal9 & Bulge & 0.97 & Disc & 0.03 & Bulge& -& - \tabularnewline
NGC6293 & - & Bulge & 0.95 & Kraken & 0.05 & Bulge& -& - \tabularnewline
NGC6256 & - & Bulge & 0.94 & Disc & 0.06 & Kraken& Kraken& - \tabularnewline
NGC6517 & - & Bulge & 0.93 & Kraken & 0.07 & Kraken& Kraken& - \tabularnewline
NGC6144 & - & Bulge & 0.89 & Disc & 0.11 & Kraken& Kraken& - \tabularnewline
VVVCL001 & - & Bulge & 0.89 & Kraken & 0.11 & -& -& - \tabularnewline
NGC6723 & - & Bulge & 0.83 & Disc & 0.17 & Bulge& -& Bulge \tabularnewline
VVVCL002 & - & Bulge & 0.80 & Kraken & 0.19 & -& -& - \tabularnewline
NGC6171 & M107 & Bulge & 0.75 & Disc & 0.25 & Bulge& -& Bulge \tabularnewline
NGC6093 & M80 & Bulge & 0.70 & Disc & 0.16 & Kraken& Kraken& - \tabularnewline
Gran1 & - & Bulge & 0.66 & Kraken & 0.33 & -& -& - \tabularnewline
NGC6838 & M71 & Disc & 1.00 & - & - & Disc& -& Disc \tabularnewline
NGC5927 & - & Disc & 1.00 & - & - & Disc& -& - \tabularnewline
NGC104 & 47Tuc & Disc & 1.00 & - & - & Disc& -& Disc \tabularnewline
NGC6496 & - & Disc & 1.00 & - & - & Disc& -& - \tabularnewline
ESO93 & - & Disc & 1.00 & - & - & -& -& - \tabularnewline
NGC6362 & - & Disc & 1.00 & - & - & Disc& -& - \tabularnewline
NGC6366 & - & Disc & 1.00 & - & - & Disc& -& - \tabularnewline
NGC6352 & - & Disc & 1.00 & - & - & Disc& -& - \tabularnewline
BH176 & - & Disc & 1.00 & - & - & Disc& -& - \tabularnewline
Pal10 & - & Disc & 0.99 & - & - & Disc& -& Disc \tabularnewline
E3 & - & Disc & 0.99 & Ungr & 0.01 & Helmi/? & -& - \tabularnewline
NGC6218 & M12 & Disc & 0.98 & Bulge & 0.01 & Disc& -& Disc \tabularnewline
NGC6441 & - & Disc & 0.97 & Bulge & 0.03 & Kraken& -& Kraken \tabularnewline
Pal11 & - & Disc & 0.71 & GEn & 0.28 & Disc& -& - \tabularnewline
 \hline  
    \end{tabularx}
    \end{center}
    \end{table}
    
 \begin{table}
	\centering
    \setlength\extrarowheight{1pt}
    \rowcolors{2}{gray!25}{white}
    \begin{center}
    \begin{tabularx}{\columnwidth}{l l c c c c c r r} 
    \hline
    \rowcolor{white}
    Name & Alternative & Main Group & Prob & Alt Group & Alt Prob & M19 & F20 & H20 \tabularnewline[.1cm] 
\hline  
IC1276 & Pal7 & Disc & 0.68 & Kraken & 0.28 & Disc& -& - \tabularnewline
Lynga7 & BH184 & Disc & 0.62 & Bulge & 0.37 & Disc& -& - \tabularnewline
Pfleiderer & - & Disc & 0.38 & GEn & 0.32 & -& -& - \tabularnewline
NGC6205 & M13 & GEn & 1.00 & - & - & GEn& GEn& GEn \tabularnewline
NGC362 & - & GEn & 1.00 & - & - & GEn& GEn& GEn \tabularnewline
NGC6779 & M56 & GEn & 1.00 & - & - & GEn& GEn& - \tabularnewline
NGC7089 & M2 & GEn & 1.00 & - & - & GEn& GEn& GEn \tabularnewline
NGC2298 & - & GEn & 1.00 & - & - & GEn& GEn& - \tabularnewline
NGC1851 & - & GEn & 1.00 & - & - & GEn& GEn& GEn \tabularnewline
NGC2808 & - & GEn & 1.00 & - & - & GEn& GEn& GEn \tabularnewline
NGC7099 & M30 & GEn & 1.00 & - & - & GEn& GEn& - \tabularnewline
NGC6341 & M92 & GEn & 1.00 & - & - & GEn& GEn& GEn \tabularnewline
NGC5286 & - & GEn & 1.00 & - & - & GEn& GEn& - \tabularnewline
NGC1261 & - & GEn & 1.00 & - & - & GEn& GEn& - \tabularnewline
ESO-SC06 & ESO280 & GEn & 1.00 & - & - & GEn& -& - \tabularnewline
NGC288 & - & GEn & 1.00 & - & - & GEn& GEn& GEn \tabularnewline
NGC5139 & oCen & GEn & 1.00 & - & - & GEn/Seq & Seq& - \tabularnewline
NGC6864 & M75 & GEn & 1.00 & - & - & GEn& GEn& - \tabularnewline
NGC5897 & - & GEn & 0.99 & Disc & 0.01 & GEn& GEn& - \tabularnewline
NGC6235 & - & GEn & 0.95 & Kraken & 0.05 & GEn& GEn& - \tabularnewline
Ryu879 & RLGC2 & GEn & 0.90 & Kraken & 0.09 & -& -& - \tabularnewline
BH140 & - & GEn & 0.90 & Disc & 0.06 & -& -& - \tabularnewline
NGC6656 & M22 & GEn & 0.85 & Disc & 0.15 & Disc& -& Disc \tabularnewline
NGC7078 & M15 & GEn & 0.83 & Disc & 0.17 & Disc& -& Disc \tabularnewline
IC1257 & - & GEn & 0.66 & Helmi & 0.34 & GEn& GEn& - \tabularnewline
NGC5904 & M5 & Helmi & 1.00 & - & - & Helmi/GEn & Helmi& GEn/Helmi  \tabularnewline
NGC4147 & - & Helmi & 1.00 & - & - & GEn& GEn& - \tabularnewline
NGC5634 & - & Helmi & 1.00 & - & - & Helmi/GEn & Helmi& - \tabularnewline
NGC5272 & M3 & Helmi & 1.00 & - & - & Helmi& Helmi& Helmi \tabularnewline
NGC5053 & - & Helmi & 1.00 & - & - & Helmi& Helmi& Helmi \tabularnewline
Pal5 & - & Helmi & 1.00 & - & - & Helmi/? & Helmi& Helmi \tabularnewline
NGC7492 & - & Helmi & 1.00 & - & - & GEn& GEn& - \tabularnewline
NGC5024 & M53 & Helmi & 1.00 & - & - & Helmi& Helmi& Helmi \tabularnewline
NGC6229 & - & Helmi & 1.00 & - & - & GEn& GEn& GEn \tabularnewline
NGC4590 & M68 & Helmi & 1.00 & - & - & Helmi& Helmi& Helmi \tabularnewline
NGC6981 & M72 & Helmi & 0.99 & GEn & 0.01 & Helmi& Helmi& - \tabularnewline
Rup106 & - & Helmi & 0.93 & Ungr & 0.07 & Helmi/? & Helmi& - \tabularnewline
NGC6584 & - & Helmi & 0.92 & GEn & 0.08 & Ungr& Ungr& - \tabularnewline
Bliss1 & - & Helmi & 0.88 & Ungr & 0.12 & -& -& - \tabularnewline
NGC6426 & - & Helmi & 0.73 & GEn & 0.27 & Ungr& Ungr& - \tabularnewline
NGC1904 & M79 & Helmi & 0.59 & GEn & 0.40 & GEn& GEn& GEn \tabularnewline
NGC6254 & M10 & Kraken & 1.00 & - & - & Kraken& Kraken& Kraken \tabularnewline
NGC6712 & - & Kraken & 1.00 & - & - & Kraken& Kraken& - \tabularnewline
NGC6544 & - & Kraken & 1.00 & - & - & Kraken& Kraken& Kraken \tabularnewline
NGC5946 & - & Kraken & 1.00 & - & - & Kraken& Kraken& - \tabularnewline
NGC6121 & M4 & Kraken & 1.00 & - & - & Kraken& -& Kraken \tabularnewline
NGC6809 & M55 & Kraken & 1.00 & - & - & Kraken& Kraken& Kraken \tabularnewline
NGC4833 & - & Kraken & 1.00 & - & - & GEn& GEn& - \tabularnewline
NGC6681 & M70 & Kraken & 1.00 & - & - & Kraken& Kraken& - \tabularnewline
NGC6287 & - & Kraken & 1.00 & - & - & Kraken& Kraken& - \tabularnewline
NGC5986 & - & Kraken & 1.00 & - & - & Kraken& Kraken& - \tabularnewline
NGC6541 & - & Kraken & 0.99 & - & - & Kraken& Kraken& - \tabularnewline
Terzan10 & - & Kraken & 0.99 & - & - & GEn& GEn& - \tabularnewline
NGC6752 & - & Kraken & 0.99 & GEn & 0.01 & Disc& -& Disc \tabularnewline
NGC6749 & - & Kraken & 0.99 & Disc & 0.01 & Disc& -& - \tabularnewline
NGC6760 & - & Kraken & 0.99 & Disc & 0.01 & Disc& -& Disc \tabularnewline
UKS1 & - & Kraken & 0.99 & GEn & 0.01 & -& -& - \tabularnewline
NGC6284 & - & Kraken & 0.99 & GEn & 0.01 & GEn& GEn& - \tabularnewline
Mercer5 & - & Kraken & 0.98 & Disc & 0.02 & -& -& - \tabularnewline
NGC6397 & - & Kraken & 0.98 & Disc & 0.01 & Disc& -& Disc \tabularnewline

 \hline  
    \end{tabularx}
    \end{center}
    \end{table}
    
 \begin{table}
	\centering
    \setlength\extrarowheight{1pt}
    \rowcolors{2}{gray!25}{white}
    \begin{center}
    \begin{tabularx}{\columnwidth}{l l c c c c c r r} 
    \hline
    \rowcolor{white}
    Name & Alternative & Main Group & Prob & Alt Group & Alt Prob & M19 & F20 & H20 \tabularnewline[.1cm] 
\hline  %
Terzan3 & - & Kraken & 0.98 & Disc & 0.02 & Disc& -& - \tabularnewline
FSR1716 & - & Kraken & 0.97 & Disc & 0.03 & Disc& -& - \tabularnewline
FSR1735 & - & Kraken & 0.97 & Bulge & 0.02 & Kraken& Kraken& - \tabularnewline
NGC6539 & - & Kraken & 0.97 & Disc & 0.02 & Bulge& -& Bulge \tabularnewline
Ton2 & Pismis26 & Kraken & 0.96 & Bulge & 0.03 & Kraken& Kraken& - \tabularnewline
Terzan12 & - & Kraken & 0.96 & Bulge & 0.03 & Disc& -& - \tabularnewline
NGC6402 & M14 & Kraken & 0.94 & Bulge & 0.05 & Kraken& Kraken& - \tabularnewline
Pal8 & - & Kraken & 0.94 & Bulge & 0.05 & Disc& -& - \tabularnewline
NGC6139 & - & Kraken & 0.94 & Bulge & 0.05 & Kraken& Kraken& - \tabularnewline
Djorg1 & - & Kraken & 0.94 & GEn & 0.05 & GEn& GEn& - \tabularnewline
NGC6553 & - & Kraken & 0.93 & Disc & 0.07 & Bulge& -& Bulge \tabularnewline
NGC6316 & - & Kraken & 0.92 & Bulge & 0.07 & Bulge& -& - \tabularnewline
NGC4372 & - & Kraken & 0.92 & Disc & 0.04 & Disc& -& - \tabularnewline
NGC6273 & M19 & Kraken & 0.90 & Bulge & 0.10 & Kraken& Kraken& - \tabularnewline
BH261 & AL3 & Kraken & 0.85 & Bulge & 0.12 & Bulge& -& - \tabularnewline
NGC6569 & - & Kraken & 0.84 & Bulge & 0.12 & Bulge& -& - \tabularnewline
NGC6333 & M9 & Kraken & 0.84 & GEn & 0.16 & Kraken& Kraken& - \tabularnewline
NGC6356 & - & Kraken & 0.83 & GEn & 0.15 & Disc& -& - \tabularnewline
Arp2 & - & Sag & 1.00 & - & - & Sag& Sag& - \tabularnewline
NGC6715 & M54 & Sag & 1.00 & - & - & Sag& Sag& - \tabularnewline
Terzan8 & - & Sag & 1.00 & - & - & Sag& Sag& - \tabularnewline
Terzan7 & - & Sag & 1.00 & - & - & Sag& Sag& - \tabularnewline
Pal12 & - & Sag & 1.00 & - & - & Sag& Sag& - \tabularnewline
Whiting1 & - & Sag & 1.00 & - & - & Sag& Sag& - \tabularnewline
Munoz1 & - & Sag & 0.99 & Ungr & 0.01 & -& -& - \tabularnewline
Kim3 & - & Sag & 0.97 & Ungr & 0.03 & -& -& - \tabularnewline
Ko1 & - & Sag & 0.71 & Ungr & 0.29 & -& -& - \tabularnewline
NGC5466 & - & Seq & 1.00 & - & - & Seq& Seq& Seq \tabularnewline
NGC6101 & - & Seq & 1.00 & - & - & Seq/GEn & Seq& - \tabularnewline
NGC7006 & - & Seq & 1.00 & - & - & Seq& Seq& - \tabularnewline
NGC3201 & - & Seq & 1.00 & - & - & Seq/GEn & Seq& Seq \tabularnewline
IC4499 & - & Seq & 1.00 & - & - & Seq& Seq& - \tabularnewline
Pal13 & - & Seq & 0.99 & Ungr & 0.01 & Seq& Seq& - \tabularnewline
NGC5694 & - & Seq & 0.98 & Ungr & 0.02 & Ungr& Ungr& - \tabularnewline
Pal15 & - & Seq & 0.96 & Ungr & 0.04 & GEn/? & GEn& - \tabularnewline
AM4 & - & Seq & 0.83 & Ungr & 0.15 & -& Sag& - \tabularnewline
Ryu059 & RLGC1 & Ungr & 1.00 & - & - & -& -& - \tabularnewline
Ko2 & - & Ungr & 1.00 & - & - & -& -& - \tabularnewline
Pal3 & - & Ungr & 1.00 & - & - & Ungr& Ungr& - \tabularnewline
NGC6934 & - & Ungr & 1.00 & - & - & Ungr& Ungr& - \tabularnewline
Crater & - & Ungr & 1.00 & - & - & Ungr& Ungr& - \tabularnewline
Pyxis & - & Ungr & 1.00 & - & - & Ungr& Ungr& - \tabularnewline
Segue3 & - & Ungr & 1.00 & - & - & -& -& - \tabularnewline
Pal14 & - & Ungr & 1.00 & - & - & Ungr& Ungr& - \tabularnewline
AM1 & - & Ungr & 1.00 & - & - & Ungr& Ungr& - \tabularnewline
Eridanus & - & Ungr & 1.00 & - & - & Ungr& Ungr& - \tabularnewline
Pal4 & - & Ungr & 1.00 & - & - & Ungr& Ungr& - \tabularnewline
Pal1 & - & Ungr & 1.00 & - & - & Disc& GEn& - \tabularnewline
NGC5824 & - & Ungr & 0.99 & Helmi & 0.01 & Sag& Sag& - \tabularnewline
NGC2419 & - & Ungr & 0.98 & Seq & 0.02 & Sag& Sag& - \tabularnewline
Laevens3 & - & Ungr & 0.94 & Seq & 0.06 & -& -& - \tabularnewline
Pal2 & - & Ungr & 0.81 & Seq & 0.14 & GEn& GEn& - \tabularnewline
FSR1758 & - & Ungr & 0.62 & Seq & 0.28 & Seq& Seq& - \tabularnewline
 \hline  
    \end{tabularx}
    \end{center}
    \end{table}

\twocolumn


\section{Gaussian Fitting of Small Groups}
\label{Appendix:DegenFit}
In our model, in principle, all data points contribute to each component,
although some points can have very low responsibilities.
On average, each component fits $N_{\mathrm{points}} = W_{c}\times N_{\mathrm{total}}$ points,
where $N_{\mathrm{total}}$ is the total number of GCs.
If the weight of the component is such that $N_{\mathrm{points}}<n_{\mathrm{dim}}$,
where $n_{\mathrm{dim}}$ is the number of dimensions of the space,
then $\Sigma$ tends to become degenerate within machine precision.
This causes the responsibility of the GCs to tend to one and the fit is unable to improve.
To prevent this, after calculating the covariance matrix, we change the
$\lfloor n_{\mathrm{dim}}-N_{\mathrm{points}} \rfloor$ smallest eigenvalues to half of the smallest non-degenerate value.
If $N_{\mathrm{points}}$ drops below $1.5$, we set the eigenvalue to be $0.05$
(note that internally the space is scaled by the $25-75\%$ range to be dimensionless).
If $N_{\mathrm{points}}$ drops below $0.5$, the cluster is then considered extinct,
and the normalisation weight is set to zero.
We note that the weights of the MW groups generally do not decrease sufficiently when
performing the multi-component fit to cause this issue.
This affects only a few groups from the mock samples,
but is nevertheless important to include to accurately fit the groups.

\Change{\section{Misclassifying the Initial Groups}}
\label{Appendix:InitialGroups}

\Change{Here we study how sensitive our GC grouping algorithm is on the initial groupings
used as the starting point of our iterative method.
We have explored this by selecting a subset of GCs and by changing their label to another group.
When misclassifying GCs in observations, they are not assigned to a random group,
but actually to a neighbouring group.
To identify in a simple way the closest incorrect neighbour for each GC we proceed by calculating
the best fitting distributions in a non-iterative way.
This corresponds to applying the maximization step of our algorithm,
where the responsibilities are calculated using the true GC labels,
and then updating the responsibilities using these new best fitting distributions.
Then, we assign to the misclassified GCs the label associated to the most likely group that was not their true group.
We study what is the impact of such an initial mislabelling of GC groups as a function of the misclassified fraction.
In particular, we are interested in testing the effect of starting with  33\% of the GCs incorrectly identified, approximately in line with the average final purity and completeness of our methodology.
}

\begin{figure}
\centering
\includegraphics[width=.9\columnwidth]{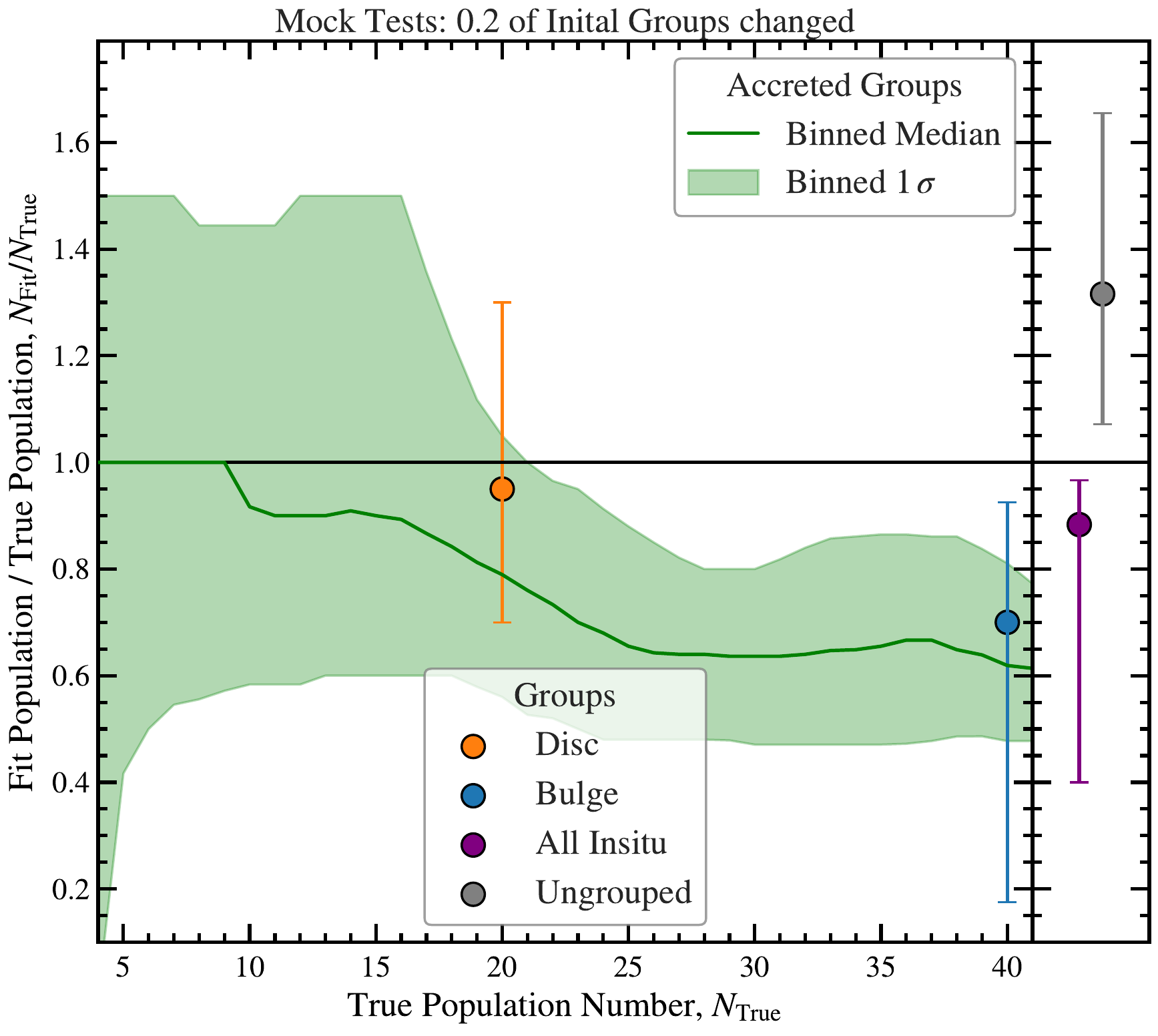}
\includegraphics[width=.9\columnwidth]{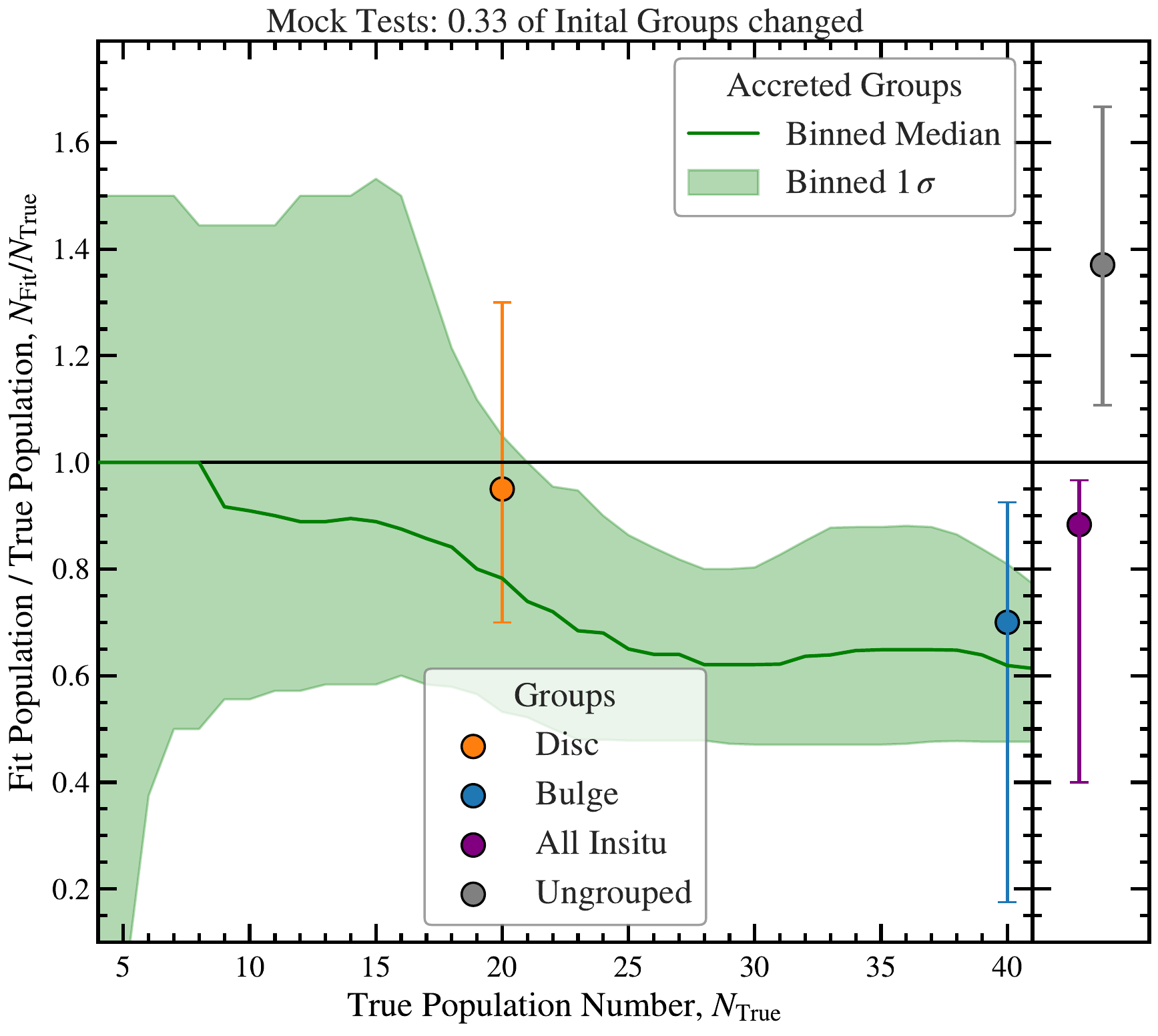}
\includegraphics[width=.9\columnwidth]{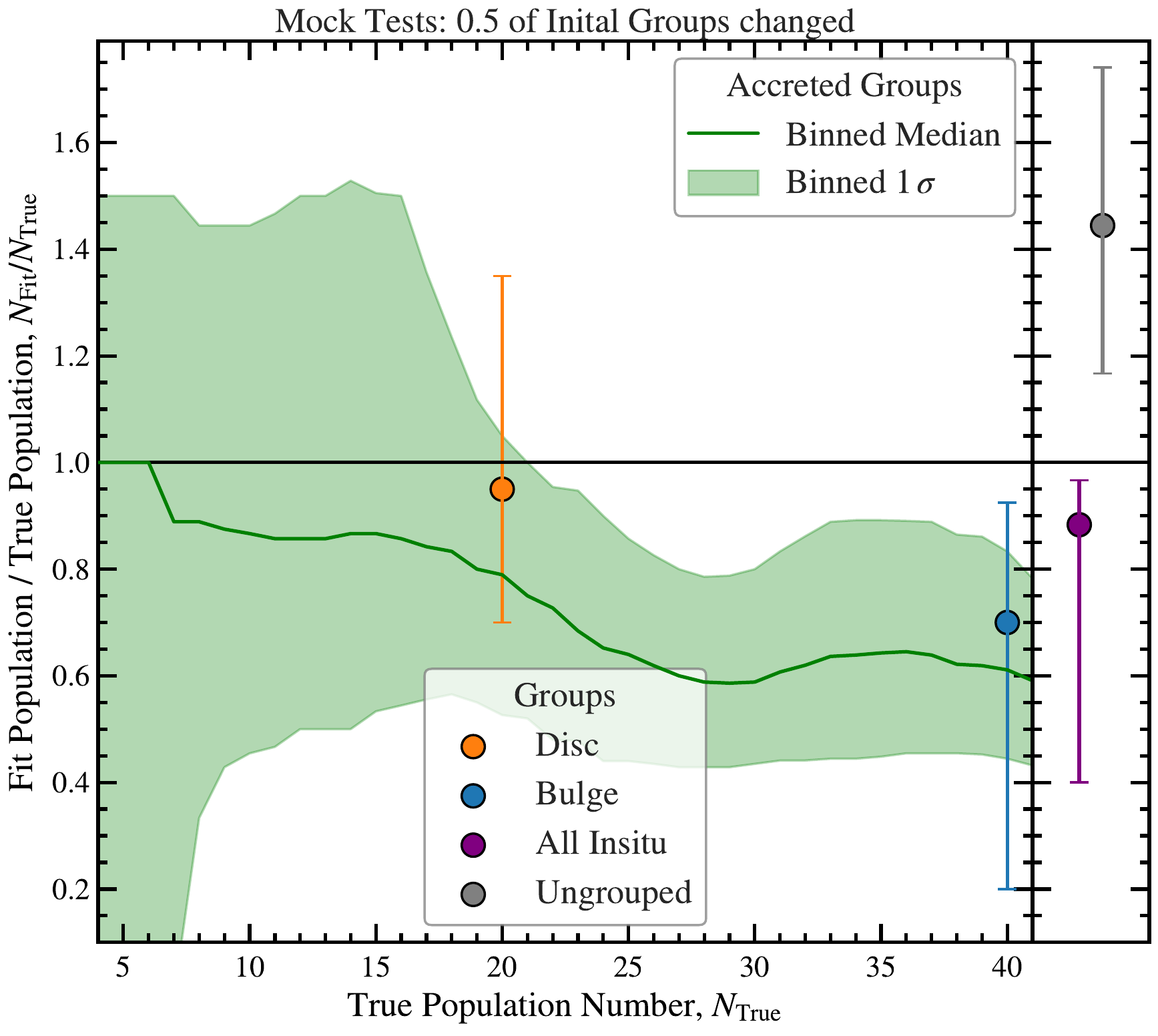}
    \caption{
    \Change{Like Fig. \ref{fig:Mock_NtrueNfit},
	    but now showing how the ratio $N_{\mathrm{fit}}/N_{\mathrm{true}}$ as a function of $N_{\mathrm{true}}$
	    depends on the fraction of mislabelled GCs present in the groupings used to
	    initialise our iterative clustering algorithm.
	    The tests are done on the \auriga{} mock GC catalogues.
    In each panel, from top to bottom, we reassign a fraction of respectively 0.2, 0.33, and 0.5 of the GCs from their true group to the next closest group.}
	}
	\label{fig:InitialCompare}
\end{figure}

\Change{We redo most of the analysis presented in Section \ref{sec:Testing_on_mocks}
but now starting from initial labels that have various fractions of misclassified GCs.
To start with, we study how the ratio, $N_{\mathrm{fit}}/N_{\mathrm{true}}$,
of the recovered to true group richness changes when mislabelled 20, 33 and 50\% of the GCs.
This is shown in Fig. \ref{fig:InitialCompare}.
We can see that increasing the fraction of GCs with incorrect initial groups increases the error spread gradually.
The greatest effects can be seen in the smallest groups.
This makes intuitive sense:
changing a single GC in a poor group represents a more significant change than for a richer group.
Furthermore, this large change can lead to the group no longer being well modelled in dynamical space,
potentially leading the group to go extinct as the method iterates.
The larger groups are, in general, robust to even large changes,
as enough of the true members remain for the average position of the group to be found and the majority of the group recovered.}

\Change{However, it is not until the change reaches the 50\% level that the median trend is changed,
and then only for small  
groups.
For a 33\% misclassification fraction, roughly the uncertainty resulting from our clustering process,
there is little change in the distribution of $N_{\mathrm{fit}}/N_{\mathrm{true}}$
compared to the case of the true starting groups.
While not shown, the trends in recovered group purity and completeness are similarly robust to initially mislabelled GCs,
especially for a 33\% or lower misclassification fraction.}

\Change{To conclude, this shows that our methodology is insensitive to potential misclassifications as large as 33\%
(and even 50\%) of the GC groups used to initialise our clustering algorithm.
Whilst the test performed here is not a direct equivalent to any mislabelling present in the
literature groupings used to initialise our method when applied to the MW data,
we believe that it indicates that our methodology and results are robust in application to the MW.
}

\section{Redshift Dependence of the SMHM}
\label{Appendix:zSMHM}
\begin{figure}
\includegraphics[width=\columnwidth]{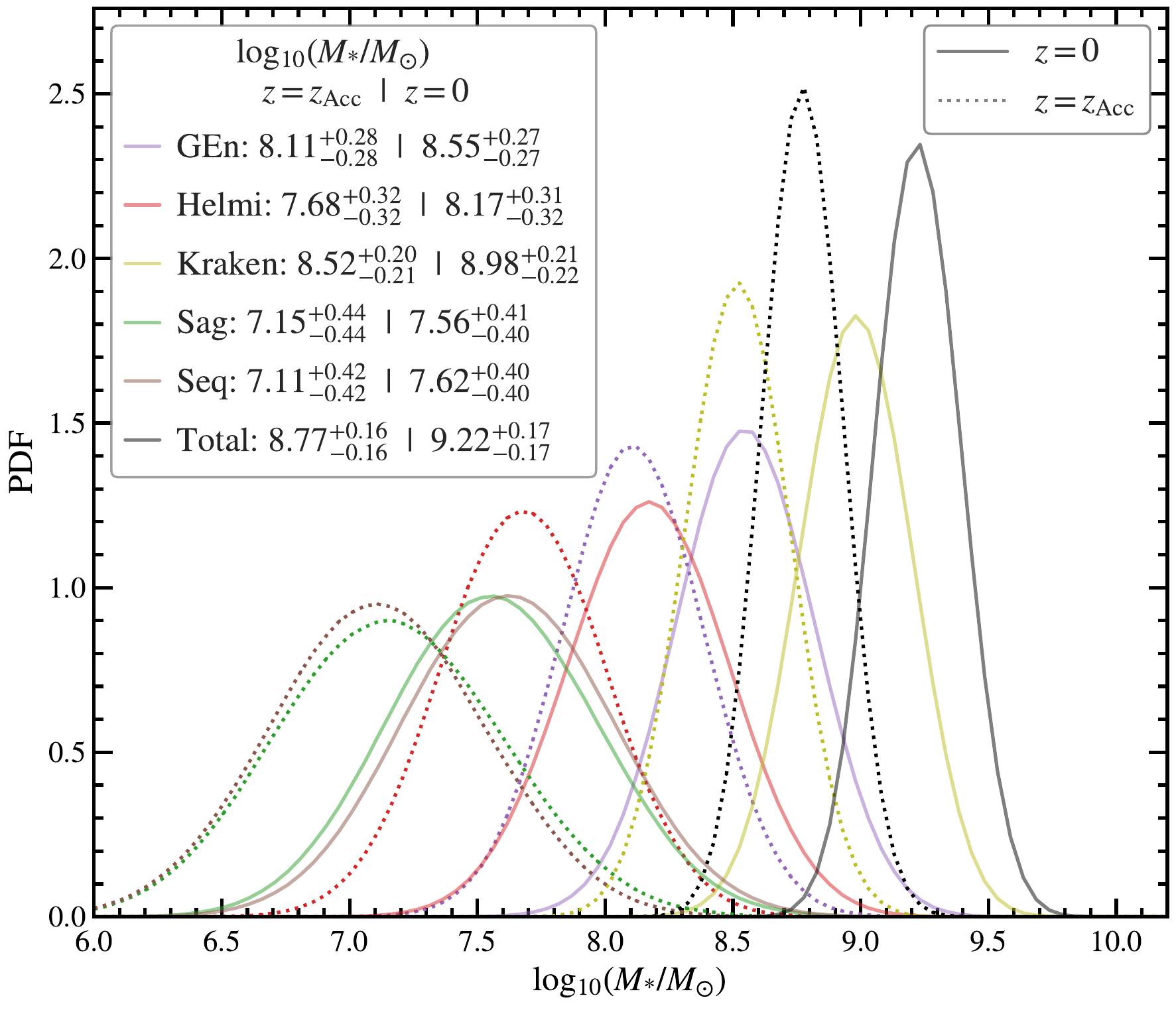}
\smallspace{}
\caption{
	The likelihoods of the stellar mass of accreted galaxies,
	calculated by assuming the  stellar mass-halo mass relation of \protect\cite{behroozi_universemachine_2019}
	to transform the halo mass PDFs of Fig.~\ref{fig:MW-Mass}.
	This relationship is dependent on redshift.
	In this figure, we compare the results found by assuming the present day, $z=0$,
	relationship (as shown in Fig.~\ref{fig:MW-StellarMass}).
	We use estimates of the approximate accretion time of the groups found in the literature (see the main text).
	Due to the large systematic uncertainties on these accretion times,
	we consider these results as demonstrative of the effect of redshift dependence.
	}
	\label{fig:MW-StellarMass_z0}
\end{figure}

In the main paper, when using the SMHM relation of \cite{behroozi_universemachine_2019}, we assumed the present day,
$z=0$, relation (see Fig.~\ref{fig:MW-StellarMass}).
Alternatively, it is reasonable to assume that star formation in the accreted satellites approximately stops upon accretion.
We consider the approximate infall time estimates from \cite{kruijssen_kraken_2020}:
Kraken, $z_{\mathrm{Acc}}=2.26$; Helmi Streams, $z_{\mathrm{Acc}}=1.75$;
Sequoia, $z_{\mathrm{Acc}}=1.46$; \GES{}, $z_{\mathrm{Acc}}=1.35$; and Sagittarius, $z_{\mathrm{Acc}}=0.76$.
The resulting PDFs, and original $z=0$ estimates, can be seen in Fig. C1. 

The effect of truncating the star formation histories in accreted galaxies lowers the inferred stellar masses.
This truncation has the greatest effect on older mergers, such as Kraken (with approximately a three times decrease in stellar mass).
The total mass is approximately reduced by a factor of $2.5$.
This effect is considerable
but depends on the highly uncertain accretion times and the considerable systematic uncertainty of the SMHM relation at high redshift.
\bsp
\label{lastpage}
\end{document}